\documentclass[5p,times]{elsarticle}
\usepackage{amssymb}
\usepackage{amsmath}
\usepackage{lscape}
\usepackage{makecell}
\usepackage{flushend}
\usepackage{microtype}

\journal{New Astronomy Reviews}

\usepackage{ifthen}
\usepackage{upgreek}
\usepackage{nicefrac}
\usepackage{xspace}
\usepackage{physics}
\usepackage{changes}
\usepackage{bm}
\usepackage{ulem}
\usepackage{url} 
\usepackage{hyperref}
\usepackage{breakurl} 
\usepackage{aas_macros}

\newcommand{\unit}[1]{\ensuremath{\,\mathrm{#1}}}

\newcommand{\ergs}{\ensuremath{~\mathrm{erg\,s^{-1}}}}

\newcommand{\be}{\begin{equation}}
\newcommand{\ee}{\end{equation}}
\newcommand{\ba}{\begin{eqnarray}}
\newcommand{\ea}{\end{eqnarray}}

\newcommand{\ve}{\ensuremath{\varepsilon}}

\newcommand{\mysub}[1]{\ensuremath{_{\mathrm{#1}}}}

\newcommand{\myerror}[2][NONE]{%
  \ifthenelse { \equal {#1} {NONE} } %
  {\ensuremath{\pm #2}}%
  {\ensuremath{_{-#1}^{#2}}}%
}

\newcommand{\rg}{\ensuremath{r_{\mathcal{G}}}}
\newcommand{\rsch}{\ensuremath{R_{\mathcal{SCH}}}}
\newcommand{\D}{\ensuremath{\mathcal{D}}}
\newcommand{\graya}{gamma-ray\xspace}
\newcommand{\gray}{gamma rays\xspace}

\newcommand{\hs}{H.E.S.S.\xspace}
\newcommand{\mg}{MAGIC\xspace}

\definecolor{dg}{rgb}{0.0, 0.6, 0.1}
\definecolor{ed}{rgb}{1.0, 0.6, 0.1}

\definechangesauthor[name={Mitya}, color=orange]{DK}
\definechangesauthor[name={Dani}, color=dg]{DH}

\makeatletter
\def\Dani{\def\xx{DH}\@ifnextchar[{\@mwith}{\@mwithout}}
\def\Mitya{\def\xx{DK}\@ifnextchar[{\@mwith}{\@mwithout}}
\def\@mwith[#1]#2{\replaced[id=\xx]{#2}{#1}}
\def\@mwithout#1{\added[id=\xx]{#1}}
\makeatother

\graphicspath{{./}{figures/}}
\begin{document}

\fontsize{9pt}{11pt}\selectfont

\begin{frontmatter}
\title{Extreme Transients in Gamma Rays} 
\author[ICE]{Daniela Hadasch}
\author[IHEP,CRRC]{Dmitriy Khangulyan}
\affiliation[ICE]{organization={Institute of Space Sciences (ICE-CSIC)},
            addressline={Campus UAB, Carrer de Can Magrans s/n},
            city={Cerdanyola del Vallés},
            postcode={E-08193},
            state={Catalonia},
            country={Spain}}
\affiliation[IHEP]{organization={Institute of High Energy Physics, Chinese Academy of Sciences},
            addressline={19B Yuquan Road},
            city={Shijingshan District},
            postcode={100049},
            state={Beijing},
            country={PRC}}
\affiliation[CRRC]{organization={Tianfu Cosmic-Ray Research Center},
            addressline={Kezhi Road 1500},
            city={Chengdu},
            postcode={610213},
            state={Sichuan},
            country={PRC}}
\begin{abstract}
Extreme \graya transients represent some of the most energetic and physically constraining phenomena in high-energy astrophysics. They are characterized by rapid, large-amplitude variability and by physical conditions approaching fundamental limits on particle acceleration, cooling, and compactness. In this review, we focus on transients detected above $\sim 100$~MeV and define \emph{extreme} events as either those involving catastrophic transformations of astrophysical systems (such as stellar explosions, compact-object mergers, and tidal-disruption events) or those exhibiting evidence for particle acceleration operating in an extreme regime. These systems are powered by the rapid release of gravitational, magnetic, nuclear, or kinetic energy, with shocks and magnetic reconnection playing a central role in producing ultra-relativistic particle populations and non-thermal radiation.

We summarize observational and theoretical diagnostics that constrain the size, magnetization, and Lorentz factor of the emitting region, including variability timescales, luminosity--timescale correlations, and spectral evolution across the MeV--TeV domain. We further review the complementary capabilities of space-borne \graya instruments, ground-based Cherenkov and air-shower observatories in detecting short-lived, high-energy outbursts. Extreme transient classes discussed include \graya bursts, novae, rapidly variable emission from extragalactic and Galactic jets. Also, because of its extreme aspects, we include flaring emission detected from the Crab Nebula. While each type of these flares poses interesting challenges for phenomenology and theory of these sources,  together, these events form the landscape of extreme \graya variability. 
\end{abstract}
\end{frontmatter}
\section{Introduction} \label{sec:intro}\nopagebreak
\subsection{Scope of this review}
While the detection and study of \graya transients started with the onset of \graya astronomy \citep[see, e.g.,][]{1997asxo.proc...21G,2001ApJ...559..187K}, the most robust results have been obtained over the last two decades with the Fermi space telescope and the current generation of Cherenkov experiments, \hs, \mg, and VERITAS. With the first systematic detections, several types of sources emerged as standard sites of \graya transient activity, most notably Gamma-Ray Bursts (GRBs), relativistic jets from Active Galactic Nuclei (AGN), and, to some extent, novae. In other cases -- most prominently in binary systems -- transient emission has so far been detected only from a few individual representatives \cite{2007ApJ...665L..51A,2009Sci...326.1512F,2010ApJ...712L..10S,2011ApJ...736L..11A,2016A&A...596A..55Z}, which has led primarily to modeling of these specific events rather than to the establishment of well-defined new classes of transients.

As a result, the present literature treats some transient classes in great detail. This includes, for example, observational and theoretical reviews of GRBs \cite{2024Univ...10...57V}, summaries of AGN flaring activity (e.g., \cite{2018MNRAS.477.4257C}), and comparisons of competing theoretical models (e.g., \cite{2017ApJ...841...61A}). Other transient phenomena have also been discussed in dedicated reviews, often motivated by similarities in their phenomenology to more established classes. For instance, GRBs and tidal disruption events share certain observational characteristics (see, e.g., \cite{2017SSRv..207...63W}), while novae have attracted attention as scaled-down analogues of supernova explosions (e.g., \cite{2021ARA&A..59..391C}). These works have provided detailed observational summaries (see in particular \cite{2024Univ...10..163C}) and, in many cases, source-specific theoretical interpretations.

The phenomenology and modeling of individual transient sources can be highly complex. However, the most extreme transients often exhibit clear signatures that the physical processes responsible for the detected emission operate close to fundamental limits. Consequently, many of the challenges in interpreting these events can be addressed at a more basic level by considering general constraints that cut across traditional source classifications. In particular, extreme \graya transients are increasingly recognized as events that approach fundamental physical limits imposed by causality, particle acceleration rates, radiative cooling, energy density, and intrinsic opacity, rather than being defined solely by their astrophysical origin.

The goal of this review is therefore not to provide an exhaustive census of transient source classes, but to synthesize observational and theoretical results from a constraint-driven perspective. We focus on how variability timescales, luminosities, spectral cutoffs, and multi-band correlations can be used to infer the size, magnetization, and Lorentz factor of the emitting region, and to identify regimes in which particle acceleration and energy dissipation operate close to their theoretical limits.

In this sense, our review complements existing source-oriented studies by emphasizing the common physical framework underlying diverse extreme events -- from gamma-ray bursts and novae to flares in relativistic jets and pulsar wind nebulae -- and by highlighting the role of current and next-generation gamma-ray instruments in probing these limits.

In the following, we therefore introduce the physical criteria that define \emph{extreme} \graya transients, focusing on events detected at energies $\gtrsim 100,\mathrm{MeV}$ and on the fundamental limits imposed by causality, particle acceleration, radiative cooling, energy density, and intrinsic opacity.

\subsection{Defining extreme gamma-ray transients}
Astronomical transients are events characterized by a sudden and significant change in brightness or spectral properties, occurring over a timescale much shorter than the typical variability of the source before and after the event. These events are often associated with a catastrophic transformation of the object, for example a merger of a binary system, or a rapid release of energy that was accumulated over a period much longer than the transient event. This distinguishes transients from variable sources that exhibit regular periodicity or long-term variability patterns. In practice, the ability to detect transients is constrained by the cadence, sensitivity, and coverage of observational surveys. While transients can be identified purely through observational signatures, defining \emph{extreme transients} requires additional phenomenological and theoretical considerations. In this manuscript, we focus on extreme transients~---~events that either involve a catastrophic transformation of the astrophysical object or provide evidence that the underlying physical processes operate in an extreme regime. Furthermore, we limit our consideration to transients detectable in the high-energy \graya band, \(\hbar \omega\gtrsim 100\unit{MeV}\).

Transients involving catastrophic transformations of astrophysical objects include various manifestations of stellar collapse \cite{1993apj...405..273w}, binary mergers \cite{1989Natur.340..126E}, and, to some extent, tidal disruption events (TDEs) \cite{1975Natur.254..295H,1988Natur.333..523R}. In the first two cases, the emitting object itself undergoes a fundamental transformation~---~either by collapsing into a compact remnant or merging into a more uniform object. In contrast, during a TDE, the black hole responsible for the disruption remains essentially unchanged; the catastrophic transformation affects only the star that ventures too close. The gravitational energy of the disrupted stellar debris is rapidly released and powers the TDE transient. This energy was stored in the system, as the gravitational energy of the star, for an extremely long period and is suddenly liberated due to a coincidental close encounter~---~making TDEs conceptually similar to other transients powered by a rapid release of energy accumulated over long timescales.

For example, recurrent nova explosions are driven by nuclear energy stored in hydrogen gas that accumulates on a white dwarf over extended periods \cite{1972ApJ...176..169S}. Once ignition occurs, the hydrogen envelope is expelled, but the white dwarf remains largely unaffected. In addition to gravitational and nuclear energy, magnetic energy can also power extreme transients. The most dramatic realization of this occurs in magnetar giant flares, which involve large-scale reconfiguration of the magnetospheric field and potentially a structural transformation in the neutron star's crust \cite{1995MNRAS.275..255T}. Although the magnetar itself survives the flare, its internal state or the configuration of its magnetosphere may be significantly altered.

Releases of gravitational energy during episodes of enhanced accretion, magnetic energy through reconnection events, and bulk kinetic energy at shock fronts are commonly associated with astrophysical transients. In the context of \graya transients, the latter two processes --magnetic reconnection and heating at shock waves-- are particularly relevant, as they not only facilitate rapid energy release but are also closely tied to particle acceleration \cite{1965JGR....70.4219S,1978MNRAS.182..147B}. These mechanisms generate non-thermal particle distributions, often transferring a significant fraction of the released energy to ultra-relativistic particles. Such conditions are essential for producing high-energy photons and are a prerequisite for the detection of a transient event in the \graya band.

\subsection{Extreme accelerators}
In the MeV band, one may still expect a significant contribution from non-relativistic particles, for example, nuclei heated at supernovae blast wave. However, higher-energy \gray are typically produced by ensembles of relativistic particles, which follow non-thermal (i.e., thermodynamically non-equilibrium) energy distributions. Even in the case of ultra-relativistic outflows --whose thermal emission can be Doppler-boosted to higher frequencies-- a non-thermal component is usually present. This implies that particle acceleration is an essential ingredient of \graya transients.

Spectral and temporal data obtained during a transient event, together with a physical model of the source, often allow one to constrain properties of the underlying acceleration mechanism. Based on the revealed properties a transient may appear as an event powered by an \emph{extreme accelerator}.

A key characteristic of any acceleration process is the energy gain rate\cite{1984ARA&A..22..425H}, which is conventionally expressed as the acceleration time:
\be
t\mysub{acc}=\frac{E}{\dot{E}}\,.
\ee
Here, \(E\) is the particle energy, and \(\dot{E}\) is the average rate of energy change for particles with energy \(E\) in the acceleration region. The energy gain rate for any acceleration process depends on the macroscopic physical conditions in the acceleration site and by the specific realization of microscopic processes that lead to increasing particle energy. There are some fundamental arguments, which define the maximum possible rate for given physical conditions in the source. If the anticipated acceleration rate during a transient event approaches this limit, we expect the event is powered by an \emph{extreme accelerator}.

Among the critical parameters that limit the acceleration rate are the accelerator size, \(\mathcal{R}\), and its magnetic field, \(\mathcal{B}\). Magnetic fields are believed to be a fundamental component of space plasmas. However, the magnetic force does not change the particle’s energy. Energy gain arises from electric fields, \(\mathcal{E}\), as the Coulomb (electric) force can perform work on the particle. In the electromagnetic context, the instantaneous rate of energy change for a particle is given by:
\be
\dot{\gamma}=\frac{q}{mc^2}{\bm v}\cdot{\mathcal{\bm E}}\,,
\ee
where \(m\) and \(q\) are the particle’s mass and charge; \(\bm v\) and \(\gamma\) its velocity and Lorentz factor.  As different particles can interact with electric field of different strength and may have different directions of their velocities,  in the astrophysical context, this rate should be averaged over the ensemble of particles having the same energy in the accelerator: 
\be
\dot{E}=\overline{mc^2\dot{\gamma}}=q\overline{{\bm v} \cdot {\mathcal{\bm E}}}\,.
\ee

\begin{figure*}[t]
\centering
\includegraphics{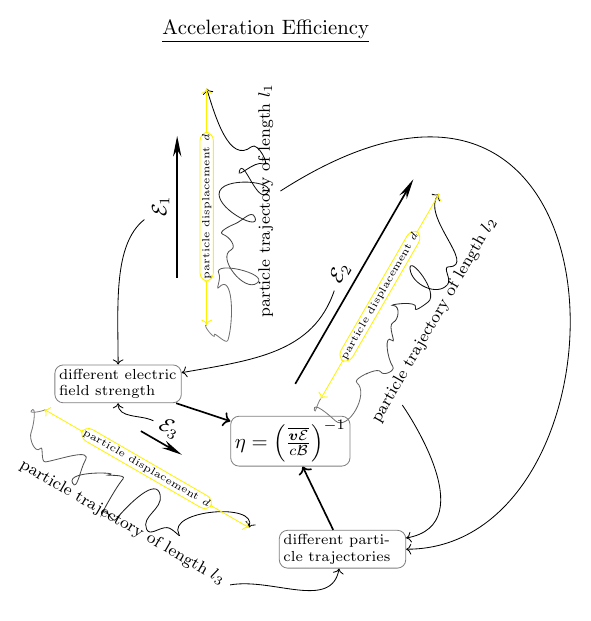}
\caption{A sketch that illustrates the factors contributing to the efficiency of the particle acceleration,  often defined via a phenomenological parameter \(\eta\).}\label{fig:accel_sketch}
\end{figure*}

In space plasmas, the electric field is often not a convenient or stable parameter because it quickly vanishes in the co-moving
frame of the plasma due to high conductivity. In contrast, the magnetic field in the co-moving frame is a robust and
commonly used characteristic. For this reason, acceleration time is often expressed in terms of the magnetic field
strength:
\be
t\mysub{acc}=\eta \frac{\rg}{c}
\ee
where \(\rg=E/(q\mathcal{B})\) is the particle gyro-radius in the source typical magnetic field, and \(\eta\) is a dimensionless parameter quantifying the efficiency of acceleration:
\be
\eta = \qty(\frac{\overline{{\bm v} \cdot {\mathcal{\bm E}}}}{c\mathcal{B}})^{-1}\,.
\ee
This parameter reflects the ratio of the effective accelerating electric field, \(\mathcal{E}\), to the magnetic field \(\mathcal{B}\), as well as the average angle between the particle trajectory and the direction of \(\mathcal{\bm E}\) (see sketch in Fig.~\ref{fig:accel_sketch}).

In most astrophysical environments, we expect the electric field to be smaller than the characteristic magnetic field, \(\mathcal{E}<\mathcal{B}\). Also particles move along complex trajectories such that their velocity directions change strongly relative to the electric field, thus  averaging yields \(\overline{\bm v \cdot\mathcal{\bm E}}\ll c\mathcal{E}\). Therefore, the acceleration efficiency parameter \(\eta\) is typically large, \(\eta\gg1\).

If the observed variability or spectral properties of a transient imply an acceleration process with a relatively small \(\eta\) parameter --~say, \(\eta\lesssim10^2\)~-- this can be interpreted as evidence for the operation of an \emph{extreme acceleration process} and the event can be considered as an \emph{extreme transient}.

The determination of the parameter $\eta$ from first principles is a challenging task, as it requires detailed knowledge
of the acceleration mechanism and of the physical conditions at the emission site (see \cite{1983RPPh...46..973D} for a
consideration of the diffusive shock acceleration case). Under certain circumstances---in particular when the maximum
particle energy is set by the balance between the acceleration rate and synchrotron cooling---the position of the cutoff
in the synchrotron spectrum is determined by the acceleration efficiency and becomes independent of the magnetic-field
strength. For simplicity, let us assume that the electrons interact with a magnetic field of fixed strength
$\mathcal{B}$ and with random pitch angles \cite{2010PhRvD..82d3002A}. In this case, the synchrotron cooling time is
given by

\be
  t\mysub{syn} = \frac{E_e}{\abs{\dot{E}\qty(E_e)}} = \frac{3}{4}\frac{m_e^2c^3}{\sigma\mysub{T} E_e ({\cal B}^2/8\pi)}\,,
\ee
where \(m_e\) and \(\sigma\mysub{T}\) are electron mass and Thomson cross-section.
Balancing this time scale to the acceleration time, one obtains the maximum energy of electrons:
\be
E\mysub{max} = \frac{3}{2}m_ec^2\sqrt{\frac{m_e^2c^4}{e^3{\cal B}\eta}}\,,
\ee
where \(e\) is electron charge.
For this electron energy, the critical frequency, \(\omega_c\),  appear to be independent of the magnetic field strength for this electron energy:
\be
\hbar \omega_c = \frac{3}{2} \frac{ E_e^2 e {\cal B}\hbar}{m_e^3c^5} = \frac{27}{8} \frac{m_ec^2}{\alpha \eta}= \frac{236\unit{MeV}}{\eta}\,,
\ee
here $\alpha$ denotes the fine-structure constant and \(\hbar\) is Planck constant. In the case of emission in a
chaotic magnetic field in the co-moving frame, the spectral maximum appears at
$0.23\,\hbar\omega_{\rm c} \approx 50~\mathrm{MeV}/\eta$, while the maximum of
the spectral energy distribution occurs at
$1.15\,\hbar\omega_{\rm c} \approx 270~\mathrm{MeV}/\eta$. For the limiting value
$\eta \rightarrow 1$, these limits are referred to as the synchrotron burn-off
limits; an extension of the synchrotron component up to this regime indicates
the operation of an extremely efficient particle-acceleration process.

\subsection{Extremely fast variability}
Another fundamental reason to classify an event as an extreme transient is its duration when it approaches fundamental limits. Among such limits, the most basic one is the \emph{light crossing time}. For a source with size \(\mathcal{R}\), the minimum timescale is \(\mathcal{R}/c\). There are several lines of arguments that help to obtain constraints on the source size. For example, typically the source size should be significantly larger than the gyro-radius of emitting particles: \(\mathcal{R}\gg \rg\), thus we generally expect that  \(t\mysub{var}\gg \rg/c\). If the variability timescale approaches this limit \(t\mysub{var}\sim \rg/c\), the transient can be considered as an extreme one. As it was illustrated above, the ratio of the particle gyro-radius to light speed can be treated as the shortest acceleration time. Thus this constraint on the variability is tightly connected to the limitation arising from the maximum acceleration rate. 

The gyro-radius is not the only length constraint for transient events. Transients are often linked to outflows launched by accreting black holes. In this case, the characteristic hydrodynamic scale is the Schwarzschild radius, \(\rsch\), of the black hole, and one may expect that this scale is preserved also in the emission detected during related transients: \(t\mysub{var}\gg \rsch/c\) \cite{1974ApJ...192L...3E}. While following detection of ultra-fast flares from AGN jets with \hs and \mg several different scenarios were suggested to interpret this type of phenomena, there is still no consensus on that matters. Thus, transients which show variability on a scale comparable to the black hole horizon light crossing time,  \(t\mysub{var}\lesssim\rsch/c\), should be considered as transients with \emph{extremely short} variability.

Similarly, transient can be classified as an extreme if the variability time scale approaches the anticipated cooling time, \(t\mysub{var}\lesssim t\mysub{cool}\), where \(t\mysub{cool}\) is radiative or non-radiative (adiabatic, escape, decay, etc) time scale for particles responsible for the emission.

\subsection{Extreme energetics}
Another fundamental aspect, which needs to be taken into account, in the context of the size of the region producing the transient emission is its energy content \cite{2008MNRAS.384L..19B}. The variability time scale and the characteristic luminosity, \(L\), determine a lower limit on the dissipated energy, \(E\mysub{dis}\gtrsim L t\mysub{var}\). As the volume of the production region is constrained by  the variability time, \(\mathcal{V}\mysub{dis}\sim\mathcal{R}^3\lesssim \qty(ct\mysub{var})^3\), we obtain a lower limit on the density of the dissipated energy
\be
w\mysub{dis} \gtrsim \frac{L}{c^3t\mysub{var}^2}\,.
\ee
Since only a modest part of the energy can be transferred to the emission, typical energy density of the source, \(w_0\), implies a lower limit on the transient variability time:
\be
t\mysub{var}\gg \sqrt{\frac{L}{c^3w_0}}\,.
\ee
Transients, in which the variability time scale approaches this limit, may require unrealistic assumptions regarding the energy content, thus should be considered as \emph{extreme transients}.

The above estimate tests the observational properties of a transient, such as its variability time scale and luminosity, against a phenomenological parameter, namely the energy density at the production site. There is a related constraint, which, however, does not depend on any phenomenological assumptions. The constraint arises from the intrinsic transparency of the production site\cite{1983MNRAS.205..593G}. The two-photon pair production cross section reaches its maximum value of \(0.2\sigma\mysub{T}\) when the product of the energies of the interacting photons is approximately \(0.9\unit{MeV^2}\). The reaction threshold lies approximately a factor of four below this value,  and above the maximum the cross section decreases approximately as inverse product of energies. The constraint imposed by production region transparency is least model dependent when the detected photons are in the MeV band, thus the attenuation occurs on the very same photons. In this case, the density of the photons should satisfy
\be
\sigma\mysub{T}\mathcal{R}n\mysub{ph}\lesssim 5\,,
\ee
and the flare luminosity thus
\be
L\sim 4\pi \mathcal{R}^2c n\mysub{ph}\ve \ll \frac{20\pi c^2 t\mysub{var}\ve}{\sigma\mysub{T}}\,,
\ee
where \(\ve\approx1\unit{MeV}\) is photon energy. Solving this relation for the shortest variability time one obtains 
\be
t\mysub{var}\gg \frac{\sigma\mysub{T}L}{20\pi c^2 \ve}\,.
\ee
We note that if a flare is detected at energies noticeably different from MeV, then one needs to implement an assumption regarding the flare spectrum at lower energies to correctly evaluate the opacity.

\subsection{Relativistic effects}\label{sec:intro_rel}
Variability and apparent luminosity of transients can be strongly affected by relativistic effects. While for certain constraints relativistic motion alone does not provide a panacea, in some other cases it may help relax the requirements for the transient production site. Among the former case, it is believed that outflow relativistic motion has a small impact on the rigidity of the constraint imposed by the black hole horizon crossing time --- as the physical scale is determined in the frame where the black hole is at rest (thus one needs to correct only for the cosmological redshift). By contrast, the estimate for the energy density is very sensitive to the relativistic motion: for a blob-type production site \(L\mysub{ap}=\D^4L\) and \(t\mysub{ap}=t\mysub{var}/\D\) (here \(\D\) is Doppler factor, \(L\mysub{ap}\) and \(t\mysub{ap}\) are apparent luminosity and variability), thus \(w\mysub{dis}\propto  L\mysub{ap}/\qty(\D^3t\mysub{ap})^2\). As in astrophysical outflow the Doppler factor can be very significant, \(\D\sim 10^2\), the relaxation of the requirements for the energy density is drastic, even if one accounts for \(w\mysub{dis}\sim w_0/\D^2\).  

Constraints on the intrinsic opacity of the transient site depend strongly on the relativistic motion:
\be
t\mysub{var}\gg \frac{\sigma\mysub{T}L}{20\pi c^2 \ve \D^4}\,,
\ee
where one also needs to account for the shift of the threshold energy by a factor of \(\D\). For sources featuring ultra-relativistic outflows, \(\D\gtrsim10^2\), photons detected at energies \(\ve \approx 100 \qty(\D/10^2)\unit{MeV}\) can be used to constrain the size without involving additional assumptions. 

\subsection{Gamma-ray transients on cosmological distances}
Observations of AGN flares have already put constraints on the size of the transient ``active zone'' comparable to the Schwarzschild radius of the central black hole \cite{2007ApJ...664L..71A,2014A&A...563A..91A}.
Lower-energy instruments can be used to study black hole systems on that scale in our Galaxy and a few nearby supermassive black holes (SMBHs). However, \graya observations offer a unique opportunity to investigate processes near SMBHs even in sources at cosmological distances. If one adopts an \(\ve^{-2}\) spectrum in the range from \(\ve\mysub{min}\) to \(\ve\mysub{max}\) and the Eddington luminosity as the reference value,
\be
L\mysub{edd} = 1.4\times10^{44}\qty(\frac{M\mysub{bh}}{10^6M_\odot})\unit{erg\,s^{-1}}\,,
L\mysub{edd} = \frac{4\pi G m_pc}{\sigma_T} M\mysub{bh}
\ee
then from a source at a distance of \(d\) the photon flux per unit time/energy/area (\(S\)) is
\be
\begin{split}
  \dv{N}{S\dd{t}\dd{\ve}} &\sim \frac{L\mysub{edd}}{4\pi d^2 \ln(\ve\mysub{max}/\ve\mysub{min})}\frac{1}{\ve^2}\,,\\
  &\sim 2\times10^{-7} \qty(\frac{M\mysub{bh}}{10^7M_\odot})\qty(\frac{d \ve}{\unit{Gpc\,GeV}})^{-2}\unit{erg^{-1}\,cm^{-2}\,s^{-1}}\,.
\end{split}
\ee
To determine the requirements for the signal detection, it is necessary to compare it to the background level, which in the \graya band (between \(10\unit{MeV}\) and \(100\unit{GeV}\)) can be roughly approximated \cite{2015ApJ...799...86A} as
\be
\dv{N\mysub{bg}}{S\dd{t}\dd{\ve}}\approx 3\times10^{-9}\qty(\frac{\ve}{1\unit{GeV}})^{-2.3}\qty(\frac{\mathrm{PSF}}{10^{-5}\unit{sr}})\unit{erg^{-1}\,cm^{-2}\,s^{-1}}\,.
\ee
Therefore, such flares can be detected in a background free regime with instruments having a decent angular resolution of \(\approx0.1^\circ\) for events occurring up to several \unit{Gpc}.

To study variability on \(\rsch/c=2GM\mysub{bh}/c^3\) scale, it is necessary to detect a significant number of photons, say \(\mathcal{N}\mysub{ph}\sim10\). Thus, we can estimate a number of detectable photon as 
\be
\begin{split}
  \frac{\ve\rsch}{c}\dv{N}{S\dd{t}\dd{\ve}}   &=\dv{N}{S}   \sim \frac{2m_p\ve}{\sigma_T c^2\ln(\ve\mysub{max}/\ve\mysub{min})}\qty(\frac{GM\mysub{bh}}{d\ve})^2\\
                                                           &\sim 3\times10^{-8}\unit{cm^{-2}} \qty(\frac{d}{1\unit{Gpc}})^{-2}\qty(\frac{M\mysub{bh}}{10^7M_\odot})^{2}\qty(\frac{\ve}{1\unit{GeV}})^{-1}
\end{split}
\ee

Space-born \graya detectors have a typical collection area of \(10^4\unit{cm^2}\), thus they do not allow studying extreme transients on the SMBH horizon scale. Ground based instruments have a significantly larger collection area, \(10^{8-10}\unit{cm^2}\), thus hypothetically allow studying transient events on the scale of black hole horizon. More specifically, an instrument with performance similar to the future Cherenkov Telescope Array Observatory (CTAO) at \(100\unit{GeV}\), should provide a collection area of \(10^9\unit{cm^2}\), thus study of SMBH on the horizon scale is possible up to the distance of  \(100\unit{Mpc}\qty(M\mysub{bh}/10^7M_\odot)\). Here we assumed that detection of \(10-30\) photons is sufficient to assert variability.
If there is a Cherenkov telescope with the threshold as low as \(1\unit{GeV}\), then one will be able to detect transients on \unit{Gpc} distances, effectively increasing the accessible region of the Universe by three orders of magnitude {(see, e.g., \cite{2005astro.ph.11139A} for a qualitative discussion of such instruments)}. Here, we would like also to note that while for sub-TeV photons the attenuation on extragalactic background light (EBL) is not critical for sources at \(d\lesssim \unit{Gpc}\), for studying variability of sources at high energies, \(\ve\gtrsim 10\unit{TeV}\), may significantly affected by EBL absorption.

\subsection{Outlook}
In summary, extreme high-energy transients probe the limits imposed by basic physical principles rather than by source-specific phenomenology alone. Constraints arising from causality, particle acceleration rates, energy density, radiative cooling, and intrinsic opacity jointly define a narrow region of parameter space in which such events can occur. When observed variability timescales, luminosities, or inferred acceleration efficiencies approach these bounds, the transient must be powered by processes operating in an extreme regime, independent of the particular astrophysical environment. Relativistic motion can relax some of these constraints, most notably those related to energy density and internal opacity, but cannot eliminate them entirely. As a result, \graya observations of rapid, luminous transients provide a uniquely sensitive probe of fundamental plasma processes, particle acceleration mechanisms, and energy dissipation under extreme conditions. Identifying and characterizing such events therefore offers a powerful pathway toward testing the physical limits of astrophysical accelerators and deepening our understanding of the most violent phenomena in the Universe.

\section{Instruments for the Study of Transient Sources} 

Investigating transient sources demands instruments capable of capturing the universe’s most energetic radiation. By probing the high-energy (HE; 100 MeV–100 GeV), the very-high-energy (VHE; 100\,GeV–100\,TeV) and the ultra-high-energy (UHE; E$>$100\,TeV) \graya regimes, we can explore the underlying mechanisms of explosive astrophysical events and the extreme physical conditions that drive them. 

Detection of transient phenomena in the \graya domain is
fundamentally constrained by a combination of instrumental and
astrophysical factors. These include the collection area, field of
view, duty cycle, angular resolution, and the level and character of
background noise. Their interplay determines the sensitivity of an
instrument to transient events across the MeV–PeV energy
range.

At photon energies below several tens of GeV, \graya instruments
must be space-borne to avoid atmospheric absorption. Their effective
collection area is intrinsically limited by payload constraints and
typically does not exceed $10^4\unit{cm^2}$. While such instruments offer wide
fields of view of order one steradian and nearly continuous sky
coverage, their small collection areas restrict photon statistics for
short or faint transients. This limitation becomes particularly severe
at higher energies, where photon fluxes decline steeply. In contrast,
ground-based detectors operating above tens of GeV exploit the
atmosphere as a calorimeter and achieve enormous effective collection
areas of $10^8$~--~$10^{10}\unit{cm^2}$. This difference of multiple orders of
magnitude is critical for detecting short-duration events, as the
number of collected photons scales directly with the product of the
collection area and the transient duration. In additional to statistical constraints on detectability of transients, there are also ``hardware'' limitations. As space-borne instruments
directly detect \graya emission they potentially can detect very
short, sub-second or even ms, transients. For ground-based telescopes
the detection time is typically limited by seconds.

The ability to detect unpredictable events also depends strongly on sky coverage and duty cycle. Space-borne telescopes,
with their large fields of view and continuous operation, can effectively monitor the entire sky with short cadence and
thus are well suited for discovery of transient phenomena. Ground-based instruments, however, are subject to geometrical
and operational limitations. Imaging atmospheric Cherenkov telescopes (IACTs) have relatively narrow fields of view (10
–- 100 deg${}^2$, note that deg${}^2 = 3\times10^{-4}$ steradian) and a duty cycle limited to dark, moonless nights,
typically below 15\%. As a result, their capacity for serendipitous transient detection is small, and rapid follow-up
often depends on external triggers. Moreover, source visibility for IACTs is seasonal, leading to months-long gaps in
sky coverage. Extensive air-shower arrays, in contrast, operate with nearly 100\% duty cycles and fields of view
approaching two steradians, providing continuous monitoring of a large fraction of the sky. Their sensitivity peaks in
the multi-TeV domain, where attenuation by the extragalactic background light (EBL) severely limits the observable
volume, confining their transient detections mainly to Galactic sources or the nearest galaxies.

Background noise constitutes another major limiting factor. At low
energies, the dominant background is instrumental, arising from
cosmic-ray interactions with spacecraft materials that generate
secondary \gray or activate detector components. This effect is
particularly severe in the MeV band, where Compton-scattering--based
instruments suffer from both strong instrumental backgrounds and
limited event reconstruction accuracy. In the GeV-TeV range,
astrophysical diffuse backgrounds dominate --- emission from the Galactic
plane and an isotropic extragalactic component --- both of which decline
rapidly with increasing energy, somewhat improving detectability of
bright transients. For ground-based detectors, the primary background
originates from extensive air showers initiated by cosmic rays. Since
hadronic showers outnumber gamma-ray--induced ones by several orders of
magnitude, the ability to discriminate between them is crucial. IACTs
achieve efficient background rejection through image-shape analysis,
whereas water-Cherenkov arrays provide coarser discrimination. Hybrid
systems, such as those incorporating muon detectors, offer the most
effective suppression of hadronic contamination.

The impact of the background also depends on the instrument’s angular resolution, as the integrated background within
the source region scales with the size of the point-spread function (PSF) or angular resolution. Space-borne MeV
instruments typically have degree-scale PSFs, while GeV telescopes reach sub-degree precision, and ground-based
detectors achieve arcminute-scale (IACTs) or $\sim0.1^\circ$ (air-shower arrays) resolutions. The background-limited
sensitivity of a transient search therefore improves rapidly with energy, owing both to better angular resolution and to
the declining intensity of diffuse emission.

Overall, the detection of \graya transients is governed by a
balance between collection area, temporal coverage, and background
control. Space-borne detectors provide continuous all-sky monitoring
but are limited in instantaneous sensitivity; ground-based facilities
achieve enormous collection areas but operate with restricted sky and
time coverage. Together they provide complementary capabilities:
wide-field, high-duty-cycle surveillance from space to discover and
localize transient events, and large-area, high-sensitivity
ground-based observations to resolve and characterize the most
energetic episodes.

\subsection{High-energy detectors}

\textbf{Imaging Compton Telescope - COMPTEL}
One of the instruments aboard the Compton Gamma Ray Observatory (CGRO) was a gamma-ray imaging system operating from 1991 to 2000. This Imaging Compton Telescope (COMPTEL), sensitive to photons between roughly 800\,keV and 30\,MeV, covered approximately one steradian at a time. Depending on photon energy and incident angle, it achieved an angular resolution between 1$^\circ$ and 2$^\circ$. Its energy resolution surpasses 10\% (FWHM), allowing for detailed spectral analysis of features such as nuclear gamma-ray lines. The system’s effective area, which depends on energy and data selection criteria, typically fell between 10\,cm$^2$ and 50\,cm$^2$. Its timing capability, with a precision of 0.125 milliseconds, also enabled it to analyze pulsed emission. 
Strictly speaking, COMPTEL was a medium-energy gamma-ray detector, not a high-energy detector, but it still covered a crucial energy range for astrophysical processes, especially nuclear lines and certain transient phenomena \citep{NTRS19920012642}.

\textbf{Fermi Gamma-ray Space Telescope - Fermi}
Launched in 2008, the Fermi Gamma-ray Space Telescope has played a transformative role in high-energy astrophysics. It carries two key instruments: the Large Area Telescope (LAT) and the Gamma-ray Burst Monitor (GBM).

The LAT is designed to detect \gray from 20 MeV up to over 300 GeV. With its wide field of view, it can scan the entire sky approximately every three hours—an essential capability for monitoring transient and variable sources. The maximum effective area is about $\sim$9000\,cm$^2$, achieved in the 1–10\,GeV energy range. The angular resolution of \emph{Fermi}-LAT (68\% containment radius) improves with energy down to less than 0.1$^\circ$ for energies above 100\,GeV. Complementing the LAT, the GBM is optimized for detecting \gray bursts (GRBs) and other short-lived phenomena in the lower energy range from a few keV to 40 MeV. This broad energy coverage makes \emph{Fermi} a powerful tool for studying both prompt and long-duration \graya events. For technical specifications, see \citep{Thompson2022}.

\textbf{Astro-rivelatore Gamma a Immagini LEggero - AGILE}
The Italian space agency operated the AGILE satellite including a Gamma Ray Imaging Detector (GRID; 30\,MeV–50\,GeV), a hard X-ray detector SuperAGILE (SA; 18–60\,keV) and a Mini-Calorimeter (MCAL; 350\,keV–100\,MeV) during 2007-2024. This satellite enabled an angular resolution characterized by a 68\% containment radius down to 0.8$^\circ$ for 1\,GeV gamma rays and a few arc\-minutes for X-rays and offered fields of view of 2.5 and 1 steradian, respectively. After two years of a pointing strategy, the AGILE satellite operated in a “spinning observing mode” from 2009 on. The AGILE-GRID, operating in spinning mode, was able to survey about 80\% of the sky each day, with an average effective area of $\sim 400\,\mathrm{cm}^2$ at 400~MeV. A summary of the mission can be found in \citep{2009A&A...502..995T}.

\subsection{Very-high-energy detectors}

\textbf{\mg Telescopes}
The Major Atmospheric Gamma Imaging Cherenkov (\mg) telescopes consist of two 17-meter-diameter instruments located at the Roque de los Muchachos Observatory on La Palma, Canary Islands (28\({}^\circ\)45\({}'\)43\({}''\)N, 17\({}^\circ\)53\({}'\)24\({}''\)W, 2.2\,km  a.s.l.). The project started with one telescope in 2004 and became a stereoscopic system in 2009. These telescopes are designed to observe VHE \gray, covering an energy range from approximately 30\,GeV up to several tens of TeV. For further technical details, see \cite{2016APh....72...76A}.

\textbf{\hs (High Energy Stereoscopic System)}
The \hs array sits 1.8\,km  a.s.l. 
in Namibia’s Khomas Highlands (23\({}^\circ\)16\({}'\)18\({}''\) S, 16\({}^\circ\)30\({}'\)00\({}''\) E). Phase I comprised four 12\,m Cherenkov telescopes arranged in a square; Phase II added a 28\,m centre-dish, extending the array’s collecting area and lowering its energy threshold. 
{H.E.S.S. operates at energies above $\sim 100$~GeV, with the addition of the large CT5 telescope in Phase II lowering the energy threshold to $\sim 80$~GeV, depending on the analysis and observing conditions (see \citep{2006A&A...457..899A, vanEldik:2016Ns} for technical specifications).}

\textbf{VERITAS (Very Energetic Radiation Imaging Telescope Array System)}
VERITAS operates at the Fred Lawrence Whipple Observatory on Mount Hopkins, Arizona (31\({}^\circ\)40\({}'\)30\({}''\) N, 110\({}^\circ\)57\({}'\)07\({}''\) W, 1.3\,km a.s.l.). The instrument consists of four 12\,m imaging atmospheric Cherenkov telescopes, providing sensitivity to \gray between $\sim$90\,GeV and 50\,TeV. Performance was enhanced by relocating one telescope in 2009 to optimise array geometry and by upgrading the camera systems in 2012. Detailed descriptions are given in \citep{Park:20161B}.

\textbf{The High Altitude Water Cherenkov (HAWC) experiment}
HAWC is a gamma-ray observatory located at 4.1\,km a.s.l. near the Sierra Negra volcano in Mexico (18\({}^\circ\)59\({}'\)41\({}''\)N, 97\({}^\circ\)18\({}'\)30.6\({}''\)W). It uses water Cherenkov detectors to study high-energy cosmic rays and gamma rays. {With improved reconstruction algorithms, HAWC is sensitive to gamma rays from $\sim 300$~GeV to several hundred TeV, with sensitivity improving significantly toward the TeV range \cite{2024ApJ...972..144A}.}

HAWC consists of 300 large water tanks, each containing 188,000 liters of water and four photomultiplier tubes. The experiment has a field of view (2 sr). Continuous operation allows it to survey the sky and observe transient events like gamma-ray bursts and AGN flares. Technical description can be found in \citep{2023NIMPA105268253A}.

\subsection{Ultra-high-energy detector}

\textbf{The Large High Altitude Air Shower Observatory (LHAASO)}
LHAASO is situated on Mt. Haizi in Daocheng, Sichuan Province, China, at an altitude of 4.41\,km  a.s.l. (29\({}^\circ\)21\({}'\)27.56\({}''\)N, 100\({}^\circ\)08\({}'\)19.66\({}''\)E). The observatory has several detector components. 
The Water Cherenkov Detector Array (WCDA) is a survey instrument with a total active area of 78,000\,m\({}^2\) consisting of 3120 water Cherenkov detectors (WCDs) sensitive to gamma rays with energies between 100\,GeV -- 30\,TeV. The angular resolution is $<$0.2\({}^\circ\) at 10\,TeV and 1.0\({}^\circ\) at 600\,GeV.
{The KM2A array spans 1.3~km$^2$ and comprises electromagnetic particle detectors (EDs) and 1188 muon detectors (MDs), which cover a total area of $\sim 4\times10^4$~m$^2$ ($\sim$4\% of the array).}
The 18 wide field-of-view Cherenkov telescopes (WFCTAs) measure the component and energy spectrum of cosmic rays from 10\,TeV to a couple of EeV. All details can be found in \citep{2022ChPhC..46c0001M}.
%

\section{Gamma-ray bursts}

\subsection{Introduction}
Gamma-ray bursts (GRBs) were first discovered serendipitously in 1967 by military satellites designed to monitor nuclear tests in space. These instruments detected unexpected flashes of \gray with unusual temporal profiles. Early localization excluded a terrestrial or solar origin, and later observations demonstrated that GRBs are distributed across the sky  isotropically \cite{1992Natur.355..143M}, pointing to their extragalactic nature. These discoveries opened a completely new field in astrophysics, providing access to some of the most powerful transient events in the Universe.

GRB light curves generally exhibit two distinct phases: the prompt emission, associated with the activity of the central engine and internal dissipation within the relativistic jet, and the afterglow, which arises from the forward shock propagating into the circumburst medium. The prompt phase typically lasts from a fraction of a second up to several minutes, while the afterglow can be observed for days to weeks across the electromagnetic spectrum. Based on the duration of the prompt emission, GRBs are classified into two populations \cite{1993apj...413l.101k}: long GRBs (duration 
$>2$~s), thought to result predominately from the collapse of massive stars \cite{2003apj...593l..19k,2003apj...591l..17s}, and the majority of short GRBs ($<2$~s), attributed to mergers of compact object binaries \cite{2017PhRvL.119p1101A}. Since both progenitor scenarios may produce bursts with a range of durations, the observed duration alone does not provide a robust criterion for distinguishing between them, particularly for events near the nominal boundary.

For GRBs with identified host galaxies, redshifts span from \(z\approx0.009\) \cite{2017PhRvL.119p1101A} to
\(z\approx9.4\) \cite{2011apj...736....7c}, showing that GRBs have occurred since the Universe was only about
\(500\)~Myr old. Their extragalactic origin implies extreme energetics, with isotropic-equivalent energy releases of
\(10^{48}\)~--~\(10^{53}\)~erg, making GRBs the brightest explosions known. Although the apparent luminosity
overestimates the true energy output because of relativistic beaming, only exceptionally high initial bulk Lorentz
factors, \(\Gamma_0\), can bring the intrinsic source power to less extreme values. Thus, whether measured in terms of
released energy or relativistic motion, GRBs represent a class of truly extreme astrophysical explosions. The intrinsic
explosion energy scales as \(\Gamma_0^{-2}\) relative to the apparent isotropic value; hence, for an observed isotropic
equivalent energy of \(10^{54}\unit{erg}\) and a characteristic \(\Gamma_0 = 300\), the true energy is approximately
\(
E\mysub{true} \simeq 10^{49}\qty(\frac{\Gamma_0}{300})^{-2}\unit{erg}\,,
\)
marking GRBs as extreme both in total released energy and in the bulk Lorentz factors of their outflows.

According to the criteria discussed above, the collapse of a progenitor object into a black hole alone suffices to classify GRBs as extreme transients. However, GRBs exhibit extreme properties in several additional respects, most notably their luminosity, total energy release, and relativistic bulk motion. Extreme relativistic motion, with initial Lorentz factors 
\(\Gamma_0 \gtrsim \text{a few} \times 10^2\), is arguably the defining characteristic of GRBs. Such large bulk Lorentz factors imply strong Doppler boosting, which both shifts the emission to higher frequencies and greatly enhances the apparent luminosity of the observed radiation.

The outflow bulk Lorentz factor also determines the Lorentz factor of the blast wave propagating into the circumburst medium. Physical processes operating at this blast wave are responsible for the afterglow emission, which is detected across the entire electromagnetic spectrum, from radio wavelengths to the VHE \graya domain. Observed VHE emission extends to energies at which attenuation by the EBL becomes significant, rendering a direct observational determination of the intrinsic high-energy cutoff of the \graya spectrum effectively impossible. Nevertheless, existing observations demonstrate that the afterglow blast wave acts as a highly efficient particle accelerator.

This makes GRBs exceptional laboratories for studying particle acceleration at relativistic shocks, as they provide direct access to emission produced by particles accelerated at a relativistic shock propagating into an initially non-relativistic circumburst medium. While the formation, collimation, and stability of such ultra-relativistic outflows remain subjects of active investigation, many aspects of their dynamical evolution can be described using self-similar solutions. These solutions, in turn, enable robust estimates of the physical conditions in the emission region.

In the afterglow phase, the outflow energy is carried by a
relativistically hot blast wave sweeping through the circumburst
medium. The essential relations between the radius and bulk Lorentz factor of the blast wave can be obtained from  the energy conservation equation, which in this case is written as
\be
E = \Gamma^2 M c^2\,.
\ee
Here \(M\) is the accumulated mass of the swept-up material and \(\Gamma\) the bulk Lorentz factor of the expanding shell. For a stellar-wind environment, appropriate for long GRBs associated with the collapse of massive stars, the swept-up mass {by the blast wave of radius \(R\)} is
\be
M = \frac{\dot{M} R}{v\mysub{wind}}\,,
\ee
where \(\dot{M}\) and \(v\mysub{wind}\) are the progenitor’s mass-loss
rate and wind velocity, respectively. Although uniform-density media
are sometimes invoked to interpret certain afterglow light curves, the
stellar-wind environment remains the physically natural case for
collapsar-type GRBs and is adopted below.

Massive stars typically create wind bubbles extending over parsec
scales. Consequently, the blast wave requires months or even years in
the progenitor’s rest frame to reach the termination shock of the
stellar wind. Even accounting for relativistic time compression, the
corresponding observer-frame delay remains substantial. In studying GRB emission it is important to distinguish three time coordinates: the
progenitor-frame time \(\tau\mysub{p}\), the proper time \(\tau\) of a
comoving element of the blast wave (a non-inertial frame), and the
observer-frame time \(t\), defined relative to the arrival of photons
emitted at the instant of explosion. These times are related through
the shell radius \(R\) as 
\be
\tau\mysub{p} = \int\limits_0^R \frac{\dd{r}}{c\sqrt{1-\nicefrac{1}{\Gamma^{2}}}}\,,\quad\tau = \int\limits_0^R \frac{\dd{r}}{c\Gamma\sqrt{1-\nicefrac{1}{\Gamma^{2}}}}\,,\qq{and}t = \tau\mysub{p} - \frac{R}{c}\,,
\ee
where the blast wave Lorentz factor is (adopting the wind-like density profile) 
\be
\Gamma(r) = \sqrt{\frac{E}{M(r) c^2}} \propto R^{\nicefrac{-1}{2}} \,.
\ee
%
%
These relations allow relating the physical parameters of the expanding shell to its emission detected by a distant observer.

VHE detections typically occur at observer times ranging from minutes to hours after the GRB onset. For illustrative estimates, we adopt \(t = 600\unit{s}\), \(\Gamma_0 = 300\), and an isotropic explosion energy of \(E = 10^{53}\unit{erg}\). The corresponding ejecta mass is
\(E/(\Gamma_0 c^2) \simeq 2\times10^{-4} M_\odot E_{53}\Gamma_{2.5}^{-2}\).
During the early, so-called coasting phase within the coasting radius \(R\mysub{c}\), the ejecta expand with nearly constant velocity and is only weakly affected by the external medium. This regime holds for
\be
R \ll R\mysub{c}=\frac{E v\mysub{wind}}{\Gamma_0^2 \dot{M} c^2}\,,
\ee
which, for a typical stellar-wind mass-loss rate of \(\dot{M} = 10^{-6}\dot{M}_{-6}M_\odot\unit{yr^{-1}}\) and wind velocity \(v\mysub{wind} = 10^3v\mysub{wind,3}\unit{km\,s^{-1}}\),
yields
\be
R\mysub{c} \approx 2\times 10^{15} E_{53}
\dot{M}_{-6}^{-1} v\mysub{wind,3}\Gamma_{2.5}^{-2}\unit{cm}\,.
\ee

In the progenitor frame, the blast wave traverses this distance in approximately 20 hours. However, owing to relativistic time compression by a factor \(\sim (2\Gamma_0^2)^{-1}\), a distant observer perceives the end of the coasting phase as occurring only fractions of a second after the explosion. The coasting phase is followed by the so-called self-similar deceleration phase.

Solving the relations for the characteristic timescales in the self-similar phase under the assumption \(\Gamma\gg1\), one finds
\(\tau\mysub{p}\approx R/c\),
\(\tau\approx {2R}/{(3c\Gamma_0)}\sqrt{R/R\mysub{c}}\), and
\(t \approx {R}/{(4c\Gamma_0^2)}(R/R\mysub{c})\).
Here, for simplicity, we neglect the time difference accumulated during the coasting phase, which can nevertheless be important during the early afterglow evolution (see \cite{2024ApJ...966...31K}).

These relations outline the physical scales and timescales relevant to the production of VHE radiation in GRBs. In particular, a useful relation links the blast-wave bulk Lorentz factor to the observer time:
\be
\Gamma \approx \sqrt[4]{\frac{Ev\mysub{wind}}{4\dot{M}c^3t}},.
\ee
This expression shows that the bulk Lorentz factor depends only weakly on the model parameters, scaling as the fourth root of each. Consequently, uncertainties associated with the specific parameter values translate into relatively modest uncertainties in \(\Gamma\). This simple relation therefore provides a robust estimate of the blast-wave Lorentz factor. Combined with typical stellar-wind parameters and the Rankine--Hugoniot relations, it allows one to estimate the energy density in the emission region.

The partition of the energy in the production region among thermal and non-thermal particle populations (protons and electrons), magnetic fields, and radiation fields is a formidable problem. It requires modelling a wide range of kinetic processes operating across many orders of magnitude in scale. Fully self-consistent simulations of this kind are not yet available, and the complexity is therefore commonly encapsulated in a set of phenomenological parameters, such as the plasma magnetization and the fraction of energy transferred to non-thermal particles. Within this framework, one can predict the spectra and light curves of the synchrotron and inverse-Compton (IC) emission produced at the blast wave. By confronting these models with observational data, it is possible to infer preferred values of the underlying parameters and thereby gain insight into the physical processes operating at relativistic shocks.

Detection of VHE emission from GRBs is an essential probe of the physical conditions in the production region for a simple reason. Synchrotron radiation is expected to contribute primarily below \(\sim 10 (\Gamma/100)\unit{GeV}\), owing to the synchrotron burn-off limit. As a result, emission detected in the VHE domain is generally expected to be dominated by IC processes. A simultaneous detection of emission produced through these two distinct radiation channels provides much stronger constraints on the physical conditions in the emission region than either component alone. For this reason, the detection of GRBs in the VHE regime has long been regarded as highly desirable for advancing our theoretical understanding of GRB afterglows and particle acceleration at relativistic shocks \cite{2001AdSpR..27..813D}.


\subsection{GRBs at High Energies}
Observationally, detecting GRBs in the \graya band remains challenging. Space-based instruments, which are the primary tools for GRB discovery, have limited collection areas, restricting sensitivity especially at the highest energies. Wide-field-of-view ground-based \graya telescopes, on the other hand, probe the very-high-energy and ultra-high-energy domains but are constrained by attenuation due to the EBL, effectively limiting their detection horizon to \(z\ll1\), while the majority of the detected GRBs come from \(z\sim2\). Furthermore, IACTs, which are less prone to the EBL attenuation, are limited due to their smaller duty cycle constraining observations only during sufficiently dark sky and good weather conditions.

In the high energy \graya regime the \emph{Fermi} satellite detected 2356 GRBs with the \emph{Fermi}-GBM and 186 with the \emph{Fermi}-LAT between 2008-2018 (\cite{2020ApJ...893...46V, 2019ApJ...878...52A}).
Basically, the GBM acts as the GRB ``finder'' by detecting nearly all bursts, characterizing their prompt keV–MeV emission, and providing triggers, whereas LAT is the high-energy follow-up: it detects only the brightest $\sim$10\% of GBMs but provides crucial GeV emission data, better localization, and insight into the most energetic processes.
In the TeV \graya range there are five GRBs significantly detected  \cite{2025arXiv250804557F}.
Below, we summarize observational findings about three of these detected GRBs, which provided us with the most versatile data.

\textbf{GRB~190114C} was the first gamma-ray burst ever detected at TeV energies.
The burst was identified as a long-duration GRB by the Burst Alert Telescope
(BAT) onboard the \emph{Neil Gehrels Swift Observatory} and by the Gamma-ray Burst
Monitor (GBM) onboard \emph{Fermi} on 14 January 2019 at 20:57:03~UT (hereafter
$T_0$). Spectroscopic observations of the optical afterglow yielded a redshift of
$z = 0.4245 \pm 0.0005$, corresponding to a luminosity distance of
$\simeq 2.3$~Gpc \cite{2019Natur.575..455M}.

MAGIC observations commenced at $T_0 + 57$~s, triggered by the \emph{Swift} alert, and
continued until $T_0 + 1.6\times10^{4}$~s. A highly significant excess of
gamma-ray events was detected from the start of the observations, with a total
significance exceeding $50\sigma$ in the first 20~minutes and photon energies
extending from $\sim0.2$ to $\sim1$~TeV \cite{2019Natur.575..455M}. The intrinsic
TeV light curve, corrected for attenuation by the 
EBL, follows a smooth power-law decay $F(t)\propto t^{-1.6}$, with no evidence
for rapid variability or spectral breaks, indicating an origin in the afterglow
phase rather than in the highly variable prompt emission.

The isotropic-equivalent energy radiated in the $0.3$--$1$~TeV band between
$T_0 + 62$~s and $T_0 + 2454$~s is
$E_{\rm iso}^{0.3\text{--}1\,\mathrm{TeV}} \simeq 4\times10^{51}$~erg, with the
total TeV output potentially reaching $\sim2\times10^{52}$~erg when extrapolated
to earlier times \cite{2019Natur.575..455M}. For comparison, \emph{Fermi}-GBM
measured a prompt-emission duration of $T_{90}\simeq116$~s in the
$10$--$1000$~keV band and an isotropic-equivalent energy release of
$E_{\rm iso}^{\rm keV\text{--}MeV}\simeq3\times10^{53}$~erg, placing GRB~190114C
among the more energetic, though not exceptional, long GRBs
\cite{2019Natur.575..455M,2020ApJ...890....9A}.

Joint multiwavelength observations from radio to TeV energies revealed a double-peaked spectral energy distribution
during the early afterglow \cite{2019Natur.575..459M}, with a low-energy synchrotron component peaking in the X-ray band
and a distinct high-energy component dominating above $\sim100$~GeV (see, however,
Ref.~\cite{2023MNRAS.520..839K}). The comparable radiated power and similar temporal decay of the X-ray, GeV, and TeV light curves, revealed in this burst, seem to be a common feature in GRBs \cite{2019ApJ...878...52A,2021Sci...372.1081H}.
Hadronic scenarios, such as proton synchrotron emission, are disfavored due to the extreme
energetic requirements and low radiative efficiency.

Time-resolved spectral analyses of the prompt and early afterglow emission with
\emph{Fermi}-GBM and \emph{Fermi}-LAT revealed evidence for a high-energy cutoff in
the sub-GeV range during the prompt phase, consistent with internal
$\gamma\gamma$ pair-production opacity and implying bulk Lorentz factors of
$\Gamma \sim 300$--700 \cite{2020ApJ...890....9A,2020ApJ...903....9C}. The TeV emission detected by
MAGIC thus originates from a physically distinct emission region associated with
the external forward shock, marking the first unambiguous detection of inverse
Compton radiation from a GRB afterglow.

\textbf{GRB~190829A} represents a unique case among very-high-energy--detected
gamma-ray bursts due to its low redshift and comparatively modest isotropic
energy release. The burst was detected by \emph{Fermi}-GBM on 29 August 2019 at
19:55:53~UT (hereafter $T_0$) and classified as a long-duration GRB with a prompt
emission duration of $T_{90}\simeq63$~s in the 50--300~keV band. Spectroscopic
observations of the host galaxy yielded a redshift of
$z = 0.0785 \pm 0.0005$, making GRB~190829A one of the nearest GRBs ever observed
at TeV energies.

The H.E.S.S. telescopes detected VHE gamma-ray emission from the afterglow of
GRB~190829A between approximately 4 and 56~hours after the trigger, marking the
first {high-significance} detection of TeV emission during the late afterglow phase of a GRB {(see, however, also \cite{2019natur.575..464a}, where one reported the detection of GRB~180720B with \hs ten hours after the end of the prompt emission phase)}.
Observations were carried out on three consecutive nights, with significant
detections on each night, reaching a peak significance of $21.7\sigma$ during
the first night \cite{2021Sci...372.1081H}. The detected photon energies extended
from $\sim0.18$ to $\sim3.5$~TeV, with the intrinsic spectrum, corrected for
EBL absorption, well described by a simple power law
with photon index $\Gamma_{\rm VHE}\simeq2.1$, remarkably similar to the
simultaneously observed X-ray spectral index.

The temporal evolution of the VHE emission closely tracks that of the X-ray afterglow, with both light curves exhibiting smooth power-law decays, $F(t)\propto t^{-1.1}$, and no evidence for rapid variability. This close temporal correlation, together with the similar spectral evolution observed in the X-ray and TeV bands, indicates that both components are powered by the same underlying population of relativistic electrons. At the same time, the fact that the TeV spectrum closely follows the X-ray synchrotron component and, in particular, that the measured VHE slope is relatively hard, poses a challenge for standard one-zone SSC afterglow models. In such models, SSC emission is typically expected to exhibit significantly steeper spectra due to Klein–Nishina suppression \cite{2021Sci...372.1081H}.

The observations instead suggest a scenario in which the high-energy emission represents a continuation of the synchrotron component into the TeV regime, implying electron acceleration beyond the classical synchrotron burn-off limit and requiring unusually efficient acceleration conditions \cite{2021Sci...372.1081H}. Alternatively, if one does not attempt to reproduce the detailed spectral hardness reported by \hs, the observed VHE flux level and the detection of TeV \gray at late stages of the afterglow--when the blast-wave Lorentz factor has decreased--can be naturally explained within the framework of converter-type acceleration operating in a pair-balanced production site (i.e., in the framework of models similar to the one suggested for GRB~190114C \cite{2021apj...923..135d}).

The isotropic-equivalent energy released during the prompt phase was
$E_{\rm iso}^{10\text{--}1000\,\mathrm{keV}}\simeq2\times10^{50}$~erg, placing
GRB~190829A at the lower end of the GRB energy distribution. In comparison, the
isotropic-equivalent energy radiated in the VHE band amounted to
$E_{\rm iso}^{0.18\text{--}3.5\,\mathrm{TeV}}\simeq3.4\times10^{48}$~erg during the
first night and $\simeq7.5\times10^{47}$~erg during the second night. Despite its
lower overall energetics, the proximity of GRB~190829A and the reduced attenuation
by the EBL enabled an unprecedented characterization
of its intrinsic TeV spectrum.

Together with GRB~190114C, GRB~190829A establishes TeV emission as a common feature of GRB afterglows under favorable
conditions, while highlighting the strong diversity in energetics, temporal evolution, and possibly some uncertainty on
the involved emission mechanisms among VHE-detected GRBs.

\textbf{GRB~221009A} is an exceptionally luminous long GRB, widely referred to as the ``brightest of all time'' (BOAT) due to its
unprecedented fluence across the electromagnetic spectrum. The burst was
triggered by \emph{Fermi}/GBM on 9 October 2022 at 13:16:59.99~UT
($T_0$) and is located at a redshift of $z = 0.1505$ \cite{2023ApJ...952L..42L}.
In the 10--1000~keV band, \emph{Fermi}/GBM measured a duration of
$T_{90} \approx 327$~s, placing the event firmly in the long-GRB class.
The isotropic-equivalent energy release integrated from
$T_0 - 2.7$~s to $T_0 + 1449.5$~s was
$E_{\mathrm{iso},\,1\text{--}10\,000~\mathrm{keV}}
= (1.01 \pm 0.007)\times10^{55}$~erg, making GRB~221009A one of the most
energetic explosions ever observed \cite{2023ApJ...952L..42L}.

VHE observations of GRB~221009A marked
a watershed moment for GRB physics. LHAASO, which had the burst position within its field of view at
trigger time, detected more than $6.4\times10^{4}$ photons with energies above
200~GeV within the first $\sim3000$~s after $T_0$, with a statistical significance
exceeding $250\sigma$ \cite{2023Sci...380.1390L}. The detected photons extended
up to energies of several TeV, establishing the first unambiguous detection of a
TeV afterglow with an extensive air-shower array. The TeV emission exhibited a
rapid rise within minutes after the trigger, followed by a broken power-law
decay, with a steepening at $t \sim 650$~s. The characteristic shape of the light curve can be taken as an evidence for an explosion in a constant density medium \cite{2023Sci...380.1390L}. However, the effects related to the transition from the coasting to self-similar phase challenge this interpretation, making the scenario involving the blast wave propagation through the progenitor wind more feasible \cite{2024ApJ...966...31K}. Also the TeV light curve may contain the information regarding the time-dependent energy injection to the shell propagating through the circumburst medium \cite{2024MNRAS.530..347D}.

Complementary observations were reported by imaging atmospheric Che\-ren\-kov
telescopes at later times. H.E.S.S. conducted follow-up observations of
GRB~221009A starting at $T_0 + 4.3$~h and continuing over several nights but did
not detect significant VHE emission. Integral flux upper limits were derived at
energies above $\sim1$~TeV at the level of a few $10^{-12}$~cm$^{-2}$~s$^{-1}$,
corresponding to $\lesssim5$--10\% of the Crab Nebula flux for typical
assumptions on the spectral shape \cite{2023ApJ...946L..27A}. These limits
constrain any long-lived TeV afterglow component at late times.

In addition, the LST-1 telescope
reported a hint of VHE emission at a post-trial significance of
$\sim3$--$3.5\sigma$ during observations starting at
$T_0 + 1.33$~days ($\sim1.1\times10^{5}$~s). The excess was detected at energies
above $\sim300$~GeV and is consistent with a rapidly fading afterglow component.
However, the significance remains below the standard threshold required for a
firm detection, and the result is therefore treated as an indication rather than
a confirmed VHE signal \cite{2025ApJ...988L..42A}.

Fermi-LAT analyses of the GRB 221009A prompt and early afterglow emission indicate that the HE events above $\sim$30\,MeV are inconsistent with a pure synchrotron origin and require an additional component such as synchrotron self-Compton, with the highest-energy LAT photon ($\sim$400\,GeV) detected ~33 ks after the trigger challenging simple cascade interpretations \cite{2025ApJS..277...24A}.

The combined GeV--TeV dataset strongly favors an afterglow origin for the
highest-energy emission. Modeling of the LHAASO observations indicates that the
TeV photons can be explained by synchrotron self-Compton emission from electrons
accelerated in the external forward shock, provided that the jet has a narrow
core with a half-opening angle of $\theta_{\rm j} \simeq 0.8^\circ$ and an initial
bulk Lorentz factor exceeding $\Gamma \sim 300$ \cite{2023Sci...380.1390L}.
This structured-jet geometry naturally reconciles the extreme apparent
isotropic energy with physically plausible jet energetics and explains the
early onset and high luminosity of the TeV afterglow.

GRB~221009A thus represents the most extreme example to date of VHE and
multi-TeV emission from a GRB, bridging the gap between classical IACT detections
of TeV afterglows and the ultra-high-energy domain probed by wide-field
air-shower arrays. Its unprecedented brightness and proximity establish it as a
cornerstone event for understanding particle acceleration, radiation
mechanisms, and jet structure in relativistic explosions.

\begin{table}[t]
\centering
\caption{Key properties of GRBs detected at TeV and ultra-high energies.
All times are given relative to the trigger time $T_0$.}
\label{tab:grb_vhe_comparison}
\scriptsize
\setlength{\tabcolsep}{2pt}
\begin{tabular}{p{2cm}ccc}
\hline
 & \textbf{GRB 190114C} & \textbf{GRB 190829A} & \textbf{GRB 221009A} \\[3pt]
\hline
Redshift $z$ &
0.4245 &
0.0785 &
0.1505 \\[1pt]
\hline
$T_{90}$ (10--1000 keV)$^{\,a}$ &
$\sim116$ s &
$\sim63$ s &
$\sim327$ s \\[1pt]
\hline
$E_{\rm iso}$ (keV--MeV)$^{\,a}$ &
$\sim3\times10^{53}$ erg &
$\sim2\times10^{50}$ erg &
$\sim10^{55}$ erg \\[1pt]
\hline
TeV instrument &
MAGIC &
H.E.S.S. &
LHAASO \\[1pt]
\hline
TeV onset$^{\,b}$ &
$+57$ s &
$+4$--6 h &
$+200$--226 s \\[1pt]
\hline
TeV energy range &
0.2--1 TeV &
0.18--3.5 TeV &
$\gtrsim0.2$ TeV \\[1pt]
\hline
Peak significance &
$>50\sigma$ &
$\sim22\sigma$ &
$>250\sigma$ \\[1pt]
\hline
$E_{\rm iso}$ (TeV band)$^{\,c}$ &
$\sim4\times10^{51}$ erg &
$\sim10^{48}$ erg &
$\gtrsim10^{52}$ erg \\[1pt]
\hline
Temporal phase$^{\,d}$ &
Early afterglow &
Late afterglow &
Early afterglow \\[1pt]
\hline
suggested origin &
SSC (forward shock) &
Synch./SSC &
SSC (forward shock) \\[1pt]
\hline
Key feature &
First TeV GRB &
Nearest TeV GRB &
Brightest GRB (BOAT) \\[1pt]
\hline
\end{tabular}

\vspace{2mm}
\noindent\footnotesize{
\textbf{Notes.}
$^{a}$~$T_{90}$ and $E_{\rm iso}$ refer to the prompt emission in the
10--1000~keV (\textit{Fermi}/GBM) band, using the integration intervals reported
in the cited analyses.
$^{b}$~Time of first significant TeV detection relative to $T_0$;
for GRB~221009A the range reflects differing instrumental coverage.
$^{c}$~TeV-band $E_{\rm iso}$ values correspond to the energy bands and
time intervals used in the discovery papers and should be interpreted as
lower limits.
$^{d}$~``Early afterglow'' denotes emission detected within
$t\lesssim10^{4}$~s; ``Late afterglow'' refers to $t\gtrsim10^{4}$~s.
}
\end{table}

\subsection{Summary and Outlook}
Table~\ref{tab:grb_vhe_comparison} illustrates the diversity of very-high-energy phenomenology among the three GRBs
detected at TeV and ultra-high energies.  They reveal that very-high-energy afterglow emission is not associated with a
single phenomenological pathway but instead reflects a combination of intrinsic energetics, distance, circumburst
environment, and jet structure. GRB~190114C established the existence of a luminous TeV afterglow emerging within the
first minute after the trigger, with isotropic-equivalent energies of order $10^{51}$~erg in the sub-TeV band and
temporal and spectral evolution consistent with inverse-Compton emission from the external forward shock {(see \cite{2024MNRAS.530..347D} for a self-consistent consideration of particle acceleration-radiation for this burst)}. GRB~190829A,
although intrinsically less energetic in the keV–MeV band by more than two orders of magnitude, was detected at TeV
energies over several nights owing to its unusually small distance, exhibiting a long-lived component up to
$\sim3.5$~TeV whose spectrum closely tracks the X-ray emission and thereby probes the transition regime between
synchrotron-dominated and SSC-dominated cooling. It is important to note that the modest energetics and the late
observational time imply that the bulk Lorentz factor is small, say \(\Gamma \approx 5\), during the epoch of the emission
detection with \hs, since the intrinsic spectrum should be de-boosted, \(\hbar\omega'\approx \hbar\omega/\Gamma\sim 1\unit{TeV}\) (here we neglect the cosmological correction). Intrinsic emission at such high energy will be unavoidably absorbed on EBL if emitted from a production site moving with a higher Lorentz factor, \(\Gamma>10\), even the burst is a very nearby. Thus, the constraints on the intrinsic spectrum obtained with \hs will likely stay as the most {stringent} in foreseeable future.

GRB~221009A represents an extreme outlier, combining an unprecedented isotropic-equivalent energy release of
$\sim 10^{55}\,\mathrm{erg}$ in the keV--MeV prompt emission with ultra-high-energy afterglow emission extending into
the multi-TeV regime within minutes after the trigger.  Theoretical modelling of this event favors a structured-jet
configuration that reconciles its extreme apparent energetics with physically plausible jet power and explains both the
early onset and the high luminosity of the TeV afterglow. Detailed analysis of the TeV light curve show that there could
be evidence for a number of new effects, such as long lasting energy supply to the shell \cite{2024MNRAS.530..347D}, or shift of the reference time \cite{2024ApJ...966...31K}.

Taken together, these events demonstrate that TeV emission from GRBs may arise from distinct physical regimes --- early,
rapidly evolving SSC components in energetic bursts, late and long-lasting TeV afterglows in nearby events with modest
energetics, and ultra-energetic, structured-jet explosions bridging the TeV and multi-TeV domains.

From a theoretical perspective, these detections provide new constraints on particle acceleration and radiative processes in relativistic collisionless shocks. The combined temporal and spectral behavior indicates that TeV emission can persist from minutes to days after the trigger, with characteristic fluxes at $E\gtrsim100$~GeV ranging from $\sim10^{-11}$ to $10^{-9}$~erg~cm$^{-2}$~s$^{-1}$ during the early afterglow for nearby or extremely energetic events, and at the level of $\sim10^{-12}$~erg~cm$^{-2}$~s$^{-1}$ at late times. In several cases, the inferred evolution of the TeV spectrum requires time-dependent microphysical parameters and high maximum electron energies, $\gamma_{\max}\sim10^{7}$–$10^{8}$, while the relative importance of synchrotron and SSC components depends sensitively on the ambient density profile, shock magnetization, and Klein–Nishina suppression. These results establish TeV afterglows as a key diagnostic of the efficiency and extremity of particle acceleration in GRB external shocks.

\section{Novae}

\subsection{Introduction}

Novae are thermonuclear explosions occurring in close binary systems hosting a
white dwarf (WD) accreting hydrogen-rich material from a companion star. Once
sufficient material has accumulated on the WD surface, degenerate conditions
trigger a thermonuclear runaway, leading to the rapid fusion of hydrogen into
helium and the ejection of $\sim10^{-5}$--$10^{-4}\,M_\odot$ of material at
velocities of several hundred to a few thousand km\,s$^{-1}$. The resulting
outburst produces a sudden optical brightening by up to $\sim10^5$ in luminosity
but, unlike supernovae, does not disrupt the WD, allowing the process to repeat
over time.

Depending on the nature of the donor star, novae are commonly classified as
classical novae, where the companion is a main-sequence star and mass transfer
occurs via Roche-lobe overflow, or symbiotic novae, where the WD accretes from
the dense wind of an evolved red giant. The latter systems are comparatively
rare but are characterized by denser circumstellar environments and shorter
recurrence times. In both cases, the recurrence timescale is primarily governed
by the WD mass and the accretion rate, with recurrent novae representing systems
that erupt multiple times within a human lifetime. Please refer to \cite{2021ARA&A..59..391C} for a detailed review.

For decades, novae were primarily studied in the optical, infrared, and X-ray
bands, where they provide key insights into thermonuclear burning, mass
ejection, and binary evolution. However, they were not traditionally considered
as sites of efficient particle acceleration. This view changed fundamentally
with the discovery of transient GeV gamma-ray emission from novae by the
\textit{Fermi}-LAT.


\subsection{Novae at High Energies}

For many years, it had been hypothesized that shock acceleration within the dense wind of a red giant could accelerate particles—both protons and electrons—due to the strong magnetic fields present in such environments \cite{2007ApJ...663L.101T}. This theoretical prediction was confirmed on March 10, 2010, when Fermi-LAT detected the first nova emitting at GeV energies, V407 Cyg. The \graya outburst was observed concurrently with the optical discovery, peaking a few days later and lasting about two weeks in total (\cite{2010Sci...329..817A} and references therein).

Subsequent observations by VERITAS did not yield a detection, but the resulting upper limits provided valuable constraints on hadronic acceleration  \cite{2012ApJ...754...77A}. Since then, roughly 25 novae have been detected by Fermi-LAT\footnote{https://asd.gsfc.nasa.gov/Koji.Mukai/novae/latnovae.html}, allowing statistical characterization of their high-energy behavior. Typically, the emission lasts 5–55 days, with rise and decay times of a few days. Peak fluxes above 0.1\,GeV are of order 0.1 -- 5 $\times$ 10$^{-6}$ ph cm$^{-2}$ s$^{-1}$, and the spectra follow power laws with indices between 1.8 and 2.3, often with spectral cutoffs in the GeV range. Corresponding luminosities exceed 10$^{34}$ -- 4 $\times$ 10$^{36}$ erg s$^{-1}$ \cite{2018A&A...609A.120F}.

Distance estimates for over 400 Galactic novae have been compiled using Gaia DR3 parallaxes, showing that around 40\% reside in the Galactic bulge ($\approx$8\,kpc) and the remainder within the disk (scale height $\approx$140\,pc) \cite{2022MNRAS.517.6150S}. Among the 19 novae detected by Fermi-LAT between 2008 and 2023, most are disk sources, which tend to be both closer and optically brighter. On average, about one to two novae are detected per year at GeV energies, gradually expanding the known high-energy nova population.

\subsection{Novae at Very-High Energies}

For more than a decade, Cherenkov telescopes searched for VHE gamma-ray emission from novae without success \cite{2012ApJ...754...77A,2015A&A...582A..67A}. Such detections are crucial because GeV data alone cannot distinguish whether the emission originates from leptonic (electron) or hadronic (proton) processes.

A breakthrough occurred with the 2021 outburst of the recurrent symbiotic nova RS Ophiuchi, a well-known binary that erupts roughly every 15 years. The event was observed in the VHE range by H.E.S.S., MAGIC, and the LST-1. These detections established novae as a new class of VHE gamma-ray sources. Observations began roughly one day after the optical and GeV detections. 
{The optical and GeV light curves peaked at similar early times (around day 1), although the $\gamma$-ray peak may be slightly delayed. The GeV flux then decayed, decreasing by a factor of $\sim$2 within $\sim$2–3 days after the peak \cite{2022ApJ...935...44C}.}

In contrast, the VHE emission remained roughly constant over the first four days and then faded below detection limits within two weeks (\cite{2022Sci...376...77H,2022NatAs...6..689A,2025A&A...695A.152A} and references therein).




\begin{figure*}
  \centering
    \includegraphics[width=0.99\textwidth]{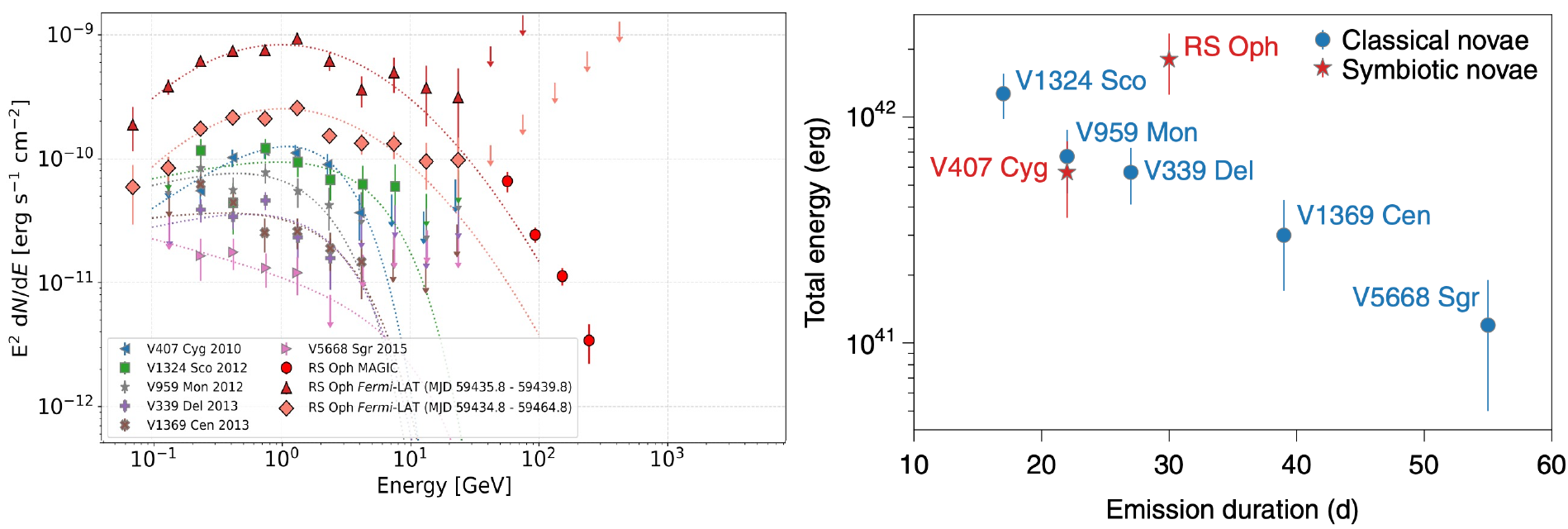}
    \caption{\textit{Left:} Comparison of the spectrum from RS Ophiuchi with spectra of other \textit{Fermi}-LAT-detected novae.  
    \textit{Right:} Total energy versus the duration of the RS Ophiuchi outburst compared to other novae detected with the \textit{Fermi}-LAT. Figures taken from \cite{2022NatAs...6..689A}.}
      \label{fig:sn_pred}
\end{figure*}

\subsection{Particle Acceleration and Radiation Mechanisms in Novae}

\subsubsection{Shock Formation and Acceleration Sites}

The gamma-ray emission from novae is widely attributed to particle acceleration
at shocks generated during the interaction of the nova ejecta with the
surrounding medium. Two main classes of shocks are expected. In classical
novae, internal shocks arise when fast outflows driven by ongoing nuclear
burning on the WD surface collide with slower material ejected earlier in the
outburst. In symbiotic novae, external shocks dominate, forming as the rapidly
expanding ejecta plow into the dense wind of the red giant companion
\cite{2007ApJ...663L.101T,2012BaltA..21...62H}.

In both cases, the resulting shocks share similarities with young supernova
remnants, albeit on much smaller spatial and temporal scales. Typical shock
velocities of $\sim10^3$~km\,s$^{-1}$ and upstream densities orders of magnitude
higher than in the interstellar medium provide favorable conditions for efficient
diffusive shock acceleration (DSA). Observational evidence for nonlinear DSA,
including shock deceleration and discrepancies between infrared- and
X-ray-derived shock velocities, was already identified during the 2006 outburst
of RS~Ophiuchi \cite{2007ApJ...663L.101T}.

\subsubsection{Leptonic and Hadronic Emission Channels}

Once accelerated, non-thermal particles can produce \gray through several
radiative channels. Relativistic electrons may emit via IC
scattering of optical and infrared photons from the nova photosphere and, in the
case of symbiotic systems, from the red giant companion. They can also contribute
via non-thermal bremsstrahlung in dense environments. However, electrons suffer
strong radiative losses, primarily due to IC cooling, which limits the maximum
energies they can reach within the short duration of a nova outburst
\cite{2022NatAs...6..689A}.


{In contrast, relativistic protons lose energy predominantly through inelastic proton--proton collisions, with characteristic timescales that are often longer than the duration of the gamma-ray emission. This reduces the efficiency of \graya emission processes, but also allows protons to be accelerated to higher energies and naturally explains the extension of the gamma-ray spectrum into the VHE regime. Hadronic interactions produce neutral pions, which promptly decay into gamma rays, yielding broad spectra that can extend from the GeV to the TeV domain. At the same time, these interactions inevitably also produce charged pions, whose decay leads to high-energy neutrinos via the $\pi^{\pm} \rightarrow \mu^{\pm} \rightarrow e^{\pm}$ chain. Therefore, $\gamma$-ray transients powered by hadronic processes are, in general, also expected to be sources of neutrinos, although the predicted fluxes are typically low and remain challenging for current detectors (see, e.g., \cite{2026NewAR.10201747R}).}

\subsubsection{Constraints from RS Ophiuchi}

The combined \textit{Fermi}-LAT, MAGIC, H.E.S.S., and LST-1 observations of the
2021 outburst of RS~Ophiuchi provide the most stringent constraints on nova
gamma-ray models to date. The smooth, single-component spectrum observed from
$\sim50$~MeV up to $\sim250$~GeV is well reproduced by hadronic models assuming
a power-law proton distribution with an exponential cutoff that increases with
time \cite{2022NatAs...6..689A,2025A&A...695A.152A}. In contrast, purely leptonic
models require ad hoc spectral breaks to counteract severe IC cooling and are
strongly disfavored by the data.

Time-dependent modeling further indicates that the VHE emission peaks a few days
after the optical and GeV maxima, consistent with a scenario in which the maximum
energy of accelerated protons increases as the shock propagates and evolves
\cite{2022Sci...376...77H}. The inferred total energy transferred to relativistic
protons in RS~Ophiuchi is of order $\sim10^{43}$--$10^{44}$~erg, corresponding to a
few percent of the kinetic energy of the ejecta.

\subsubsection{Long-term Particle Escape and Cosmic-Ray Contribution}

Beyond the prompt gamma-ray phase, a fraction of the non-thermal particle
population accelerated at nova shocks is expected to escape the expanding shell
and propagate into the surrounding medium. This process is particularly relevant
for hadrons, whose radiative loss timescales generally exceed both the duration
of the \graya emission and the dynamical timescale of the ejecta. As a result,
protons accelerated to GeV--TeV energies may survive long enough to diffuse into
the ambient circumstellar or interstellar environment.

For the 2021 outburst of RS~Ophiuchi, time-dependent modeling of the combined
\textit{Fermi}-LAT and VHE data implies a total energy transferred to accelerated
protons of order $E_{\rm p}\sim10^{43}$--$10^{44}$~erg
\cite{2022NatAs...6..689A,2025A&A...695A.152A}. While this represents only a small
fraction ($\lesssim$1\%) of the kinetic energy of the nova ejecta, it is
comparable to the energy injected into cosmic rays by individual stellar-wind
termination shocks and other localized Galactic accelerators. For a single
eruption, the contribution to the global Galactic cosmic-ray (CR) budget is
therefore negligible, amounting to $\lesssim0.1$--0.2\% when integrated over the
entire nova population.

However, recurrent novae such as RS~Ophiuchi represent a qualitatively different
case. With recurrence timescales of $\sim10$--20~yr, repeated injections of
relativistic protons over $10^4$--$10^5$~yr can lead to a cumulative CR energy
input of $\sim10^{46}$--$10^{47}$~erg within a localized region of a few parsecs.
Under reasonable assumptions for diffusion coefficients in dense stellar-wind
environments, this can result in a persistent enhancement of the CR density
relative to the Galactic background \cite{2023JHEAp..38...22B}. Such ``nova
super-remnants'' would be long-lived, even though individual \graya outbursts
remain short-lived and episodic.

From an observational perspective, this scenario implies that the most promising
signatures of nova-driven CR acceleration beyond the prompt phase may not be
transient GeV--TeV flares, but rather extended, low-surface-brightness emission
produced by escaped particles interacting with ambient gas and radiation fields.
In particular, inverse Compton emission from accumulated electrons and
$\pi^0$-decay \gray from long-lived protons could, in principle, be
detectable with next-generation instruments for nearby recurrent systems.
Conversely, the absence of such extended emission would place important
constraints on particle escape efficiencies and diffusion properties in nova
environments.


In this context, individual novae contribute less than 0.2\% to the total Galactic CR population, recurrent systems like RS Ophiuchi could locally enhance the CR density within several parsecs. For RS Ophiuchi, the estimated energy injected into accelerated protons is about $\sim$ 4.4 $\times$ 10$^{43}$, consistent with modest but non-negligible CR enrichment of its immediate environment.

\subsection{Future Prospects and Open Questions}

The detection of very-high-energy gamma rays from RS~Ophiuchi has fundamentally
changed the perspective on novae as particle accelerators, demonstrating that
nova-driven shocks can accelerate hadrons to at least TeV energies under
favorable conditions. Together with the growing population of GeV-detected
novae, these results indicate that efficient non-thermal processes may be a
generic feature of nova eruptions, while their detectability at VHE energies
depends sensitively on environmental factors such as ambient density, shock
geometry, and energetics.

A key question remains whether all novae emit GeV gamma rays. Current evidence suggests that most novae within a few kiloparsecs—up to $\approx$4\,kpc are indeed detectable by Fermi-LAT, with some more distant examples from the Galactic bulge also observed.

Another open issue concerns the uniqueness of RS Ophiuchi: why was it visible at VHE energies when others were not? Comparisons of Fermi-LAT spectra show that RS Ophiuchi’s gamma-ray flux was nearly two orders of magnitude higher than in previous events, though its spectral and temporal behavior otherwise appeared typical as can be seen in Figure~\ref{fig:sn_pred}.

The anticipated eruption of the recurrent symbiotic nova T Coronae Borealis (T CrB)  offers a unique opportunity to probe particle acceleration and radiation mechanisms in detail. T CrB erupted in 1217, 1787, 1866 and 1946, whereas the first two are based on historical reports fitting the expectations with high confidence for a T CrB eruption \cite{2023JHA....54..436S}.
There are several predictions for the next eruption ranging between 2023-2027 (see \cite{2024RNAAS...8..272S} and references therein).
A few predictions for a 2024–2025 eruption (e.g. \cite{2023MNRAS.524.3146S}) rely on similarities between the 1946 pre-eruption light curve and variations seen since 2023, interpreting these as signs of an impending outburst. However, since no pre-eruption light curves exist for the former events, it is unknown whether these light curve features are truly recurrent.
With a distance of only $\approx$900\,pc, it is considerably closer and optically brighter than RS Ophiuchi ($\approx$ 2.7\,kpc), potentially reaching V $\approx$ 2.5\,mag at peak brightness. This proximity will enable unprecedented multi-wavelength coverage and more detailed modeling of non-thermal emission processes. Coordinated global observation campaigns are in preparation.

The Galactic transients science case for the CTAO emphasizes novae as promising
time-domain targets, where enhanced sensitivity at sub-TeV energies and
real-time analysis capabilities will enable the detection of shorter-lived and
fainter VHE components than accessible today \cite{2025MNRAS.540..205A}.

More broadly, novae are now established as a new class of Galactic gamma-ray
transients that bridge the gap between stellar-scale explosions and classical
cosmic-ray accelerators such as supernova remnants. While their contribution to
the global Galactic cosmic-ray budget is likely modest, their role as localized,
recurrent sources of relativistic particles and their impact on the surrounding
medium remain open and testable questions. Addressing these issues will require
the combined capabilities of next-generation gamma-ray observatories, wide-field
optical surveys, and dedicated multiwavelength follow-up campaigns.




\section{Microquasars}

\subsection{Introduction}\label{sec:mq_intro}

Microquasars are X-ray binary systems in which a compact object ---either a stellar-mass black hole or a neutron star--- accretes matter from a companion star and launches relativistic jets. They were first recognized in the early 1990s, when the discovery of apparent superluminal radio jets in GRS~1915+105 \cite{1994Natur.371...46M} demonstrated that relativistic effects play an essential role in shaping the observed properties of Galactic jet sources. This discovery fostered the expectation that studies of these nearby systems could provide insight into the physical processes operating in quasars and other powerful extragalactic jet sources. While this analogy has not yet led to a direct breakthrough in either domain, microquasars nevertheless serve as unique nearby laboratories for investigating accretion and jet-launching processes that, in active galactic nuclei (AGNs), are accessible only at much larger distances and on far longer timescales.


The observational phenomenology of microquasars is diverse and complex, reflecting the interplay between accretion flows,
jet activity, {jet-stellar wind interaction,} and emission across the electromagnetic spectrum. By analogy with the classification commonly adopted for
X-ray binary systems, microquasars are divided into high-, intermediate-, and low-mass systems, depending on the mass of
the non-degenerate companion star. Systems with OB-type companions are classified as high-mass microquasars, while those
with less massive companions fall into the intermediate- or low-mass categories. In all cases, it is thought that the
compact object powers relativistic jets through the accretion of matter transferred from the companion star. As the
accreted material approaches the compact object, it forms an accretion disk that is typically observed in the X-ray
band. The X-ray emission often exhibits distinct spectral states—commonly referred to as ``low-hard'' and ``high-soft''
states—associated with changes in the accretion geometry and jet power \cite{2006csxs.book..157M}. Transitions between
these states are frequently accompanied by pronounced variability. In particular, a persistent jet is expected during
the low-hard state, while transient jet ejections are commonly associated with transitions from the low-hard to the
high-soft state \cite{2004MNRAS.355.1105F}. Correlations between radio and X-ray luminosities, reflecting the coupling
between accretion and ejection processes, have been proposed
\cite{2003MNRAS.344...60G,2003A&A...400.1007C}.


The formation of relativistic jets and the production of non-thermal synchrotron radiation are the key features that
distinguish microquasars from other X-ray binary systems. Non-thermal emission from microquasar jets has been resolved in
the radio band over a wide range of spatial scales \cite{2005ASPC..340..269R}, and in some cases {it is} also detected in X-rays
at large distances from the binary system \cite{2002Sci...298..196C}. These observations provide clear evidence that
efficient particle acceleration operates at different locations along microquasar jets and under a broad range of
physical conditions (see \cite{2009IJMPD..18..347B} for a review). This interpretation is further supported by detections in the \graya energy band.


{In large-scale microquasar jets, the properties of target fields are comparable to whose in the ISM \cite{2009A&A...497..325B,2011A&A...528A..89B}, leading to typical cooling times longer than a kyr time scale. Thus, } unless Galactic microquasars are capable of producing
ultra-relativistic outflows, extreme variable emission and flaring
activity must be generated on spatial scales comparable to the binary
system itself. {Flaring activity can be triggered by a change of the jet state or by a stochastic interaction of the jet with the stellar wind, in particular due to a penetration of a dense stellar wind clump \cite{2017A&A...604A..39D}.}

{On the binary scales}, transient \graya emission is expected to
arise from interactions of relativistic particles with local target
fields. For relativistic electrons, the dominant channel for \graya
production is IC scattering. Within the binary
system, the most important target photon fields are provided by the
donor star and by the accretion disk. The luminosity of the optical
companion can range from \(\sim0.1 L_\odot\) in low-mass X-ray binaries,
such as A0620$-$00, up to \(\sim10^6 L_\odot\) in extreme systems like
Cygnus~X$-$3, with stellar temperatures spanning from a few thousand to
several tens of thousands of kelvin. The luminosity of the accretion
disk is a standard proxy for the jet power, which can be close or even
excess the Edington limit.

The emission from the accretion disk approximately follows a
multicolor blackbody spectrum, peaking in the X-ray band for a
stellar-mass accreting black hole. Typically one expects a jet luminosity comparable to the energy dissipated in the disk: \(L\mysub{disk}\sim
L\mysub{jet}\). Introducing the jet magnetization parameter \(\sigma\),
the magnetic field strength in the jet can be related to the jet power
as

\be
\frac{c\beta}{4} R\mysub{jet}^2 {\cal B}\mysub{jet}^2 = \sigma L\mysub{jet}\,,
\ee
where \(R\mysub{jet}\) is the jet cross-section; \({\cal B}\mysub{jet}\) is its magnetic field; and \(\beta\) is the dimensionless jet speed.

The energy density of disk photons at a distance \(z\) along the jet
is \(w\mysub{disk}=L\mysub{disk}/(4\pi z^2c)\), while the magnetic energy
density is \(w\mysub{B}={\cal B}^2\mysub{jet}/(8\pi)\). Their ratio can be estimated as

\be
\frac{w\mysub{disk}}{w\mysub{B}} \approx \frac{\beta}{2\sigma}\qty(\frac{R\mysub{jet}}{z})^2 \qty(\frac{L\mysub{disk}}{L\mysub{jet}})\,.
\ee

This estimate shows that disk photons are unlikely to power strong
\graya flares, as synchrotron losses typically dominate,
i.e. \(w\mysub{disk}\ll w\mysub{B}\), unless the jet magnetization is
very low or the disk emission escapes though a very narrow funnel (see, e.g., \cite{2024NatAs...8.1031V}). 


We also note that the ratio of the energy densities of the target fields reflexes the ratio of the luminosities of the corresponding components only if IC scattering proceeds in the Thomson regime. The Klein-Nishina effect strongly suppresses the IC emission. For an X-ray target photon field, scattering of GeV proceeds already in the Klein-Nishina regime, provided that $\unit{keV}\times\unit{GeV}\approx 4 m_e^2c^4$.  

A similar estimate for stellar photons yields
\be
\frac{w\mysub{star}}{w\mysub{B}} \approx \frac{\beta}{2\sigma}\qty(\frac{R\mysub{jet}}{r})^2 \qty(\frac{L\mysub{star}}{L\mysub{jet}}),
\ee
where \(r\simeq\sqrt{D^2+z^2}\) is the distance from the donor star to
the emission region, \(D\) is the orbital separation, and the jet is
assumed to be approximately perpendicular to the orbital plane. If the
jet power approaches the Eddington luminosity for a stellar-mass black
hole,
\be
L\mysub{jet}\approx 4\times10^{38}\qty(\frac{M\mysub{bh}}{3M_\odot})\unit{erg\,s^{-1}},
\ee
the above ratio becomes
\be\label{eq:photon_to_B}
\frac{w\mysub{star}}{w\mysub{B}} \approx \qty(\frac{L\mysub{star}}{8\times10^{38}\unit{erg\,s^{-1}}})\qty(\frac{M\mysub{bh}}{3M_\odot})^{-1}\qty(\frac{R\mysub{jet}}{r})^2\frac{\beta}{\sigma}\,.
\ee

Relativistic effects further modify this balance {for the part of the jet located within or close to the binary system}. In the jet comoving
frame, the stellar photon energy density is enhanced by a factor
\(\sim\Gamma\mysub{jet}^2=1/(1-\beta^2)\), while the magnetic energy density is
reduced by a similar factor. As a result, the \emph{relative}
efficiency of IC cooling with respect to synchrotron cooling can be
enhanced by a factor \(\sim\Gamma\mysub{jet}^4\). However, in the
energy range where IC scattering enters the Klein--Nishina regime, the
IC cooling rate is strongly suppressed, making it difficult to
produce a {high-luminosity (i.e., comparable to the jet power)} flare in the VHE regime, particularly given that the
temperatures of the donor stars in the most prominent cases are very
high, \(\qty(3-10)\times10^4\unit{K}\) in Cyg~X$-$1 and Cyg~X$-$3,
respectively.

Notably, the scattering regime in which the Klein--Nishina effect
becomes important largely overlaps with the energy range where
\(\gamma\gamma\) absorption on stellar photons is significant.
Consequently, even though relativistic motion favors IC scattering in
principle, synchrotron losses are expected to dominate in this regime,
and strong IC-driven \graya flares in the VHE band are not anticipated.
{However, provided absence of persistent emission in the VHE band and large collection area of VHE instruments, the detection of VHE flares is still possible. }

While the above discussion concerns the processes relevant for electrons, we cannot exclude \emph{a priori} that protons are also accelerated in the jets of microquasars. The primary channels through which protons can generate \graya emission are related to the creation of neutral pions, $\pi^0$, via interactions with proton or photon targets (see, e.g., \cite{2004vhec.book.....A}). The typical cross sections for these processes are $\sigma_{pp}\approx 30\unit{mb}$ and $\sigma_{p\gamma}\approx 300\unit{\upmu b}$.
In what follows, for order-of-magnitude estimates, we adopt a fiducial normalization $\sigma\sim3\unit{mb}$, which lies between the characteristic values of the two channels.

The $pp$ process has a threshold kinetic energy of $T_{p,\mysub{th}}\simeq 280\unit{MeV}$ in the laboratory frame, corresponding to the onset of pion production. The threshold energy for the photo-meson process depends on the energy of the target photon, $\hbar\omega$, and can be estimated as
\be\label{eq:pgamma_thr}
E\mysub{p\gamma, th}\approx \frac{m_pm_\pi c^4}{\hbar\omega}
\approx 10\qty(\frac{\hbar\omega}{10\unit{eV}})^{-1}\unit{PeV}\,.
\ee

The potential of hadronic processes to generate flare-like emission is determined by the density of the corresponding target. If protons cannot lose a notable fraction of their energy while interacting with the target on a timescale comparable to the observed variability, their emission will appear steady (and possibly spatially extended). The characteristic cooling time,
\be
t\mysub{cool} = \qty(n\mysub{t}\sigma f c)^{-1}\,,
\ee
where $n\mysub{t}$ is the density of the target and $f\approx 0.1$--$0.5$ is the fraction of energy lost by a proton per interaction, should be sufficiently short to produce flaring emission. This requirement implies
\be
n\mysub{t} \gtrsim 3\times10^{11}
\qty(\frac{t\mysub{var}}{1\unit{day}})^{-1}
\qty(\frac{\sigma}{3\unit{mb}})^{-1}
\unit{cm^{-3}}\,.
\ee

The main source of target protons in microquasars is the stellar wind, which at a distance $r$ from the star has a characteristic density
\be\label{eq:pp_target}
n\mysub{wind}\approx 3\times10^{10}
\qty(\frac{\dot{M}\mysub{wind}}{10^{-6}M_\odot\unit{yr^{-1}}})
\qty(\frac{v\mysub{wind}}{10^3\unit{km\,s^{-1}}})^{-1}
\qty(\frac{r}{10^{12}\unit{cm}})^{-2}
\unit{cm^{-3}}\,.
\ee
This density is marginally sufficient to cool protons on a day timescale. The main challenge in this scenario is the confinement of protons within a region of size $\sim10^{12}\unit{cm}$, corresponding to only $\sim30$ light-seconds, for a sufficiently long time. This estimate indicates that rather special conditions may be required to generate \graya flares via the $pp$ process in microquasars. {For typical stellar wind speed of \(10^3\unit{km\,s^{-1}}\), plasma travels approximately \(10^{13}\unit{cm}\) per day bringing particles to regions with lower density of the target and making the cooling even less efficient. Thus, producing flaring emission by the $pp$ process may require very efficient non-radiative cooling and consequently very powerful injection of relativistic protons.} 

The photo-meson process is characterized by a smaller cross section; however, the photon target can be much more abundant. For a source with luminosity $L_*\sim10^{38}\unit{erg\,s^{-1}}$ and a characteristic photon energy of $\hbar\omega_*\sim10\unit{eV}$, the density of the photon target at distance $r$ is
\be\label{eq:pgamma_target}
n\mysub{ph}\approx 2\times10^{13}
\qty(\frac{L_*}{10^{38}\unit{erg\,s^{-1}}})
\qty(\frac{\hbar\omega_*}{10\unit{eV}})^{-1}
\qty(\frac{r}{10^{12}\unit{cm}})^{-2}
\unit{cm^{-3}}\,.
\ee
{Although we deal with order-of-magnitude estimates in Eqs.~\ref{eq:pp_target} and ~\ref{eq:pgamma_target}, the } higher density of the photon target makes this channel more favorable for producing \graya flares in microquasars {compared to inelastic $pp$ interactions, provided that protons are accelerated to sufficiently high energies to overcome the photo-meson threshold. It is important to note here that the threshold energy for the $p\gamma$ channel is typically significantly higher than the threshold energy for the $pp$ process (see Eq.~\ref{eq:pgamma_thr}),  thus such hypothetical flares should be accompanied by the emission of lower energy protons via the $pp$ channel}.

The phenomenological arguments presented above suggest that, if relativistic particles are accelerated in microquasar jets, these systems may produce flaring emission in the HE band via the IC process and in the UHE domain via the photo-meson channel. In the VHE band, flaring emission may be produced either via the IC mechanism---with systems hosting cooler companion stars being more favorable---or via the $pp$ channel, provided that unusually high target densities are present.


While early experiments in the \graya band reported a number of binary systems emitting VHE and even UHE \gray (see
\cite{1985ICRC....9..407H} for a brief discussion), these systems remained undetected for a long time by newer generations
of \graya instruments (see, e.g., \cite{2018A&A...612A..10H,2018A&A...612A..14M}). Only recently has a significant number of
Galactic microquasars been reliably detected through observations with HAWC, H.E.S.S., and LHAASO
\cite{2018Natur.562...82A,2024arXiv241008988L,2024ApJ...976...30A,2024Sci...383..402H,2025arXiv251110537A}. These detections
are predominantly associated with large-scale jets, where the long radiative cooling times imply the absence of rapid
flaring activity. Nevertheless, several microquasars have exhibited flaring-like behavior in the \graya band
\cite{2007ApJ...665L..51A,2009Sci...326.1512F,2016A&A...596A..55Z}. In what follows, we focus on several representative
systems and discuss their role as potential \graya transients.

\subsection{Cygnus X--1: Transient High- and Very-High-Energy Emission}

Cygnus~X--1 is the prototypical Galactic black-hole X-ray binary and one of the
best-studied microquasars. The system consists of a stellar-mass black hole with
mass $M_{\rm BH}\simeq15\pm4\,M_\odot$ \cite{2025MNRAS.tmp.2104O} accreting from the O9.7~Iab supergiant
HDE~226868 in a 5.6-day orbit at a distance of $d\simeq2.2\pm0.2$~kpc \cite{2021Sci...371.1046M}. While Cygnus~X--1
is persistently bright in X-rays, its emission at HE and VHE is highly
intermittent and dominated by rare transient episodes rather than steady
radiation \cite{2007ApJ...665L..51A,2010ApJ...712L..10S,2016A&A...596A..55Z}.

\subsubsection{VHE flaring activity}

The first indication of VHE gamma-ray emission from Cygnus~X--1 was reported by
the MAGIC Collaboration based on observations performed in 2006 with the
single-dish MAGIC telescope \cite{2007ApJ...665L..51A}. While no steady emission was
detected over the full exposure of approximately 40~h, a statistically
significant excess was observed on 24 September 2006, coincident with the rising
phase of a strong hard X-ray flare detected by \emph{INTEGRAL}, \emph{Swift}, and
\emph{RXTE}. The excess was detected over a time interval of 154~min with a
pre-trial significance of $4.0\sigma$ ($3.2\sigma$ post-trial), and reached
$4.9\sigma$ ($4.1\sigma$ post-trial) over a shorter 79-min interval. The inferred
VHE spectrum during this episode is compatible with a power-law shape extending
above 100~GeV.

This event remains the only reported hint of TeV emission from Cygnus~X--1.
Subsequent and substantially deeper observations with the MAGIC telescopes,
covering $\sim97$~h of data collected between 2007 and 2014 in both hard and soft
X-ray states, yielded no significant VHE detection \cite{2017MNRAS.472.3474A}. These data
allowed MAGIC to place stringent upper limits on steady emission above 200~GeV at
the level of $2.6\times10^{-12}$~ph~cm$^{-2}$~s$^{-1}$ during the hard state and
$1.0\times10^{-11}$~ph~cm$^{-2}$~s$^{-1}$ during the soft state (95\% confidence
level). No evidence for orbital modulation or short-term variability was found,
even when performing nightly and phase-resolved analyses. These results rule out
persistent VHE emission at the sensitivity level of MAGIC and demonstrate that
the 2006 signal, if astrophysical, corresponds to an exceptionally rare transient
event.

\subsubsection{High-energy gamma-ray flares}

At lower energies, transient emission from Cygnus~X--1 has been firmly detected in
the HE gamma-ray band. Observations with the AGILE
satellite revealed several short-lived flares, most notably on 16 October 2009,
with an integral flux of $(2.32\pm0.66)\times10^{-6}$~ph~cm$^{-2}$~s$^{-1}$ between
0.1 and 3~GeV, lasting approximately 1--2 days and occurring during the hard
X-ray state \cite{2010ApJ...712L..10S}. Additional AGILE flares were reported in 2010,
again confined to hard or intermediate spectral states and characterized by
similar durations.

Long-term monitoring with \emph{Fermi}-LAT has established the presence of a weak
but statistically significant HE gamma-ray signal associated with Cygnus~X--1
when restricting the analysis to the hard state \cite{2016A&A...596A..55Z}. Integrating 7.5
years of PASS8 data, a detection at the $\sim7\sigma$ level was obtained above
60~MeV, with a spectrum extending up to $\sim20$~GeV and well described by a
power-law photon index $\Gamma\simeq2.3\pm0.1$. The corresponding luminosity is
$L_\gamma\simeq5\times10^{33}$~erg~s$^{-1}$. No significant emission is detected
during the soft state, for which only upper limits are derived. Earlier LAT
analyses reported marginal detections or upper limits at the $3$--$4\sigma$ level,
consistent with a highly intermittent emission scenario \cite{2013MNRAS.434.2380M}.

\subsubsection{Multiwavelength context and state dependence}

A key observational result emerging from both HE and VHE studies is the strong
association of gamma-ray activity with the hard X-ray spectral state of
Cygnus~X--1. All reported GeV flares and the persistent HE signal detected by
\emph{Fermi}-LAT occur exclusively during this state, when compact, steady radio
jets are present. In contrast, no HE or VHE emission has been detected during the
soft state, despite dedicated MAGIC observations with comparable sensitivity
\cite{2017MNRAS.472.3474A}. This phenomenological connection provides robust observational
evidence that gamma-ray production in Cygnus~X--1 is closely linked to jet activity
rather than to the accretion disc alone.

Hints of orbital modulation have been reported in the HE band, with enhanced
emission preferentially occurring near superior conjunction, although the
statistical significance remains limited \cite{2016A&A...596A..55Z}. At VHE energies, the
lack of repeated detections prevents any meaningful assessment of orbital effects.

\subsubsection{Extension to multi-TeV energies}

Independent constraints at multi-TeV energies have been provided by HAWC, which
reported no significant steady emission from Cygnus~X--1 and derived stringent
upper limits above $\sim1$~TeV based on 1038 days of observations
\cite{2021ApJ...912L...4A}. These limits further disfavour persistent VHE emission and
support the conclusion that any TeV activity from Cygnus~X--1 must be rare and
transient.

Recent observations by the LHAASO experiment have reported gamma-ray emission from
the Cygnus region extending to energies of order $\sim100$~TeV, with a
significance of $4.4\sigma$ above 25~TeV \cite{2024arXiv241008988L}. While the spatial
association and temporal behaviour of this ultrahigh-energy emission remain under
active investigation, these observations indicate that particle acceleration to
extreme energies is possible in the broader Cygnus environment. The connection
between this emission and the transient MeV--TeV activity of Cygnus~X--1 is not yet
established observationally.





\subsection{Cygnus~X--3: Transient High- and Very-High-Energy Emission}

Cygnus~X--3 is a high-mass X-ray binary and microquasar located in the Galactic
plane at a distance of $9.67^{+0.53}_{-0.48}~\mathrm{kpc}$ \cite{2023ApJ...959...85R}. The system consists of a compact
object of still-uncertain nature, whose dynamically inferred mass is compatible
with either a neutron star or a stellar-mass black hole \cite{2010ApJ...718..488S,2013MNRAS.429L.104Z}, orbiting a
Wolf--Rayet companion star with an orbital period of 4.8~h \cite{2013MNRAS.429L.104Z}. The Wolf--Rayet donor is estimated to have a mass
in the range $8$--$14~M_{\odot}$ \cite{2017MNRAS.472.2181K}.

The system is characterized by extreme multiwavelength
variability and is among the most active Galactic sources at radio and
HE \graya energies. In contrast to its pronounced
HE activity, Cygnus~X--3 has not been firmly detected at VHE, despite extensive observational campaigns. Observations with LHAASO revealed, however, flaring like emission in the UHE band, making Cygnus~X--3 the first UHE transient source \cite{2025arXiv251216638L}.

\subsubsection{High-energy gamma-ray flares}

Cygnus~X--3 is the first microquasar for which variable HE gamma-ray emission has
been firmly established. The \emph{Fermi}-LAT discovery of modulated emission
above 100~MeV demonstrated orbital modulation at the 4.8~h period, securing the
association with the binary system \cite{2009Sci...326.1512F}. The detected emission
extends up to several tens of GeV and exhibits strong variability on timescales
of hours to days.

Long-term monitoring with AGILE revealed a series of intense HE gamma-ray flares
between 2007 and 2009 \cite{2012A&A...545A.110P}. Seven major events were detected,
each lasting typically one to two days, with photon fluxes above 100~MeV
reaching $(1$--$3)\times10^{-6}$~ph~cm$^{-2}$~s$^{-1}$. These flares are not
randomly distributed in source states: they occur preferentially during soft
X-ray states characterized by suppressed hard X-ray emission and are often
observed shortly before the onset of major radio flares.

A tight phenomenological connection between HE gamma-ray activity and radio
behavior has been established through coordinated multiwavelength campaigns.
In particular, transitions into or out of the radio-quenched state frequently
precede strong GeV flares, while extended quenched intervals are devoid of
detectable HE emission \cite{2012MNRAS.421.2947C}. The peak isotropic luminosity during
GeV flares reaches $L_\gamma\sim10^{36}$~erg~s$^{-1}$, making Cygnus~X--3 one of
the most luminous Galactic gamma-ray transients.

\subsubsection{VHE observations and non-detections}

Despite its pronounced HE flaring activity, Cygnus~X--3 has not been confirmed
as a VHE gamma-ray emitter. MAGIC observations conducted between 2006 and 2009,
including periods contemporaneous with strong GeV flares, yielded no significant
signal above 250~GeV \cite{2010ApJ...721..843A}. An upper limit on the integral flux
above 250~GeV of $2.2\times10^{-12}$~ph~cm$^{-2}$~s$^{-1}$ (95\% C.L.) was derived,
representing the most stringent constraint at the time.

VERITAS observations between 2007 and 2011, covering multiple radio and X-ray
states and totaling approximately 44~h of exposure, likewise resulted in no VHE
detection \cite{2013ApJ...779..150A}. The corresponding upper limits constrain
steady or recurrent TeV emission to well below the Crab Nebula flux level,
including during epochs of confirmed GeV activity.

More recently, the MAGIC Collaboration presented the results of an extensive
long-term campaign spanning 2013--2024, comprising $\sim130$~h of observations
\cite{BarriosJiménez:2025yf}. This dataset represents the deepest and most complete
VHE study of Cygnus~X--3 to date and yields the strongest upper limits between
100~GeV and a few TeV. No evidence for transient or persistent VHE emission is
found on nightly, orbital-phase-resolved, or state-selected timescales.

\subsubsection{Multi-TeV and PeV constraints}

At multi-TeV energies, wide-field air-shower observatories provide the most
stringent constraints on long-lived emission from Cygnus~X--3. Earlier searches
with HAWC, based on several years of observations, did not reveal any
statistically significant steady emission above $\sim1$~TeV and resulted in
upper limits that constrained persistent VHE components
\cite{2021ApJ...912L...4A}. Due to its integration times and analysis strategy,
HAWC is primarily sensitive to steady or long-duration emission and has limited
sensitivity to short-lived or highly intermittent flares.

This picture has been fundamentally revised by recent observations with
LHAASO. Using data from the KM2A array, LHAASO reported the detection of
variable gamma-ray emission from Cygnus~X--3 extending from $\sim60$~TeV up to
$3.7$~PeV, with a peak detection significance of $9.6\sigma$ for photons above
$0.1$~PeV \cite{2025arXiv251216638L}. The emission is strongly variable on month-long
timescales and is detected exclusively during GeV high states identified by
\emph{Fermi}-LAT. When restricting the analysis to these active periods, the
significance of the PeV signal increases to $11.5\sigma$, while no significant
excess is observed during quiescent states.

The intrinsic spectral energy distribution during high states follows a hard
power law with photon index $\Gamma = 2.18 \pm 0.14$ above $0.1$~PeV, making
Cygnus~X--3 the hardest ultra-high-energy gamma-ray source detected by LHAASO to
date. After correcting for $\gamma\gamma$ absorption on the cosmic microwave
background and interstellar radiation fields, the spectrum exhibits a pronounced
hardening above $\sim1$~PeV. Five photons with energies exceeding $1$~PeV were
detected within $\sim10$ arcmin of the source position, including events at
$3.08 \pm 0.34$~PeV and $3.73 \pm 0.41$~PeV, representing the highest-energy
photons ever associated with an X-ray binary.

The observed variability constrains the size of the emission region to
${\cal R} \lesssim 2 \times 10^{17}$~cm for the month-scale activity, while the presence
of a $3.2\sigma$ indication of orbital modulation at the 4.8-hour period further
supports an origin within or very close to the binary system itself. Taken
together, these results demonstrate that Cygnus~X--3 does not exhibit persistent
multi-TeV emission, consistent with HAWC upper limits, but instead operates as a
highly transient ultra-high-energy gamma-ray source capable of accelerating
particles to tens of PeV during episodic flaring states.

\subsubsection{Phenomenological picture}

Observationally, Cygnus~X--3 exhibits a strongly energy-dependent and highly
transient gamma-ray phenomenology. At high energies, the source
displays recurrent, day-scale flares with high luminosities,
$L_\gamma \sim 10^{36}$~erg~s$^{-1}$, tightly linked to accretion-state transitions
and radio-jet activity. These GeV flares are well established by \emph{Fermi}-LAT
and AGILE observations and, in some cases, show orbital modulation at the
4.8-hour binary period.

At VHE, dedicated observations with MAGIC
and VERITAS have yielded only non-detections, even during periods of intense GeV
activity, placing stringent upper limits on any persistent or recurrent emission
in this energy range. However, this absence of detections at $\sim100$~GeV does
not extend to higher energies. Recent LHAASO observations have revealed strongly
variable gamma-ray emission from Cygnus~X--3 at multi-TeV to PeV energies, detected
exclusively during GeV high states and evolving on month-long timescales. This
demonstrates that Cygnus~X--3 is capable of accelerating particles to at least
tens of PeV, despite the lack of detectable emission at intermediate VHE
energies.

Taken together, these results indicate that \graya emission from Cygnus~X--3
is neither steady nor broadband, but instead occurs in distinct energy bands
under specific physical conditions. The phenomenology suggests efficient
particle acceleration during episodic flaring states, with emission suppressed
or absorbed at $\sim100$~GeV while re-emerging at multi-TeV and PeV energies.
Cygnus~X--3 therefore represents an extreme example of a Galactic transient
source in which GeV, VHE, and UHE emission probe fundamentally
different acceleration regimes.

\subsection{V404 Cygni and Scorpius~X--1}

In addition to Cygnus~X--1 and Cygnus~X--3, two further Galactic microquasars,
V404~Cygni and Scorpius~X--1, have been extensively investigated as potential
sources of transient HE and VHE \graya emission.
V404~Cygni is a low-mass X-ray binary hosting a $\sim9\,M_\odot$ black hole and is
characterized by rare but extremely luminous outbursts, most notably the June
2015 event, during which the source reached X-ray fluxes exceeding tens of
Crab units and exhibited violent multiwavelength variability. During this
outburst, AGILE reported a transient enhancement below $\sim400$~MeV at a
significance of $\sim4\sigma$, temporally coincident with strong radio and X-ray
activity \cite{2017ApJ...839...84P}. Similar low-significance excesses were also reported
in early \emph{Fermi}-LAT analyses. However, subsequent re-analyses using updated
background models and source catalogs demonstrated that these apparent GeV
signals were most likely due to source confusion with a nearby bright blazar
and to limitations of earlier diffuse emission models, yielding no statistically
significant HE emission attributable to V404~Cygni \cite{2021MNRAS.506.6029H}. At higher
energies, both MAGIC and VERITAS conducted observations during the 2015 outburst
and reported only upper limits above $\sim100$~GeV, constraining any VHE emission
to fluxes below a few percent of the Crab Nebula on hour-scale timescales
\cite{2017MNRAS.471.1688A,2016ApJ...831..113A}. These results indicate that, despite its extreme
accretion and jet activity, V404~Cygni does not produce detectable VHE emission,
or that such emission is either strongly suppressed or confined to very brief
intervals below current instrumental sensitivity.

Scorpius~X--1, the prototypical Z-type neutron-star low-mass X-ray binary, provides
a complementary case. The system launches persistent radio jets whose power is
strongly modulated by its X-ray spectral state, with the so-called Horizontal
Branch associated with enhanced non-thermal activity. Motivated by this behavior,
MAGIC performed dedicated observations contemporaneous with \emph{RXTE} coverage
to target periods when a powerful jet was present. No significant VHE signal was
detected, and stringent upper limits above 300~GeV were derived at the level of
$(2$--$3)\times10^{-12}\,\mathrm{cm^{-2}\,s^{-1}}$ \cite{2011ApJ...735L...5A}. These limits
imply that the ratio of TeV luminosity to jet power is
$L_{\mathrm{VHE}}/L_{\mathrm{j}}\lesssim10^{-3}$, placing strong constraints on
particle acceleration efficiency in the jets of neutron-star microquasars.
Together, the observational results for V404~Cygni and Scorpius~X--1 reinforce a
picture in which pronounced accretion and jet activity alone are insufficient to
guarantee detectable VHE emission, underscoring the highly selective and transient
nature of gamma-ray production in Galactic microquasars.

\subsection{Summary and Outlook: Microquasars}

Observations of Galactic microquasars demonstrate that \graya emission from
accreting compact binaries is intrinsically transient, strongly state-dependent,
and highly energy selective. Cygnus~X--1 and Cygnus~X--3 exemplify two distinct but
complementary manifestations of this behavior. Cygnus~X--1 exhibits weak and
intermittent emission at HE, detectable only during specific hard or
intermediate accretion states and occasionally appearing as day-scale GeV flares.
At VHE energies, a single night-long TeV flare reported in 2006 remains the
only indication of emission above 100~GeV, with extensive follow-up observations
placing stringent constraints on any persistent or recurrent component. These
results indicate that extreme particle acceleration in Cygnus~X--1, if present,
occurs only episodically and under narrowly defined physical conditions.

Cygnus~X--3, in contrast, is characterized by frequent, luminous GeV flares with peak luminosities of order
$10^{36}$~erg~s$^{-1}$, closely linked to accretion-state transitions and radio-jet activity. While no emission has been
detected at energies around $\sim100$~GeV despite deep observations with MAGIC and VERITAS, recent LHAASO observations
have revealed strongly variable gamma-ray emission extending into the multi-TeV and PeV regime during GeV high
states. This striking energy-dependent phenomenology demonstrates that efficient particle acceleration can occur in
Cygnus~X--3 under favorable conditions, while emission absorption, geometric effects, or absence of suitable radiation
processes suppress emission at intermediate VHE energies.

Table~\ref{tab:microquasar_comparison} highlights the contrasting gamma-ray
behavior of these two archetypal systems: Cygnus~X--1 shows rare, extreme
transient activity with otherwise negligible gamma-ray output, whereas
Cygnus~X--3 operates as a recurrent GeV flaring source capable of producing
ultra-high-energy emission on longer timescales. Together, they illustrate that
\graya production in microquasars is neither steady nor universal, but instead
depends sensitively on the nature of the compact object, the accretion regime,
and the surrounding environment.

While observations in the \graya band point to the operation of extreme acceleration processes in microquasar jets on binary-system scales, some of the detected features can be explained using relatively simple arguments.
First, efficient \graya flaring via IC scattering is possible primarily in systems with very luminous optical companions, {smaller binary separations,} at least mildly relativistic jets, relatively low jet magnetization, {and smaller BH masses (see Eq.~\ref{eq:photon_to_B})}. This provides a natural qualitative explanation for the much more frequent GeV flaring activity observed in Cygnus~X$-$3 compared to Cygnus~X$-$1 {consistently with claimed small mass of the relativistic companion \cite{2013MNRAS.429L.104Z}}. In systems with less luminous donor stars, the production of detectable \graya flares becomes increasingly unfavorable.

Furthermore, the HE energy band appears to be the primary domain in which IC processes can generate intense flaring emission. In the VHE domain, the Klein--Nishina effect suppresses the efficiency of this channel, making IC-driven flaring activity significantly less efficient.  In this context, the possible detection of a VHE flare from Cygnus~X$-$1 with the MAGIC telescopes appears to be a rather extreme event. Detecting VHE flares from Cygnus~X$-$3 is expected to be particularly challenging because of the higher temperature of its companion star: for a photon field with characteristic temperature of \(\sim100\unit{kK}\), scattering by \(\sim100\unit{GeV}\) electrons proceeds deep in the Klein--Nishina regime. {Furthermore, even if the conditions in the production region appeared to be favorable for production of VHE emission, the \(\gamma\gamma\) attenuation of TeV \gray is expected be severe in binary systems with luminous stars \cite{1994ApJS...92..567M}, dramatically hardening detecting these systems in the VHE regime.}

Interactions of protons within the binary system are relatively inefficient via the $pp$ channel due to insufficient target density, making this process poorly suited for generating flare-like emission. The photo-meson process, on the other hand, can be sufficiently efficient in systems with very luminous stars, provided that protons are accelerated beyond the relevant threshold energy. Even for very high-temperature stellar photon fields, this requires acceleration of protons to PeV energies. Although this may appear highly improbable, there is observational evidence for such processes, as suggested by the LHAASO detection of Cygnus~X$-$3 in the UHE regime.


From an observational perspective, these findings underscore the importance of
time-domain gamma-ray astronomy. The CTAO will be
particularly well suited to probing this transient parameter space. Quantitative
studies indicate that CTAO will be capable of detecting short-lived VHE flares
from Cygnus~X--1 at flux levels of a few percent of the Crab Nebula on
minute-to-hour timescales, while its improved sensitivity and rapid response
capabilities will enable systematic searches for sub-hour VHE counterparts to
GeV flares in Cygnus~X--3. Establishing how common such events are, and identifying
the physical conditions under which they occur, remains a key challenge for the
next generation of \graya observatories.

\begin{table*}[t]
\centering
\caption{Comparative observational properties of gamma-ray emission from the
microquasars Cygnus~X--1 and Cygnus~X--3.}
\label{tab:microquasar_comparison}
\scriptsize
\setlength{\tabcolsep}{5pt}
\begin{tabular}{p{4.2cm}p{5.5cm}p{5.5cm}}
\hline
 & \textbf{Cygnus~X--1} & \textbf{Cygnus~X--3} \\[1pt]
\hline
Binary type &
High-mass X-ray binary with O-type supergiant &
High-mass X-ray binary with Wolf--Rayet companion \\[1pt]
\hline
Orbital period &
5.6~d &
4.8~h \\[1pt]
\hline
Dominant accretion state for $\gamma$-ray activity &
Hard / intermediate state &
Soft X-ray states and transitions out of radio-quenched phases \\[1pt]
\hline
HE emission ($E>100$~MeV) &
Weak and intermittent; detected only when selecting hard-state data with
$L_\gamma \sim 5\times10^{33}$~erg~s$^{-1}$; rare day-scale flares reported by
AGILE &
Recurrent, bright flares with durations of $\sim1$--2~d and peak fluxes
$(1$--$3)\times10^{-6}$~ph~cm$^{-2}$~s$^{-1}$; orbital modulation detected;
$L_\gamma \sim 10^{36}$~erg~s$^{-1}$ \\[1pt]
\hline
VHE emission ($E>100$~GeV) &
Single night-long TeV flare reported by MAGIC in 2006
($\sim4\sigma$); no confirmation in later observations &
No detection at $\sim100$~GeV; only upper limits from MAGIC and VERITAS, but
transient multi-TeV/PeV emission detected by LHAASO during GeV high states \\[1pt]
\hline
Variability timescales &
Minutes (TeV flare), hours--days (GeV) &
Hours (orbital modulation) to days (GeV flares); month-scale variability at
multi-TeV/PeV energies \\[1pt]
\hline
Long-term VHE constraints &
Deep MAGIC upper limits from 2007--2014; no persistent emission; multi-TeV upper
limits from HAWC &
MAGIC and VERITAS upper limits at $\sim100$~GeV; no steady emission; transient
ultra-high-energy emission detected by LHAASO; steady multi-TeV emission
constrained by HAWC \\[1pt]
\hline
Overall \graya phenomenology &
Rare, extreme transient events; otherwise \graya quiet &
Frequent, luminous GeV flares with strong state dependence; suppressed at
$\sim100$~GeV but exhibiting transient multi-TeV/PeV emission \\[1pt]
\hline
Implications for CTAO &
Sensitivity to rare, minute-scale VHE flares; rapid response essential &
Potential to probe short-lived VHE counterparts of GeV flares and intermediate-
energy gaps; requires fast follow-up and real-time analysis \\[1pt]
\hline
\end{tabular}
\end{table*}


\section{Active Galactic Nuclei}

\subsection{Introduction}
Active galactic nuclei (AGNs) are among the most luminous and
energetically extreme astrophysical objects in the Universe. Their
enormous apparent luminosities—reaching up to \(10^4\) times that of a
typical galaxy—are generated within regions that are exceedingly
compact on galactic scales, with characteristic sizes of order
\(\sim1\unit{pc^3}\). Resolving such spatial scales directly is
possible only in a handful of exceptional nearby sources and primarily
at radio frequencies. In these cases, very-long-baseline
interferometry reveals compact, variable jet structures that often
exhibit apparent superluminal motion, providing direct evidence for
relativistic bulk flows. Even when AGNs cannot be spatially resolved,
their presence is revealed through pronounced variability in their
integrated emission, detected across the electromagnetic
spectrum. This variability generally becomes more prominent at shorter
wavelengths, with higher-energy bands displaying larger amplitudes and
shorter characteristic timescales.

AGN activity is a relatively common phenomenon and manifests itself
across a wide range of astrophysical source classes, including
radio-loud and radio-quiet quasars, radio galaxies, Seyfert galaxies,
BL Lacertae objects, and flat-spectrum radio quasars (FSRQs;
historically also referred to as optically violently variable
quasars). Among these classes, BL Lac objects and FSRQs exhibit the
most pronounced and rapid variability. These sources are therefore
commonly unified under the class of blazars, interpreted as AGNs whose
relativistic jets are oriented at small angles to the line of
sight. In this configuration, relativistic beaming strongly amplifies
both the observed luminosity and variability. While the bulk of the
variable emission in blazars is generally attributed to processes
occurring within the jet, it cannot be excluded a priori that, in at
least some cases, a contribution originates from regions closer to the
central engine.

AGN jets produce broadband non-thermal radiation extending from radio
frequencies to the \graya band, implying efficient particle
acceleration under extreme physical conditions. In the \graya domain,
AGNs represent one of the dominant classes of persistent extragalactic
emitters, and they also constitute an important population of
recurrent high-energy transients due to their strong and often rapid
variability. However, our ability to detect variable \graya emission
from AGNs is subject to several fundamental limitations. First, the
cosmic history of AGN activity exhibits a strong redshift dependence,
with the peak of AGN and quasar activity occurring at
\(z\approx2.5\). At such distances, the detection of variability in
the \graya band is constrained by the limited collection area of
space-borne instruments operating in the GeV range, as well as by
severe attenuation of higher-energy photons due to \(\gamma\gamma\)
interactions with the EBL. In addition to
this extragalactic absorption, internal attenuation must also be
considered: AGNs are compact sources of intense broadband radiation,
providing dense photon fields that can lead to significant \(\gamma\gamma\)
absorption within the source itself.

Despite these challenges, observations of AGNs in the \graya
band --particularly of blazars-- have demonstrated that they are capable
of producing rapid and luminous high-energy flares. One of the most
extreme properties of \graya flares detected from some AGNs is their
very short duration, in several cases approaching the minimum
variability timescale set by the light-crossing time of a region
comparable in size to the central black-hole event horizon. These properties
establish AGN jets --and possibly the immediate vicinity of the central
engine-- as important laboratories for studying particle acceleration,
radiation processes, and extreme variability in relativistic outflows.

\subsection{PKS~2155$-$304: a prototype of extreme VHE transients} \label{ssec:pks2155}

PKS~2155$-$304 ($z=0.116$) represents the most extreme known case of transient VHE emission from a blazar jet. This classification is primarily based on the giant flaring activity observed by \hs in July 2006, during which the source reached peak flux levels exceeding the quiescent emission by more than an order of magnitude and exhibited variability on minute timescales \cite{2007ApJ...664L..71A}.

During the main flare night of 28 July 2006, PKS~2155$-$304 displayed flux doubling times as short as $t_{\rm var}\simeq2$--$3$~min at energies above a few hundred GeV, with peak integral fluxes reaching $(5$--$10)\times$ the Crab Nebula flux above 200~GeV \cite{2007ApJ...664L..71A}. The VHE light curves showed pronounced intra-night substructure, with multiple sharp rises and decays occurring on comparable timescales. Such rapid variability implies an extremely compact emission region. Using causality arguments, the characteristic size of the emitting region is constrained by
\begin{equation}
{\cal R} \lesssim \frac{c\, t_{\rm var}\, \D}{1+z}
         \simeq (3\text{--}5)\times10^{12}\,\D~{\rm cm},
\end{equation}
where $\D$ is the Doppler factor. For Doppler factors typically inferred for high-frequency-peaked BL~Lac objects, $\D \sim 10$--$20$, this yields ${\cal R} \sim 10^{13}$--$10^{14}$~cm, which is several orders of magnitude smaller than the transverse scale of the parsec-scale jet. 

Conversely, associating the emission region with a jet cross section of order 
$R \sim 10^{16}\,\mathrm{cm}$ would require unrealistically large Doppler factors, 
$\D \sim 10^{3}$, far exceeding values inferred from low-state SSC modeling 
and strengthening the case for compact sub-structures or multi-zone emission 
during the 2006 flares.

In addition to the extreme temporal behavior, the 2006 flares revealed distinctive statistical properties. The distribution of VHE fluxes was shown to follow a lognormal form rather than a normal distribution \cite{2010A&A...520A..83H}. This indicates that the variability is governed by multiplicative processes, pointing to an origin in global instabilities or modulations rather than a superposition of independent emitting zones. The corresponding power spectral density is well described by red-noise behavior over a broad range of timescales, further supporting the presence of correlated variability processes \cite{2010A&A...520A..83H}.

Multiwavelength observations obtained during the flaring episode provided additional constraints. While the VHE emission reached unprecedented levels, contemporaneous X-ray observations showed enhanced activity but did not exhibit a strictly linear correlation with the TeV flux \cite{2012A&A...539A.149H}. In particular, periods of super-linear VHE--X-ray correlations were reported, implying changes in radiative efficiency or in the underlying particle distribution. At lower energies, including optical and radio bands, variability amplitudes were comparatively modest, indicating a strong energy dependence of the transient behavior and suggesting that the most extreme variability is confined to the highest-energy particles.

Long-term monitoring of PKS~2155$-$304 with H.E.S.S. between 2005 and 2007 demonstrated that the giant flares represent rare, extreme episodes superimposed on a substantially lower baseline emission \cite{2010A&A...520A..83H}. Outside major flaring periods, the source typically resides in a low or intermediate VHE state, characterized by reduced flux levels and longer characteristic variability timescales. This picture was further refined by joint observations with \textit{Fermi}-LAT, which enabled systematic comparisons between HE and VHE variability over multi-year timescales \cite{2017A&A...598A..39H}. These studies showed that, while variability is present across the entire \graya band, the most rapid and highest-amplitude transients are predominantly confined to the VHE regime, whereas the GeV emission exhibits smoother temporal behavior.

Dedicated multiwavelength campaigns during low states reinforced this dichotomy. 
In quiescent phases, the broadband spectral energy distribution of PKS~2155$-$304 
can be satisfactorily reproduced within a standard one-zone synchrotron 
self-Compton framework using moderate Doppler factors of 
$\D \simeq 15$--$25$, magnetic field strengths 
${\cal B} \simeq 0.1$--$0.3~{\rm G}$, and characteristic emission-region sizes 
${\cal R} \simeq (3$--$10)\times10^{15}~{\rm cm}$ \cite{2010A&A...520A..83H,2017A&A...598A..39H,2009ApJ...696L.150A}. 
The electron population is typically described by a broken power-law with 
break Lorentz factors $\gamma_{\rm br}\sim(3$--$8)\times10^{4}$ and maximum 
energies $\gamma_{\rm max}\sim10^{5}$--$10^{6}$. 
Within this parameter space, both the synchrotron peak at 
$\nu_{\rm syn}\sim10^{16}$--$10^{17}$~Hz and the inverse-Compton peak at 
$\nu_{\rm IC}\sim10^{25}$--$10^{26}$~Hz are reproduced without invoking 
extreme Doppler boosting or unusually low magnetic fields, in clear contrast to the physical conditions inferred during the 2006 giant flaring episodes \cite{2009ApJ...696L.150A}, {indicating that transient episodes probe a fundamentally different physical regime than the persistent emission.}

Within this parameter space, the inferred particle and magnetic energy densities 
are close to equipartition and the associated jet power remains modest, 
$P_{\rm jet}\sim10^{44}$--$10^{45}$~erg~s$^{-1}$. 
Recent MAGIC and LST observations demonstrate that such low-state emission can be 
monitored with sufficient sensitivity and temporal coverage to search for weak 
or short-lived deviations from this baseline \cite{2025arXiv251004803N}.



Complementary optical polarization measurements reveal rapid changes in polarization degree and angle during active periods, indicating a dynamically evolving magnetic field structure and supporting scenarios in which localized dissipation events, such as magnetic reconnection, play a key role in powering extreme flares \cite{2026JHEAp..5000472B}.

Overall, PKS~2155$-$304 provides the clearest observational evidence that blazar jets can produce VHE \graya emission on timescales of only a few minutes, {despite their intrinsically large spatial scales.} 
The 2006 giant flares therefore define a benchmark for extreme transient behavior in relativistic jets and serve as a reference point for interpreting \graya variability across different source classes.

{Building on this, the phenomenology of PKS 2155-304 demonstrates that} blazar jets are capable of producing extremely luminous \graya
outbursts on minute timescales, with variability amplitudes and temporal structures that challenge standard emission
scenarios. Importantly, these extreme transients are largely confined to the VHE regime and occur episodically,
superimposed on comparatively quiescent long-term behavior. This observational picture contrasts sharply with that of
flat-spectrum radio quasars, where \graya activity is dominated by recurrent, often long-lived flares peaking at GeV
energies and shaped by the presence of dense external radiation fields. In this sense, PKS 2155-304 provides a benchmark
for the most extreme end of rapid VHE variability, while FSRQs such as 3C 454.3 probe a complementary regime of
high-energy transients governed by different radiative environments and characteristic timescales.

\subsection{3C~454.3: Extreme GeV Transients in a Powerful FSRQ}

The flat-spectrum radio quasar 3C~454.3 ($z=0.859$) represents the most luminous
and repeatedly flaring \graya source observed to date and serves as the
archetype of extreme transient activity in the GeV regime. Since the advent of
space-based \graya observatories, and in particular during the
\emph{Fermi}-LAT era, 3C~454.3 has exhibited a sequence of major outbursts
characterized by unprecedented flux levels, sustained high states, and complex
temporal substructure \cite{2010ApJ...718..455S,2011ApJ...736L..38V,2011ApJ...733L..26A}.

The first indications of extreme activity were reported by \emph{AGILE} during
the 2007–2009 observing seasons, when the source reached daily-averaged fluxes
above 100~MeV of $F_{>100\,{\rm MeV}}\sim(1$--$3)\times10^{-5}$~ph~cm$^{-2}$~s$^{-1}$
\cite{2010ApJ...712..405V,2010ApJ...718..455S}. This behavior culminated in the historic
2009 December and 2010 November flares observed by \emph{Fermi}-LAT, during which
the daily-averaged flux exceeded
$F_{>100\,{\rm MeV}}\simeq(5$--$9)\times10^{-5}$~ph~cm$^{-2}$~s$^{-1}$, temporarily
outshining the Vela pulsar and establishing 3C~454.3 as the brightest persistent
\graya source in the sky \cite{2011ApJ...733L..26A,2011ApJ...736L..38V}. On shorter
integration timescales of a few hours, peak fluxes as high as
$F_{>100\,{\rm MeV}}\simeq8.5\times10^{-5}$~ph~cm$^{-2}$~s$^{-1}$ were measured
\cite{2011ApJ...733L..26A}.

At the distance of 3C~454.3, these fluxes correspond to apparent isotropic
\graya luminosities of
$L_{\gamma}^{\rm iso}\sim10^{49}$--$10^{50}$~erg~s$^{-1}$, exceeding by several
orders of magnitude the bolometric luminosities inferred for high-frequency-
peaked BL~Lac objects such as PKS~2155$-$304 \cite{2010ApJ...718..455S,2011ApJ...736L..38V}.
Unlike TeV-dominated sources, however, the transient behavior of 3C~454.3 {appears in the HE band}, with the bulk of the radiative output concentrated
below a few tens of GeV.

Despite generally longer variability timescales than those observed in TeV
blazars, 3C~454.3 also exhibits episodes of genuinely rapid intrinsic
variability. Temporal analyses of the 2010 and 2014 outbursts revealed flux doubling times as short as t$_{\rm var}\simeq0.7$--$1.5$~h in the source
frame \cite{2011ApJ...733L..26A,2016MNRAS.458..354C}. Applying the same causality
constraint used for PKS~2155$-$304, these variability timescales imply
\[
{\cal R} \lesssim (4\text{--}9)\times10^{13}\,\D~\mathrm{cm},
\]
which corresponds to a characteristic emission-region size of
$R\sim10^{15}$~cm for typical Doppler factors $\D\simeq15$--30. Such compact
dimensions are more than an order of magnitude smaller than the characteristic
radius of the broad-line region (BLR), indicating that a substantial fraction of
the variable \graya output must originate within a dense external
radiation environment rather than on parsec scales.

Independent geometric constraints arise from the detection of very high-energy
LAT photons during major outbursts. In the same flaring episodes, photons with
energies up to $E_\gamma\simeq35$--50~GeV were detected
\cite{2011ApJ...733L..26A,2016MNRAS.458..354C}. To avoid strong
$\gamma\gamma$ attenuation on BLR photons, such emission must be produced at
distances comparable to or beyond the outer BLR boundary,
$r\gtrsim(1$--$2)\,R_{\rm BLR}$, corresponding to spatial scales of
$\sim10^{17}$--$10^{18}$~cm. Taken together, these constraints imply that the
most powerful flares in 3C~454.3 likely involve multiple emitting zones or
radially stratified dissipation regions, with fast variability imprinted by
compact substructures embedded within a more extended GeV-emitting outflow.

Long-term monitoring of 3C~454.3 over nearly a decade reveals that extreme
outbursts are not isolated events but recur frequently. An analysis of
\emph{Fermi}-LAT data between 2008 and 2017 identifies at least five major
high-activity periods, each lasting from several weeks to a few months and
often exhibiting pronounced substructure on timescales of hours to days
\cite{2020ApJS..248....8D}. The duty cycle of elevated \graya emission is therefore
substantially higher than in TeV-dominated BL~Lac objects, emphasizing that the
transient behavior of 3C~454.3 is governed by sustained high-luminosity episodes
rather than rare, ultra-fast flares.

At very high energies, MAGIC observations of 3C~454.3 obtained during
major GeV flaring states did not yield a significant detection, and only
upper limits were reported above $\sim50$--100~GeV, consistent with a
strong spectral break or cutoff between the GeV and TeV regimes
\cite{2009A&A...498...83A}. This behavior is consistent with strong attenuation of
\gray above $\sim50$--100~GeV due to interactions with ambient photon
fields, reinforcing the conclusion that the transient phenomenology of 3C~454.3
is intrinsically GeV-dominated.

In summary, 3C~454.3 exemplifies a class of \graya transients fundamentally
distinct from minute-scale TeV flares observed in HBLs. Its observational
hallmarks are extreme apparent luminosities, recurrent and long-lived GeV
outbursts, and variability timescales ranging from hours to months. Together,
these properties define the FSRQ regime of high-energy transients and provide a
quantitative counterpoint to the compact, VHE-dominated flaring behavior
observed in sources such as PKS~2155$-$304.

\subsection{IC~310: Horizon-Scale VHE Transients}

IC~310 is a nearby active galaxy at redshift $z=0.0189$ located in the Perseus
cluster and occupies a transitional position between BL~Lac objects and FR~I
radio galaxies. VHE \graya emission from IC~310 was
initially discovered by MAGIC during observations of the Perseus region,
revealing a hard spectrum extending from $\sim150$~GeV to several TeV with a
photon index of $\Gamma\simeq2.0$ and clear variability on timescales ranging
from days to months \cite{2010ApJ...723L.207A,2017A&A...603A..25A}. Subsequent observations
demonstrated that IC~310 exhibits some of the most extreme transient behavior
known at VHE energies.

The most remarkable event occurred during a bright flare detected on
12--13 November 2012, when MAGIC measured flux variability on timescales of only
a few minutes. In particular, flux doubling times as short as
t$_{\rm obs}\lesssim4.8$~min (observer frame) were observed at energies above several hundred
GeV, with statistically significant variability detected within individual
nights \cite{2014Sci...346.1080A}. Using the causality argument introduced earlier,
these timescales imply an extremely compact emitting region with a
characteristic size
${\cal R} \lesssim 8.6\times10^{13}\,\D~\mathrm{cm}$.

In contrast to blazars, very long baseline interferometry observations constrain
the jet viewing angle of IC~310 to $\theta\simeq10^\circ$--$20^\circ$, implying
only modest Doppler boosting with $\D\lesssim4$ \cite{2014A&A...563A..91A}.
Even adopting the upper end of this range yields an emitting-region size of
${\cal R}\lesssim3\times10^{14}$~cm. For a central black hole mass of
$M_{\rm BH}\simeq(2\text{--}3)\times10^{8}\,M_\odot$, corresponding to a gravitational
radius
\begin{equation}
\rsch = \frac{G M_{\rm BH}}{c^{2}}
          \simeq (3\text{--}4.5)\times10^{13}~\mathrm{cm},
\end{equation}
the inferred size of the VHE emission region is constrained to be comparable to,
or smaller than, the event-horizon scale,
${\cal R} \lesssim (0.2\text{--}0.3)\,\rsch$ \cite{2014Sci...346.1080A}. This conclusion is
robust against reasonable uncertainties in $\D$ and places IC~310 in a
qualitatively different regime from highly beamed TeV blazars.

At lower energies, long-term monitoring with \emph{Fermi}-LAT reveals pronounced
spectral variability in the GeV band. Distinct spectral states are observed,
including a hard flaring state and a much softer quiescent state, with the peak
of the inverse-Compton component shifting by more than five orders of magnitude
in energy between these states \cite{2019MNRAS.485.3277G}. However, the limited photon
statistics at GeV energies prevent resolving variability on timescales
comparable to the minute-scale TeV flares observed by MAGIC.

Time-averaged multiwavelength observations indicate that the broadband spectral
energy distribution of IC~310 during low states can be reproduced within
standard one-zone synchrotron self-Compton scenarios using moderate Doppler
factors and magnetic field strengths \cite{2017A&A...603A..25A}. In particular, the
low-state SED is well described with Doppler factors of $\D \simeq 3$--5,
magnetic field strengths ${\cal B} \simeq 0.1$--0.5~G, and characteristic emission-region
sizes ${\cal R} \simeq (1$--$5)\times10^{16}$~cm. The underlying electron population is
typically modeled with a broken power-law distribution, with break Lorentz
factors $\gamma_{\rm br}\sim(1$--$3)\times10^{4}$ and maximum energies
$\gamma_{\rm max}\sim10^{5}$. Within this parameter range, the synchrotron
component peaks at $\nu_{\rm syn}\sim10^{15}$--$10^{16}$~Hz, while the inverse
Compton component peaks at $\nu_{\rm IC}\sim10^{23}$--$10^{24}$~Hz, consistent
with the observed low-state emission from radio to TeV energies.

Compared to the extreme 2012 TeV flaring episode, for which the observed
variability timescale implies an emitting-region size of
${\cal R} \lesssim 10^{14}$~cm, the characteristic size inferred for the low-state
emission is larger by roughly two orders of magnitude. This pronounced
difference demonstrates that the persistent and flaring emissions of IC~310
originate in physically distinct regions. In particular, the combination of
minute-scale TeV variability, modest Doppler boosting, and horizon-scale
emission-region sizes during flares places severe constraints on standard
shock-in-jet interpretations and motivates alternative acceleration scenarios
operating in compact regions near the black hole \cite{2014Sci...346.1080A}.

Overall, IC~310 represents the most extreme known example of transient VHE
emission from an active galaxy, probing spatial and temporal scales that are
otherwise accessible only in compact Galactic systems. Its observed properties
define a fundamental boundary for models of particle acceleration and \graya production in relativistic outflows.

\begin{table}[t]
\centering
\caption{Observational properties of extreme \graya transients
during their most extreme flaring episodes.}
\label{tab:agn_par}
\scriptsize  
\begin{tabular}{p{1.5cm}ccc}
\hline
 & PKS~2155$-$304 & 3C~454.3 & IC~310 \\[1pt]
\hline

$z$
& 0.116 & 0.859 & 0.0189 \\[1pt]
\hline

Energy band
& TeV & GeV & TeV \\[1pt]
\hline

$t_{\rm var}$
& 2--3 min
& 0.7--1.5 h
& $\lesssim 5$ min \\[1pt]
\hline

${\cal R}_{\rm flare}$ (cm)
& $\lesssim 10^{13}$--$10^{14}$
& $\lesssim 10^{15}$
& $\lesssim 10^{14}$ \\[1pt]
\hline

$\D$
& \makecell{15--25 (low)\\$\gtrsim 50$ (flare)}
& 15--30
& $\lesssim 4$ \\[1pt]
\hline

$L_\gamma^{\rm iso}$ (erg s$^{-1}$)
& $10^{46}$--$10^{47}$
& $10^{49}$--$10^{50}$
& $10^{44}$--$10^{45}$ \\[1pt]
\hline

${\cal R}_{\rm low}$ (cm)
& $10^{15}$--$10^{16}$
& $10^{16}$--$10^{17}$
& $\sim 10^{16}$ \\[1pt]
\hline

Key challenge
& \makecell{Doppler-factor\\ crisis}
& \makecell{extreme \graya\\ luminosity}
& \makecell{horizon-scale\\emission region} \\[1pt]
\hline
\end{tabular}

\vspace{1mm}
\footnotesize{
Notes: ${\cal R}$ values are causality-based upper limits from variability.
$\D$ values are inferred from independent constraints.
}
\end{table}

\subsection{Summary} 

Taken together, PKS~2155$-$304, 3C~454.3, and IC~310 {highlight the landscape} of extreme \graya
transients in active galactic nuclei. While PKS~2155$-$304 demonstrates that blazar jets can produce minute-scale
variability at TeV energies under conditions of extreme Doppler boosting, 3C~454.3 exemplifies a complementary regime in
which sustained GeV outbursts reach unprecedented apparent luminosities on longer timescales.  IC~310, in turn, occupies
the most extreme end of this sequence, combining minute-scale TeV variability with only modest Doppler factors, thereby
constraining the emission region to horizon-scale dimensions. The pronounced differences in variability timescales,
emitting-region sizes, Doppler factors, and dominant energy bands across these sources indicate that a single
phenomenological framework is insufficient to explain all observed transient behavior. Please see
table~\ref{tab:agn_par} for a summary. The physical implications of these observational constraints are very broad and
involve scenarios that considerably modify the paradigm for modeling the emission from extragalactic jets. In
particular, it was suggested that rapid and intense flares can be caused by relativistic reconnection in the jets
\cite{2009MNRAS.395L..29G}, processes taking place in the black hole magnetosphere
\cite{2016ApJ...818...50H,2011ApJ...730..123L}, or by entering of external obstacles of different nature into the jet
flow \cite{2010ApJ...724.1517B,2012ApJ...749..119B,2010A&A...522A..97A,2013MNRAS.436.3626A} (for a review see in
\cite{2017ApJ...841...61A}). 

{In this context, it is worth noting that relativistic outflows are also potential sources of high-energy neutrinos, highlighting the importance of a multi-messenger perspective. In hadronic scenarios, \gray and neutrinos are produced together through pion decay. However, the connection between \graya and neutrino emission is not necessarily straightforward. Neutrino production may occur in regions that are partially opaque to \gray, where the accompanying high-energy photons are attenuated and reprocessed before escaping. Further, \graya emission from AGN jets is typically a subject for significant attenuation on EBL. As a result, neutrino emission may not always be accompanied by a clear \graya counterpart (see, e.g., \cite{2026NewAR.10201747R}).}

{Also, it is also important to note that efficient operation of hadronic processes requires a very high density target.  As discussed in Sec.~\ref{sec:mq_intro}, hadronic scenarios for flares require target densities that are hard to achieve even in the immediate vicinity of massive stars, thus it is not straightforward to adopt such conditions for AGN jets. An important feature of AGN jets is relativistic motion. This allows relaxing the luminosity and variability constraints significantly (see Sec.\ref{sec:intro_rel}), however in the context of hadronic scenarios, the relativistic motion also brings important limitations.}

{For example, the $pp$ channel requires a high density of the background protons, a day-scale variability in the co-moving frame implies a density of \(n\mysub{t}\sim10^{10}\unit{cm^{-3}}\). The kinetic energy flux is then}
\be
S = 4\times10^{20} \qty(\frac{\Gamma}{30})^2 \qty(\frac{n\mysub{t}}{10^{10}\unit{cm^{-3}}})\unit{erg\,cm^{-2}\,s^{-1}}\,.
\ee
{For a typical production site of \({\cal R}\sim10^{14}\unit{cm}\), the jet luminosity should exceed \(10^{49}\ergs\), which may appear beyond the feasible range. If the target protons are external to the jet, then they do not give a contribution to the jet kinetic power, however the realization of the hadronic scenarios requires at least a partial jet disruption. Such scenario can be realized when stars enter the jet, and in the case of most nearby AGNs, for example M87, this can cause \graya flares powered by $pp$ interactions \cite{2010ApJ...724.1517B}. In this case a neutrino flare of a comparable luminosity is expected. }

{Realization of photomeson scenarios for relativistic jets implies different range of difficulties. While the contribution of the photon target to the jet kinetic luminosity is negligible, the target field should be also Doppler boosted and thus should appear in the SED of the source. Provided the high density of the target and relatively high energy (determined by the threshold condition, see in Sec.~\ref{sec:mq_intro}), the observations should strongly constrain the efficiency of the \(p\gamma\) channel. Of course, this constraint is alleviated if the target the field is external to the jet. Such an external field is not enhanced by the Doppler boosting,  however, in this case the field should occupy a significantly larger volume than the flare production site. The luminosity of the corresponding component is proportional to the characteristic surface thus it will be noticeably enhanced. Also in this case one should note that the Doppler boosting of the \graya emission is different from the usual \(\D^4\) factor (for derivations based on Lorentz invariant distribution function see \cite{2018MNRAS.481.1455K}).}

{While the arguments above outline some basic constraints for generating neutrino flares from AGN jets, detailed analysis of such scenario is beyond the scope of our review. However such an analysis can be found in dedicated reviews (see, e.g., \cite{2026NewAR.10201747R}). }


\section{Crab flares \& Giant Pulses}

\subsection{Introduction}
The Crab Pulsar and its surrounding pulsar-wind nebula (PWN) remain among the most intensively studied laboratories for extreme particle acceleration. Despite its long-standing use as a flux standard in \graya astronomy, the Crab system exhibits unexpected variability on timescales ranging from milliseconds (Giant Radio Pulses, GRPs) to hours (GeV flares).

GRPs consist of microsecond -- to nanosecond -- duration bursts whose ensemble energy and intensity statistics follow a power-law distribution, making them one of the most extreme manifestations of coherent emission known. Their correlation with specific phases of the pulsed emission phasogram suggests an origin in the pulsar magnetosphere. GRPs therefore offer a unique probe of magnetospheric processes that may also drive high-energy radiation. However, the likely coherent nature of the radiation mechanism responsible for GRPs implies a much fainter counterpart emission in incoherent radiation channels. Thus, it is not certain that this phenomenon has any detectable counterpart in the \graya band, even if the magnetospheric emission extends beyond GeV energies.

In contrast to GRPs, the Crab flares are detected in the GeV band and are likely to originate in the nebula, i.e., at distances of order a fraction of a parsec from the central pulsar. PWNe are a common phenomenon observed around many pulsars, resulting from the interaction of the pulsar ultrarelativistic wind with the interstellar medium (ISM) or with the supernova remnant of the progenitor star. Provided that the pulsar wind termination shock operates as an exceptionally efficient particle accelerator, relativistic electrons in PWNe encounter nearly ideal conditions for producing high-energy emission via synchrotron and inverse Compton (IC) processes. The ratio of the fluxes in these two channels is determined by the ratio of the corresponding energy densities of the magnetic field and the target photon fields, respectively.

At large distances from the pulsar, the magnetic field strength is typically modest, \(\sim300\unit{\upmu G}\) in the case of the Crab Nebula and significantly weaker in many other PWNe. For example, H.E.S.S. and Suzaku observations revealed a magnetic field strength of \(\sim5\unit{\upmu G}\) in the Vela~X PWN \cite{2019A&A...627A.100H}. The corresponding magnetic-field energy density ranges from \(\gtrsim10^3\unit{eV\,cm^{-3}}\) in compact PWNe around young, powerful pulsars to \(\lesssim1\unit{eV\,cm^{-3}}\) in evolved PWNe associated with less energetic pulsars.

The typical energy density of background photon fields in the Milky Way Galaxy is \(\sim1\unit{eV\,cm^{-3}}\) \cite{2017MNRAS.470.2539P}, making the IC process an important channel for the cooling of ultrarelativistic electrons. The contribution from locally generated target photon fields, i.e., the synchrotron-self Compton (SSC) process, is expected to be significant only in very bright and compact PWNe. In practice, even in the case of the Crab Nebula, the SSC contribution is comparable to that produced by scattering on the diffuse background photon fields \cite{1996MNRAS.278..525A}.

For a characteristic target-field energy density of \(w\mysub{tf}\sim1\unit{eV\,cm^{-3}}\), the radiative cooling time of GeV and TeV electrons is extremely long:
\begin{equation}
t\mysub{rad} = \frac{E_e}{\abs{\dot{E}\qty(E_e)}} = \frac{3}{4}\frac{m_e^2c^3}{\sigma\mysub{T} E_e w\mysub{tf}}
\approx 0.3 \qty(\frac{E_e}{1\unit{TeV}})^{-1}\qty(\frac{w\mysub{tf}}{1\unit{eV\,cm^{-3}}})^{-1}\unit{Myr},
\end{equation}
rendering the IC component in PWNe effectively steady. Owing to this property, together with its high \graya flux, the Crab Nebula is widely regarded as the standard ``steady'' calibrator in \graya astronomy.

\subsection{Crab flares at high energies}
This view changed dramatically when AGILE reported an unexpectedly intense flare in 2010 \cite{2011Sci...331..736T}, soon followed by confirmation from Fermi-LAT of large-amplitude, short-duration outbursts that exceeded the nebula’s quiescent $>$100\,MeV flux by factors of several\cite{2011Sci...331..739A}. The seminal reports of the 2011 flare demonstrated variability on day and even sub-day timescales. The long cooling time of electrons generating emission in the GeV band through the IC scattering, excludes contribution of this mechanism to the flaring component, leaving the synchrotron as the only plausible processes. The detection of the flaring emission therefore revealed synchrotron photons reaching several hundred MeV, above what is typically allowed by classical synchrotron burn-off limits unless this component is strongly Doppler boosted.

Even, if one accepts operation of an extremely efficient acceleration mechanism, capable of overtaking the synchrotron losses, the short sub-day variability time scale implies a magnetic field significantly exceeding the values typical for the Crab Nebula.
Indeed, the strength of the magnetic field and electron energy determine the position of the spectral peak:
\begin{equation}
  \hbar\omega \approx 1.15 \frac{3 E_e^2 \hbar e {\cal B} }{ 2 m_e^3c^5} \approx 80 \qty(\frac{E_e}{1\unit{PeV}})^2\qty(\frac{\cal B}{1\unit{mG}})\unit{MeV}\,.
\end{equation}
Thus, to produce a synchrotron component peaking at \(\hbar \omega\sim300\unit{MeV}\), the following combination of electron energy and magnetic field strength is required: \(\qty(E_e/1\unit{PeV})^2 \qty({\cal B}/1\unit{mG})\approx 4\). On the other hand, the observed sub-day variability implies a condition on the cooling time, \(t\mysub{syn}\lesssim 12\unit{h}\), where the synchrotron cooling time is
\begin{equation}
  t\mysub{syn} \approx  \frac{3}{4}\frac{m_e^2c^3}{\sigma\mysub{T} E_e w\mysub{B}}=4\times10^5 \qty(\frac{E_e}{1\unit{PeV}})^{-1}\qty(\frac{\cal B}{1\unit{mG}})^{-2}\unit{s}\,.
\end{equation}
Here \(w\mysub{B} = {\cal B}^2/(8\pi)\) is the energy density in the magnetic field. 
Thus we obtain that \(\qty(E_e/1\unit{PeV}) \qty({\cal B}/1\unit{mG})^2\gtrsim 10\). Combining these two conditions, we obtain \({\cal B}\gtrsim 3\unit{mG}\). Even the limiting value in this relation exceeds by an order of the magnitude the equipartition strength, \({\cal B}_0= \sqrt{L\mysub{sd}/(c R\mysub{ts}^2)}\approx 300\unit{\upmu G}\), for the radius of the termination shock \(R\mysub{ts}=0.15\unit{pc}\) and spindown luminosity of \(L\mysub{sd}=5\times10^{38}\unit{erg\,s^{-1}}\).

Therefore the detection of flaring GeV emission in the Crab Nebula established the presence of an extremely efficient and localized particle-acceleration process capable of energizing electrons to PeV scales in mili-Gauss magnetic field. The peaking frequency of the flaring component reaching \(\hbar\omega\gtrsim 100\unit{MeV}\) implies an efficiency of \(\eta<2\) (we note here a factor of \(\approx 4\) higher position of the single-particle synchrotron spectrum in the spectral energy distribution compared to the position of the spectral peak). According to the definition adopted above, \(\eta < 10^2\), the Crab flares require extreme acceleration unless the flare production site moves with an extreme bulk Lorentz factor \(\Gamma\gtrsim 50\). 

Follow-up analyses showed that flaring behaviour was not restricted to the brightest events. A comprehensive treatment of LAT data by \cite{2012ApJ...749...26B} identified multiple episodes of enhanced emission between 2007 and 2011, including several “moderate” events that, while less dramatic than the 2011 April flare, displayed similar spectral hardening and temporal evolution. These results made clear that the flares constitute a recurrent though irregular phenomenon rather than an isolated anomaly.  Systematic searches extended to later years identified additional flares and a substantial population of lower-intensity “small flares.” The study of \cite{2020ApJ...897...33A} provided the most comprehensive accounting of such events, combining Bayesian-block detection with profile modeling to construct a catalog of statistically significant enhancements. The authors demonstrated that these smaller outbursts share the same spectral trends as the major flares, most prominently the harder-when-brighter evolution that is characteristic of rapidly accelerated synchrotron populations. 

The spectral properties of the brightest flares pose stringent theoretical constraints. The April 2011 event exhibited a synchrotron cutoff energy of approximately 375 MeV, with measurable emission extending to several hundred MeV. This requires either local magnetic fields of order a milligauss — well above the average nebular field — or strong Doppler boosting due to relativistic bulk motion. Both interpretations imply compact emission zones, a conclusion strongly supported by the short variability timescales of only a few hours measured in the most active intervals \cite{2012ApJ...749...26B}.  High-resolution optical and X-ray observations have suggested candidate morphological structures, such as the inner knot and moving wisps, but no definitive spatial counterpart to a GeV flare has yet been identified. As summarised by \cite{2014RPPh...77f6901B}, the flares challenge standard MHD flow models and likely implicate explosive magnetic-reconnection events in highly inhomogeneous regions of the pulsar-wind termination shock \cite{2011ApJ...737L..40U,2012ApJ...746..148C}. 

\subsection{Crab flares at very high energies}

During the 4-14 March 2013 Fermi-LAT/AGILE GeV flare, the flux above 100\,MeV increased by a factor of six. \hs observed the Crab on five consecutive nights during the flare and found no significant VHE variability, placing 95\% C.L. limits of $<$63\% flux change above 1 TeV and $<$78\% above 5 TeV. These limits imply that the synchrotron-flare electrons do not substantially enhance the inverse-Compton component at TeV energies, consistent with the idea that flares originate in regions of strong magnetic field. Furthermore, electrons responsible for the flare emission cool synchrotronically before reaching energies required for efficient production of TeV emission \cite{2014A&A...562L...4H}.

VERITAS also observed the March 2013 flare and similarly reported no TeV flux enhancement. Assuming a linear connection between the GeV and TeV flux increase, VERITAS constrained any proportionality to be $<$8.6 $\times$ 10$^{-3}$ (95\% C.L.). This stringent limit robustly excludes flare models predicting strong correlated IC emission, reinforcing the conclusion that the flare originates from compact high-magnetic-field regions where synchrotron cooling dominates \cite{2014ApJ...781L..11A}.


\subsection{Giant Radio Pulses}
The \mg Collaboration performed 16\,h of simultaneous observations of the Crab pulsar in radio (Effelsberg \& WSRT) and at E $>$ 60\,GeV. Out of 99,444 detected radio GRPs, no significant correlation with VHE photons was found across a wide variety of time windows and lags. Depending on phase selection (P1, P2) and window size, upper limits on a GRP-correlated VHE flux increase lie between 7\% and 61\% of the average Crab pulsar VHE flux (95\% C.L.). These are the most stringent VHE constraints to date \cite{2020A&A...634A..25M}.

\begin{figure}
  \centering
    \includegraphics[width=0.48\textwidth]{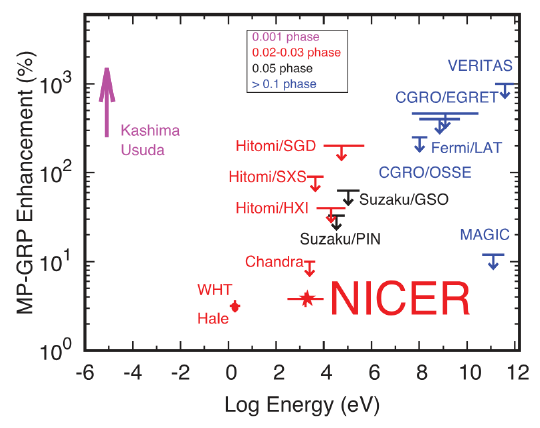}
    \caption{The enhancement fraction associated with MP-GRPs as a function of photon energy. Optical measurements from WHT and the Hale telescope, together with NICER data, yield significant detections, whereas all other measurements in the X-ray and \graya bands correspond to upper limits (downward arrows). Magenta upward arrow: detection threshold of Kashima and Usuda radio observations. Figure taken from \cite{2021Sci...372..187E}.}
      \label{fig:crab_GP}
\end{figure}

A major advance in understanding the connection between coherent and incoherent pulsar emission came from NICER’s simultaneous X-ray and radio monitoring of the Crab pulsar. Across 126 ks of overlapping NICER (0.3–10\,keV) and 2\,GHz radio observations, the campaign identified thousands of GRPs and revealed a statistically significant enhancement of the X-ray emission coincident with GRPs. NICER measured an increase of 3.8 ± 0.7\% in the pulsed X-ray flux at the phase of the main radio pulse, corresponding to a 5.4\,$\sigma$ detection.
This result demonstrates that GRPs are accompanied by small but measurable changes in the high-energy particle population within the pulsar magnetosphere. Although the X-ray enhancement is much smaller than the radio burst amplitude, it implies that the total energy output associated with GRPs is far larger than the radio alone would suggest, and that the processes generating coherent radio bursts are directly linked—at least in part—to variations in the incoherent X-ray emission region \cite{2021Sci...372..187E}. The corresponding measurements in comparison with upper limits from Fermi-LAT, MAGIC and VERITAS can be seen in Fig.~\ref{fig:crab_GP}.

\subsection{Outlook}

The rapid MeV–GeV flares of the Crab Nebula have forced a re-evaluation of longstanding assumptions concerning particle acceleration and high-energy radiation in pulsar wind nebulae. Standard MHD models predict that the synchrotron spectrum should cut off near a few tens of MeV, limited by synchrotron losses and by the achievable electric fields in the post-shock flow. Yet the flares observed since 2009 routinely reach several hundred MeV while varying on timescales of only a few hours, implying either extraordinarily efficient acceleration or strongly Doppler-boosted emission sites. As summarised in the detailed discussion of the two major Fermi–LAT and AGILE events of April 2011 and March 2013, the observed spectra, variability, and inferred sizes of the emitting regions collectively point to compact acceleration zones with magnetic fields significantly above the nebular average or to relativistic motion within strongly inhomogeneous structures of the inner nebula. The analysis of these flares, their spectral shapes, and the constraints that follow from synchrotron burn-off arguments highlight the difficulty of reconciling the observations with standard shock-acceleration scenarios alone. 

\begin{figure}
  \centering
    \includegraphics[angle=270,width=0.48\textwidth]{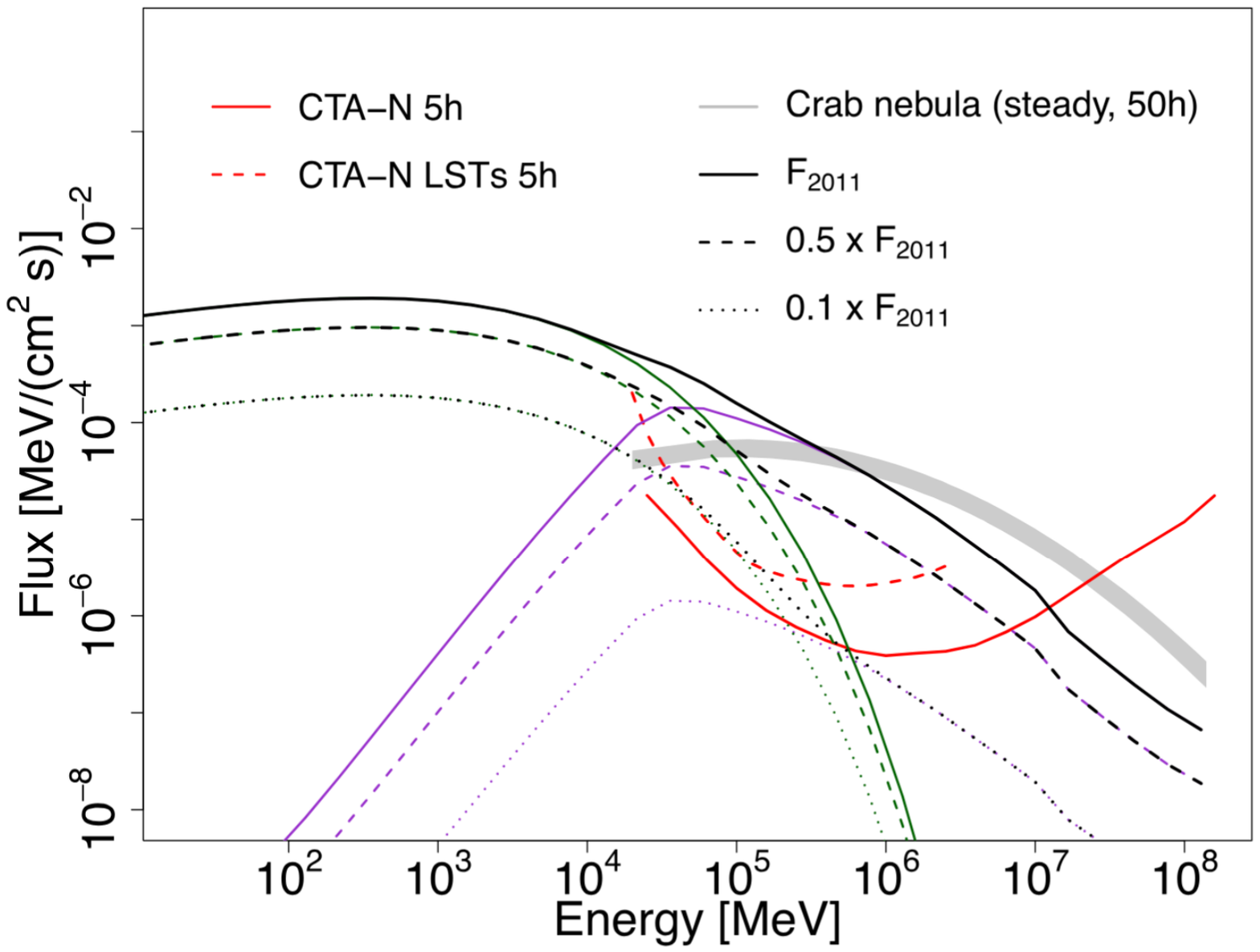}
    \caption{Synchrotron (green), IC (purple), and total (black) model spectra of the Crab Nebula for different flare scenarios. The solid curves correspond to the model fitted to the 2011 April \textit{Fermi}-LAT flare above 80~MeV assuming a particle index of 2.5, while the dashed and dotted curves represent the same model rescaled by factors of 0.5 and 0.1, respectively. All models are computed for a magnetic field of 500~$\mu$G. The red solid and dashed lines indicate the CTAO-N sensitivity (full array and LST-only configuration, respectively, for 5~h integration). The grey shaded region shows the simulated steady Crab spectrum for 50~h CTAO-N observations (3$\sigma$) \cite{2025MNRAS.540..205A}.}
      \label{fig:crab_flare}
\end{figure}

Simulations of the flares’ synchrotron and inverse-Compton (IC) components, constructed by fitting the LAT data with population models of energetic electrons and propagating these through detailed radiative transfer, reveal six key features:
\begin{enumerate}
    \item The synchrotron component responsible for the GeV emission is remarkably insensitive to the assumed magnetic field once the LAT flux is fixed, because the field merely scales the electron distribution required to reproduce the measured flux. 
    \item The IC component is extremely sensitive to the local magnetic field and to the index of the electron population. Detectability at TeV energies is therefore restricted to soft electron spectra or unusually low magnetic fields, the latter in tension with the short variability timescales unless energy budgets are increased beyond the standard nebular limits.
    \item For bright flares such as those in 2011 and 2013, variability at tens of GeV should be detectable in short exposures whenever the energy threshold is sufficiently low, while variability at TeV energies requires more constrained combinations of particle index and magnetic field.
    \item The total energy in electrons above 1\,TeV must remain below $\sim$5 $\times$ 10$^{43}$\,erg unless additional re-acceleration processes operate. This requirement eliminates a substantial region of the nominal parameter space and disfavors some of the harder spectra with weak magnetic fields (please refer to Fig.~6 in \cite{2021MNRAS.501..337M}).
    \item Jitter-radiation scenarios (see \cite{2013apj...774...61k}), which soften the curvature of the synchrotron cutoff, naturally produce high-energy power-law tails consistent with the LAT spectra above 400\,MeV and would yield substantial enhancements below 200\,GeV.
    \item Both relativistic-blob models and reconnection-layer scenarios remain viable, but they imply very different relationships between electron energy distributions, magnetic-field geometry, and the detectability of IC emission.
\end{enumerate}

Into this theoretical and observational context enters the role of CTAO. The simulations presented in \cite{2025MNRAS.540..205A} already demonstrate that flares comparable to those seen in 2011 and 2013 would be detectable below $\sim$200\,GeV for a wide range of spectral indices, provided the energy threshold is sufficiently low and exposures are on the order of tens of minutes as can be seen in Fig.~\ref{fig:crab_flare}. This is consistent with the broader conclusion that the most sensitive tests of flare physics occur below 100\,GeV, where the synchrotron tail may still contribute and where variability is most likely to be measurable over the steady nebular background. The extensive parameter surveys performed in \cite{2021MNRAS.501..337M} show that CTAO will be able to detect bright flares with steep synchrotron continua effectively independently of the magnetic field, while variability in the TeV regime—tracing the IC component—will depend strongly on the interplay between magnetic-field strength, the electron spectral index, and the total particle energy. 

Beyond these results, CTAO’s broader scientific impact on Crab flares can be described more generally. Its low-energy threshold places the array precisely at the transition between the highest-energy synchrotron emission and the onset of inverse-Compton scattering, enabling it to distinguish whether the flare spectra terminate below $\sim$20–30 GeV or instead extend into the CTAO band. If the synchrotron component continues into this regime, CTAO will measure its spectral slope and cutoff evolution with unprecedented precision, constraining the maximum electron energies and the degree of Doppler boosting that may be present. A sharp spectral cutoff below CTAO’s threshold during a bright flare, by contrast, would rule out a whole class of models relying on extreme magnetic fields or relativistic bulk motions capable of pushing the synchrotron peak to hundreds of MeV.

The short-timescale variability already seen by Fermi requires high-cadence responses, and CTAO’s rapid repointing ensures that the most dynamic intervals of the flare can be captured. The simulations in \cite{2021MNRAS.501..337M} show that a bright 2011-like flare would be detectable in well under an hour in the GeV band and, under favourable model assumptions, at TeV energies as well. Even weaker flares, corresponding to the frequently observed “small flares” in the LAT data, can produce detectable signatures below 200\,GeV if the spectrum softens or if the synchrotron tail is less curved than in standard MHD expectations. CTAO will therefore track both the brightest and the more common moderate events, establishing a statistically meaningful sample of flare and quiescent states.

Crucially, CTAO’s sensitivity in the TeV band allows it to search for the elusive IC counterpart to the GeV flares. Several reconnection-accelerated particle distributions predict an IC component that may appear contemporaneously with, or delayed relative to, the synchrotron peak. Detecting or placing strict limits on such emission will provide direct constraints on the seed-photon fields, the magnetic-field configuration, and the efficiency of particle escape from the acceleration region. Even non-detections will have interpretive power: for example, they would disfavor reconnection geometries with low effective magnetic fields or reject models requiring large populations of multi-TeV electrons during flares.

Taken together, the merged results from simulations and theoretical expectations indicate that CTAO will offer the decisive measurements needed to differentiate among competing flare models. It will either detect the high-energy tail of the synchrotron component, establish the presence or absence of an IC counterpart, or impose limits strong enough to confine the physical conditions—magnetic field, Doppler factor, particle acceleration efficiency, and geometry—to a much narrower region of parameter space. With its combination of low threshold, high sensitivity, and rapid temporal response, CTAO will therefore provide the most stringent tests yet of the extreme particle-acceleration processes that govern the Crab Nebula flares. 




\subsection{Conclusion}

Recent multi-wavelength studies of the Crab pulsar and nebula collectively establish a coherent picture of how the system reacts to rapid particle-acceleration events. Observations during major GeV flares show no accompanying variability at TeV energies, indicating that the electrons responsible for the synchrotron outbursts cool too efficiently for a measurable inverse-Compton contribution to emerge. This strongly favours models in which flares originate in compact, highly magnetized regions where synchrotron losses dominate, consistent with magnetic-reconnection scenarios.

At the magnetospheric level, GRP studies reveal a similarly constrained interplay between coherent and incoherent emission channels. MAGIC finds no VHE enhancement associated with GRPs, while NICER demonstrates that these radio bursts do leave a detectable imprint in the X-ray band. The small amplitude of the X-ray response suggests that GRPs involve localized adjustments to the energetic particle population without significantly altering the broader high-energy emission. Together, these findings imply that GRPs and nebular flares trace distinct physical processes: the former rooted in magnetospheric plasma dynamics near the light cylinder, the latter in fast, localized dissipation within the nebula.

Simulations for CTAO indicate that the next generation of \graya observations will be capable of decisively testing these interpretations. CTAO’s improved sensitivity, particularly at sub-100-GeV energies, will reveal whether subtle inverse-Compton signatures accompany future flares or GRPs, or whether their absence reflects intrinsic limits of the underlying acceleration mechanisms. By closing the observational gap between GeV and TeV energies and enabling rapid transient follow-up, CTAO will play a central role in distinguishing between competing models of reconnection, shock acceleration, and turbulence. In this way, the combined efforts of current and future instruments promise a substantially deeper understanding of how extreme particle acceleration operates within the Crab pulsar and nebula.


\section{Summary}
Extreme \graya transients constitute a distinct class of high-energy phenomena characterized by rapid, large-amplitude variability and physical conditions approaching fundamental limits on compactness, particle acceleration, and radiative efficiency. Two broad categories are identified: (i) catastrophic transformations of astrophysical systems --- such as stellar collapse, compact-object mergers, tidal-disruption events, novae, and magnetar flares --- and (ii) episodes that provide compelling evidence for particle acceleration operating in an extreme regime, typically above $\sim$100 MeV. In all cases, the transient outburst reflects the rapid release of gravitational, magnetic, nuclear, or kinetic energy that has accumulated over much longer timescales. Shock formation and magnetic reconnection emerge as the dominant channels for converting this stored energy into non-thermal particle populations and high-energy radiation.

A central theme of the review is that the ``extremeness'' of a transient is encoded in physically motivated constraints that link luminosity, variability, and emission-region compactness. Key diagnostics include light-crossing and Schwarzschild-radius limits, gyro-radius and acceleration-rate constraints, radiative cooling times, and internal opacity conditions. Variability timescales comparable to the particle gyro-time, \(\rg/c\), or to black hole horizon crossing time, $\rsch/c$, imply acceleration efficiencies $\eta \lesssim 10^{2}$ and horizon-scale emission zones, forcing models toward compact dissipation regions and non-standard acceleration scenarios. Energetic arguments further constrain the emitting volume, since $E_{\rm dis}\gtrsim L\,t_{\rm var}$ combined with ${\cal R}\lesssim c\,t_{\rm var}$ implies extreme energy densities when both luminosity and variability approach their observed limits. Relativistic motion and Doppler boosting can mitigate --- but not eliminate --- these constraints; when inferred efficiencies, luminosities, or variability scales approach such physical boundaries, the transient must be powered by processes operating in a genuinely extreme regime.

The manuscript also emphasizes the complementarity of current and forthcoming \graya observatories across the MeV--PeV domain. Space-borne detectors provide continuous monitoring and wide sky coverage up to tens of GeV, but are photon-limited for short events owing to modest collection areas. Ground-based Cherenkov and air-shower facilities achieve effective areas of $10^{8}$--$10^{10}\,\mathrm{cm^{2}}$, enabling time-resolved spectroscopy of brief VHE outbursts, including potential horizon-scale variability in the vicinity of supermassive black holes. Existing instruments (\hs, MAGIC, VERITAS, HAWC, LHAASO) already probe this regime, while next-generation facilities will expand sensitivity to short-duration, high-energy transients.

Extreme transients span both Galactic and extragalactic environments. Gamma-ray bursts exemplify catastrophic relativistic energy release on mil\-li\-se\-cond-to-week timescales; novae and recurrent novae illustrate extreme but non-destructive nuclear outbursts; giant flares trace catastrophic magnetic reconfiguration; and Crab Nebula flares demonstrate ultra-fast particle acceleration within a canonical pulsar wind nebula. Rapidly variable active galactic nuclei provide complementary laboratories: minute-scale TeV flares in blazars, ultra-luminous GeV outbursts in flat-spectrum radio quasars, and horizon-scale TeV events with modest Doppler boosting in sources such as IC~310 collectively delineate the observational parameter space in which compact dissipation, relativistic motion, and environmental photon fields interplay.

Taken together, these results show that extreme \gray transients are unified not by source class, but by their proximity to fundamental physical limits on acceleration, compactness, and dissipation. Coordinated MeV--TeV observations and rapid-response follow-up provide a unique empirical pathway to probe relativistic plasma processes, constrain the geometry and energetics of emission regions, and explore physical conditions inaccessible in steady-state high-energy sources.

\centering


\section*{Acknowledgments}
{The authors thank Valent\'i Bosch-Ramon, Evgeny Derishev, and Rasmik Mirzoyan for useful comments and suggestions.} This work has been supported by the grant PID2024-155316NB-I00 funded by MICIU /AEI /10.13039/501100011033 / FEDER, UE and CSIC PIE 202350E189. This work was also supported by the Spanish program Unidad de Excelencia Mar\'ia de Maeztu financed by MCIN/AEI/10.13039/501100011033, and by the MaX-CSIC Excellence Award MaX4-SOMMA-ICE.
Additional support was provided by the Chinese Academy of Sciences President’s International Fellowship Initiative (PIFI) Project No.:2026PVB0029. DH gratefully acknowledges Tianfu CRRC for hosting and providing an excellent research environment. DK acknowledges support by RSF grant No. 24-12-0045. The authors used AI-assisted tools for language editing and identification of grammatical and typographical errors; all scientific content, analysis, and conclusions are the author’s responsibility.






\begin{thebibliography}{170}
\expandafter\ifx\csname url\endcsname\relax
  \def\url#1{\texttt{#1}}\fi
\expandafter\ifx\csname urlprefix\endcsname\relax\def\urlprefix{URL }\fi
\expandafter\ifx\csname href\endcsname\relax
  \def\href#1#2{#2} \def\path#1{#1}\fi

\bibitem{1997asxo.proc...21G}
N.~{Gehrels}, {Gamma Ray Transients}, in: M.~{Matsuoka}, N.~{Kawai} (Eds.),
  All-Sky X-Ray Observations in the Next Decade, 1997, p.~21.

\bibitem{2001ApJ...559..187K}
H.~{Krawczynski}, R.~{Sambruna}, A.~{Kohnle} et~al., {Simultaneous X-Ray and
  TeV Gamma-Ray Observation of the TeV Blazar Markarian 421 during 2000
  February and May}, \apj 559~(1) (2001) 187--195.
\newblock \href {http://arxiv.org/abs/astro-ph/0105331}
  {\path{arXiv:astro-ph/0105331}}, \href {https://doi.org/10.1086/322364}
  {\path{doi:10.1086/322364}}.

\bibitem{2007ApJ...665L..51A}
J.~{Albert}, E.~{Aliu}, H.~{Anderhub} et~al., {Very High Energy Gamma-Ray
  Radiation from the Stellar Mass Black Hole Binary Cygnus X-1}, \apjl 665~(1)
  (2007) L51--L54.
\newblock \href {http://arxiv.org/abs/0706.1505} {\path{arXiv:0706.1505}},
  \href {https://doi.org/10.1086/521145} {\path{doi:10.1086/521145}}.

\bibitem{2009Sci...326.1512F}
{Fermi LAT Collaboration}, A.~A. {Abdo}, M.~{Ackermann} et~al., {Modulated
  High-Energy Gamma-Ray Emission from the Microquasar Cygnus X-3}, Science
  326~(5959) (2009) 1512.
\newblock \href {https://doi.org/10.1126/science.1182174}
  {\path{doi:10.1126/science.1182174}}.

\bibitem{2010ApJ...712L..10S}
S.~{Sabatini}, M.~{Tavani}, E.~{Striani} et~al., {Episodic Transient Gamma-ray
  Emission from the Microquasar Cygnus X-1}, \apjl 712~(1) (2010) L10--L15.
\newblock \href {http://arxiv.org/abs/1002.4967} {\path{arXiv:1002.4967}},
  \href {https://doi.org/10.1088/2041-8205/712/1/L10}
  {\path{doi:10.1088/2041-8205/712/1/L10}}.

\bibitem{2011ApJ...736L..11A}
A.~A. {Abdo}, M.~{Ackermann}, M.~{Ajello} et~al., {Discovery of High-energy
  Gamma-ray Emission from the Binary System PSR B1259-63/LS 2883 around
  Periastron with Fermi}, \apjl 736~(1) (2011) L11.
\newblock \href {http://arxiv.org/abs/1103.4108} {\path{arXiv:1103.4108}},
  \href {https://doi.org/10.1088/2041-8205/736/1/L11}
  {\path{doi:10.1088/2041-8205/736/1/L11}}.

\bibitem{2016A&A...596A..55Z}
R.~{Zanin}, A.~{Fern{\'a}ndez-Barral}, E.~{de O{\~n}a Wilhelmi},
  F.~{Aharonian}, O.~{Blanch}, V.~{Bosch-Ramon}, D.~{Galindo}, {Gamma rays
  detected from Cygnus X-1 with likely jet origin}, \aap 596 (2016) A55.
\newblock \href {http://arxiv.org/abs/1605.05914} {\path{arXiv:1605.05914}},
  \href {https://doi.org/10.1051/0004-6361/201628917}
  {\path{doi:10.1051/0004-6361/201628917}}.

\bibitem{2024Univ...10...57V}
A.~A. {Vigliano}, F.~{Longo}, {Gamma-ray Bursts: 50 Years and Counting!},
  Universe 10~(2) (2024) 57.
\newblock \href {https://doi.org/10.3390/universe10020057}
  {\path{doi:10.3390/universe10020057}}.

\bibitem{2018MNRAS.477.4257C}
L.~{Costamante}, G.~{Bonnoli}, F.~{Tavecchio}, G.~{Ghisellini},
  G.~{Tagliaferri}, D.~{Khangulyan}, {The NuSTAR view on hard-TeV BL Lacs},
  \mnras 477~(3) (2018) 4257--4268.
\newblock \href {http://arxiv.org/abs/1711.06282} {\path{arXiv:1711.06282}},
  \href {https://doi.org/10.1093/mnras/sty857}
  {\path{doi:10.1093/mnras/sty857}}.

\bibitem{2017ApJ...841...61A}
F.~A. {Aharonian}, M.~V. {Barkov}, D.~{Khangulyan}, {Scenarios for Ultrafast
  Gamma-Ray Variability in AGN}, \apj 841~(1) (2017) 61.
\newblock \href {http://arxiv.org/abs/1704.08148} {\path{arXiv:1704.08148}},
  \href {https://doi.org/10.3847/1538-4357/aa7049}
  {\path{doi:10.3847/1538-4357/aa7049}}.

\bibitem{2017SSRv..207...63W}
R.~{Willingale}, P.~{M{\'e}sz{\'a}ros}, {Gamma-Ray Bursts and Fast Transients.
  Multi-wavelength Observations and Multi-messenger Signals}, \ssr 207~(1-4)
  (2017) 63--86.
\newblock \href {https://doi.org/10.1007/s11214-017-0366-4}
  {\path{doi:10.1007/s11214-017-0366-4}}.

\bibitem{2021ARA&A..59..391C}
L.~{Chomiuk}, B.~D. {Metzger}, K.~J. {Shen}, {New Insights into Classical
  Novae}, \araa 59 (2021) 391--444.
\newblock \href {http://arxiv.org/abs/2011.08751} {\path{arXiv:2011.08751}},
  \href {https://doi.org/10.1146/annurev-astro-112420-114502}
  {\path{doi:10.1146/annurev-astro-112420-114502}}.

\bibitem{2024Univ...10..163C}
A.~{Carosi}, A.~{L{\'o}pez-Oramas}, {A Very-High-Energy Gamma-Ray View of the
  Transient Sky}, Universe 10~(4) (2024) 163.
\newblock \href {http://arxiv.org/abs/2404.17480} {\path{arXiv:2404.17480}},
  \href {https://doi.org/10.3390/universe10040163}
  {\path{doi:10.3390/universe10040163}}.

\bibitem{1993apj...405..273w}
S.~E. {Woosley}, {Gamma-Ray Bursts from Stellar Mass Accretion Disks around
  Black Holes}, \apj 405 (1993) 273.
\newblock \href {https://doi.org/10.1086/172359} {\path{doi:10.1086/172359}}.

\bibitem{1989Natur.340..126E}
D.~{Eichler}, M.~{Livio}, T.~{Piran}, D.~N. {Schramm}, {Nucleosynthesis,
  neutrino bursts and {\ensuremath{\gamma}}-rays from coalescing neutron
  stars}, \nat 340~(6229) (1989) 126--128.
\newblock \href {https://doi.org/10.1038/340126a0}
  {\path{doi:10.1038/340126a0}}.

\bibitem{1975Natur.254..295H}
J.~G. {Hills}, {Possible power source of Seyfert galaxies and QSOs}, \nat
  254~(5498) (1975) 295--298.
\newblock \href {https://doi.org/10.1038/254295a0}
  {\path{doi:10.1038/254295a0}}.

\bibitem{1988Natur.333..523R}
M.~J. {Rees}, {Tidal disruption of stars by black holes of
  {}10$^{6}$-{}10$^{8}$ solar masses in nearby galaxies}, \nat 333~(6173)
  (1988) 523--528.
\newblock \href {https://doi.org/10.1038/333523a0}
  {\path{doi:10.1038/333523a0}}.

\bibitem{1972ApJ...176..169S}
S.~{Starrfield}, J.~W. {Truran}, W.~M. {Sparks}, G.~S. {Kutter}, {CNO
  Abundances and Hydrodynamic Models of the Nova Outburst}, \apj 176 (1972)
  169.
\newblock \href {https://doi.org/10.1086/151619} {\path{doi:10.1086/151619}}.

\bibitem{1995MNRAS.275..255T}
C.~{Thompson}, R.~C. {Duncan}, {The soft gamma repeaters as very strongly
  magnetized neutron stars - I. Radiative mechanism for outbursts}, \mnras
  275~(2) (1995) 255--300.
\newblock \href {https://doi.org/10.1093/mnras/275.2.255}
  {\path{doi:10.1093/mnras/275.2.255}}.

\bibitem{1965JGR....70.4219S}
T.~W. {Speiser}, {Particle Trajectories in Model Current Sheets, 1, Analytical
  Solutions}, \jgr 70~(17) (1965) 4219--4226.
\newblock \href {https://doi.org/10.1029/JZ070i017p04219}
  {\path{doi:10.1029/JZ070i017p04219}}.

\bibitem{1978MNRAS.182..147B}
A.~R. {Bell}, {The acceleration of cosmic rays in shock fronts - I.}, \mnras
  182 (1978) 147--156.
\newblock \href {https://doi.org/10.1093/mnras/182.2.147}
  {\path{doi:10.1093/mnras/182.2.147}}.

\bibitem{1984ARA&A..22..425H}
A.~M. {Hillas}, {The Origin of Ultra-High-Energy Cosmic Rays}, \araa 22 (1984)
  425--444.
\newblock \href {https://doi.org/10.1146/annurev.aa.22.090184.002233}
  {\path{doi:10.1146/annurev.aa.22.090184.002233}}.

\bibitem{1983RPPh...46..973D}
L.~O. {Drury}, {REVIEW ARTICLE: An introduction to the theory of diffusive
  shock acceleration of energetic particles in tenuous plasmas}, Reports on
  Progress in Physics 46~(8) (1983) 973--1027.
\newblock \href {https://doi.org/10.1088/0034-4885/46/8/002}
  {\path{doi:10.1088/0034-4885/46/8/002}}.

\bibitem{2010PhRvD..82d3002A}
F.~A. {Aharonian}, S.~R. {Kelner}, A.~Y. {Prosekin}, {Angular, spectral, and
  time distributions of highest energy protons and associated secondary gamma
  rays and neutrinos propagating through extragalactic magnetic and radiation
  fields}, \prd 82~(4) (2010) 043002.
\newblock \href {http://arxiv.org/abs/1006.1045} {\path{arXiv:1006.1045}},
  \href {https://doi.org/10.1103/PhysRevD.82.043002}
  {\path{doi:10.1103/PhysRevD.82.043002}}.

\bibitem{1974ApJ...192L...3E}
J.~L. {Elliot}, S.~L. {Shapiro}, {On the Variability of the Compact Nonthermal
  Sources}, \apjl 192 (1974) L3.
\newblock \href {https://doi.org/10.1086/181575} {\path{doi:10.1086/181575}}.

\bibitem{2008MNRAS.384L..19B}
M.~C. {Begelman}, A.~C. {Fabian}, M.~J. {Rees}, {Implications of very rapid TeV
  variability in blazars}, \mnras 384~(1) (2008) L19--L23.
\newblock \href {http://arxiv.org/abs/0709.0540} {\path{arXiv:0709.0540}},
  \href {https://doi.org/10.1111/j.1745-3933.2007.00413.x}
  {\path{doi:10.1111/j.1745-3933.2007.00413.x}}.

\bibitem{1983MNRAS.205..593G}
P.~W. {Guilbert}, A.~C. {Fabian}, M.~J. {Rees}, {Spectral and variability
  constraints on compact sources}, \mnras 205 (1983) 593--603.
\newblock \href {https://doi.org/10.1093/mnras/205.3.593}
  {\path{doi:10.1093/mnras/205.3.593}}.

\bibitem{2007ApJ...664L..71A}
F.~{Aharonian}, A.~G. {Akhperjanian}, A.~R. {Bazer-Bachi} et~al., {An
  Exceptional Very High Energy Gamma-Ray Flare of PKS 2155-304}, \apjl 664~(2)
  (2007) L71--L74.
\newblock \href {http://arxiv.org/abs/0706.0797} {\path{arXiv:0706.0797}},
  \href {https://doi.org/10.1086/520635} {\path{doi:10.1086/520635}}.

\bibitem{2014A&A...563A..91A}
J.~{Aleksi{\'c}}, L.~A. {Antonelli}, P.~{Antoranz} et~al., {Rapid and multiband
  variability of the TeV bright active nucleus of the galaxy IC 310}, \aap 563
  (2014) A91.
\newblock \href {http://arxiv.org/abs/1305.5147} {\path{arXiv:1305.5147}},
  \href {https://doi.org/10.1051/0004-6361/201321938}
  {\path{doi:10.1051/0004-6361/201321938}}.

\bibitem{2015ApJ...799...86A}
M.~{Ackermann}, M.~{Ajello}, A.~{Albert} et~al., {The Spectrum of Isotropic
  Diffuse Gamma-Ray Emission between 100 MeV and 820 GeV}, \apj 799~(1) (2015)
  86.
\newblock \href {http://arxiv.org/abs/1410.3696} {\path{arXiv:1410.3696}},
  \href {https://doi.org/10.1088/0004-637X/799/1/86}
  {\path{doi:10.1088/0004-637X/799/1/86}}.

\bibitem{2005astro.ph.11139A}
F.~{Aharonian}, {Next generation of IACT arrays: scientific objectives versus
  energy domains}, arXiv e-prints (2005) astro--ph/0511139\href
  {http://arxiv.org/abs/astro-ph/0511139} {\path{arXiv:astro-ph/0511139}},
  \href {https://doi.org/10.48550/arXiv.astro-ph/0511139}
  {\path{doi:10.48550/arXiv.astro-ph/0511139}}.

\bibitem{NTRS19920012642}
{den Herder, J. W. et al.},
  \href{https://ntrs.nasa.gov/citations/19920012642}{{[COMPTEL: Instrument
  description and performance]}}, Tech. Rep. NTRS 19920012642, {NASA}, nASA
  Technical Reports Server (1992).
\newline\urlprefix\url{https://ntrs.nasa.gov/citations/19920012642}

\bibitem{Thompson2022}
D.~J. {Thompson}, C.~A. {Wilson-Hodge}, {Fermi Gamma-Ray Space Telescope}, in:
  C.~{Bambi}, A.~{Sangangelo} (Eds.), Handbook of X-ray and Gamma-ray
  Astrophysics, Springer, 2022, p.~29.
\newblock \href {https://doi.org/10.1007/978-981-16-4544-0_58-1}
  {\path{doi:10.1007/978-981-16-4544-0_58-1}}.

\bibitem{2009A&A...502..995T}
M.~{Tavani}, G.~{Barbiellini}, A.~{Argan} et~al., {The AGILE Mission}, \aap
  502~(3) (2009) 995--1013.
\newblock \href {http://arxiv.org/abs/0807.4254} {\path{arXiv:0807.4254}},
  \href {https://doi.org/10.1051/0004-6361/200810527}
  {\path{doi:10.1051/0004-6361/200810527}}.

\bibitem{2016APh....72...76A}
J.~{Aleksi{\'c}}, S.~{Ansoldi}, L.~A. {Antonelli} et~al., {The major upgrade of
  the MAGIC telescopes, Part II: A performance study using observations of the
  Crab Nebula}, Astroparticle Physics 72 (2016) 76--94.
\newblock \href {http://arxiv.org/abs/1409.5594} {\path{arXiv:1409.5594}},
  \href {https://doi.org/10.1016/j.astropartphys.2015.02.005}
  {\path{doi:10.1016/j.astropartphys.2015.02.005}}.

\bibitem{2006A&A...457..899A}
F.~{Aharonian}, A.~G. {Akhperjanian}, A.~R. {Bazer-Bachi} et~al., {Observations
  of the Crab nebula with HESS}, \aap 457~(3) (2006) 899--915.
\newblock \href {http://arxiv.org/abs/astro-ph/0607333}
  {\path{arXiv:astro-ph/0607333}}, \href
  {https://doi.org/10.1051/0004-6361:20065351}
  {\path{doi:10.1051/0004-6361:20065351}}.

\bibitem{vanEldik:2016Ns}
C.~van Eldik, M.~Holler, D.~Berge, D.~Zaborov, J.-P. Lenain, V.~Marandon,
  T.~Murach, H.~Prokoph, M.~de~Naurois, R.~D. Parsons, {Observations of the
  Crab Nebula with H.E.S.S. phase II }, PoS ICRC2015 (2016) 847.
\newblock \href {https://doi.org/10.22323/1.236.0847}
  {\path{doi:10.22323/1.236.0847}}.

\bibitem{Park:20161B}
N.~Park, {Performance of the VERITAS experiment }, PoS ICRC2015 (2016) 771.
\newblock \href {https://doi.org/10.22323/1.236.0771}
  {\path{doi:10.22323/1.236.0771}}.

\bibitem{2024ApJ...972..144A}
A.~{Albert}, R.~{Alfaro}, C.~{Alvarez} et~al., {Performance of the HAWC
  Observatory and TeV Gamma-Ray Measurements of the Crab Nebula with Improved
  Extensive Air Shower Reconstruction Algorithms}, \apj 972~(2) (2024) 144.
\newblock \href {http://arxiv.org/abs/2405.06050} {\path{arXiv:2405.06050}},
  \href {https://doi.org/10.3847/1538-4357/ad5f2d}
  {\path{doi:10.3847/1538-4357/ad5f2d}}.

\bibitem{2023NIMPA105268253A}
A.~U. {Abeysekara}, A.~{Albert}, R.~{Alfaro} et~al., {The High-Altitude Water
  Cherenkov (HAWC) observatory in M{\'e}xico: The primary detector}, Nuclear
  Instruments and Methods in Physics Research A 1052 (2023) 168253.
\newblock \href {http://arxiv.org/abs/2304.00730} {\path{arXiv:2304.00730}},
  \href {https://doi.org/10.1016/j.nima.2023.168253}
  {\path{doi:10.1016/j.nima.2023.168253}}.

\bibitem{2022ChPhC..46c0001M}
X.-H. {Ma}, Y.-J. {Bi}, Z.~{Cao}, M.-J. {Chen}, S.-Z. {Chen}, Y.-D. {Cheng},
  G.-H. {Gong}, M.-H. {Gu}, H.-H. {He}, C.~{Hou}, W.-H. {Huang}, X.-T. {Huang},
  C.~{Liu}, O.~{Shchegolev}, X.-D. {Sheng}, Y.~{Stenkin}, C.-Y. {Wu}, H.-R.
  {Wu}, S.~{Wu}, G.~{Xiao}, Z.-G. {Yao}, S.-S. {Zhang}, Y.~{Zhang}, X.~{Zuo},
  {Chapter 1 LHAASO Instruments and Detector technology}, Chinese Physics C
  46~(3) (2022) 030001.
\newblock \href {https://doi.org/10.1088/1674-1137/ac3fa6}
  {\path{doi:10.1088/1674-1137/ac3fa6}}.

\bibitem{1992Natur.355..143M}
C.~A. {Meegan}, G.~J. {Fishman}, R.~B. {Wilson}, W.~S. {Paciesas}, G.~N.
  {Pendleton}, J.~M. {Horack}, M.~N. {Brock}, C.~{Kouveliotou}, {Spatial
  distribution of {\ensuremath{\gamma}}-ray bursts observed by BATSE}, \nat
  355~(6356) (1992) 143--145.
\newblock \href {https://doi.org/10.1038/355143a0}
  {\path{doi:10.1038/355143a0}}.

\bibitem{1993apj...413l.101k}
C.~{Kouveliotou}, C.~A. {Meegan}, G.~J. {Fishman}, N.~P. {Bhat}, M.~S.
  {Briggs}, T.~M. {Koshut}, W.~S. {Paciesas}, G.~N. {Pendleton},
  {Identification of Two Classes of Gamma-Ray Bursts}, \apjl 413 (1993) L101.
\newblock \href {https://doi.org/10.1086/186969} {\path{doi:10.1086/186969}}.

\bibitem{2003apj...593l..19k}
K.~S. {Kawabata}, J.~{Deng}, L.~{Wang} et~al., {On the Spectrum and
  Spectropolarimetry of Type Ic Hypernova SN 2003dh/GRB 030329}, \apjl 593~(1)
  (2003) L19--L22.
\newblock \href {http://arxiv.org/abs/astro-ph/0306155}
  {\path{arXiv:astro-ph/0306155}}, \href {https://doi.org/10.1086/378148}
  {\path{doi:10.1086/378148}}.

\bibitem{2003apj...591l..17s}
K.~Z. {Stanek}, T.~{Matheson}, P.~M. {Garnavich}, P.~{Martini}, P.~{Berlind},
  N.~{Caldwell}, P.~{Challis}, W.~R. {Brown}, R.~{Schild}, K.~{Krisciunas},
  M.~L. {Calkins}, J.~C. {Lee}, N.~{Hathi}, R.~A. {Jansen}, R.~{Windhorst},
  L.~{Echevarria}, D.~J. {Eisenstein}, B.~{Pindor}, E.~W. {Olszewski},
  P.~{Harding}, S.~T. {Holland}, D.~{Bersier}, {Spectroscopic Discovery of the
  Supernova 2003dh Associated with GRB 030329}, \apjl 591~(1) (2003) L17--L20.
\newblock \href {http://arxiv.org/abs/astro-ph/0304173}
  {\path{arXiv:astro-ph/0304173}}, \href {https://doi.org/10.1086/376976}
  {\path{doi:10.1086/376976}}.

\bibitem{2017PhRvL.119p1101A}
B.~P. {Abbott}, R.~{Abbott}, T.~D. {Abbott} et~al., {GW170817: Observation of
  Gravitational Waves from a Binary Neutron Star Inspiral}, \prl 119~(16)
  (2017) 161101.
\newblock \href {http://arxiv.org/abs/1710.05832} {\path{arXiv:1710.05832}},
  \href {https://doi.org/10.1103/PhysRevLett.119.161101}
  {\path{doi:10.1103/PhysRevLett.119.161101}}.

\bibitem{2011apj...736....7c}
A.~{Cucchiara}, A.~J. {Levan}, D.~B. {Fox} et~al., {A Photometric Redshift of z
  \raisebox{-0.5ex}\textasciitilde 9.4 for GRB 090429B}, \apj 736~(1) (2011) 7.
\newblock \href {http://arxiv.org/abs/1105.4915} {\path{arXiv:1105.4915}},
  \href {https://doi.org/10.1088/0004-637X/736/1/7}
  {\path{doi:10.1088/0004-637X/736/1/7}}.

\bibitem{2024ApJ...966...31K}
D.~{Khangulyan}, F.~{Aharonian}, A.~M. {Taylor}, {Naked Forward Shock Seen in
  the TeV Afterglow Data of GRB 221009A}, \apj 966~(1) (2024) 31.
\newblock \href {http://arxiv.org/abs/2309.00673} {\path{arXiv:2309.00673}},
  \href {https://doi.org/10.3847/1538-4357/ad3550}
  {\path{doi:10.3847/1538-4357/ad3550}}.

\bibitem{2001AdSpR..27..813D}
E.~V. {Derishev}, V.~V. {Kocharovsky}, V.~V. {Kocharovsky}, {TeV photons from
  gamma-ray bursts}, Advances in Space Research 27~(4) (2001) 813--818.
\newblock \href {https://doi.org/10.1016/S0273-1177(01)00126-0}
  {\path{doi:10.1016/S0273-1177(01)00126-0}}.

\bibitem{2020ApJ...893...46V}
A.~{von Kienlin}, C.~A. {Meegan}, W.~S. {Paciesas}, P.~N. {Bhat},
  E.~{Bissaldi}, M.~S. {Briggs}, E.~{Burns}, W.~H. {Cleveland}, M.~H. {Gibby},
  M.~M. {Giles}, A.~{Goldstein}, R.~{Hamburg}, C.~M. {Hui}, D.~{Kocevski},
  B.~{Mailyan}, C.~{Malacaria}, S.~{Poolakkil}, R.~D. {Preece}, O.~J.
  {Roberts}, P.~{Veres}, C.~A. {Wilson-Hodge}, {The Fourth Fermi-GBM Gamma-Ray
  Burst Catalog: A Decade of Data}, \apj 893~(1) (2020) 46.
\newblock \href {http://arxiv.org/abs/2002.11460} {\path{arXiv:2002.11460}},
  \href {https://doi.org/10.3847/1538-4357/ab7a18}
  {\path{doi:10.3847/1538-4357/ab7a18}}.

\bibitem{2019ApJ...878...52A}
M.~{Ajello}, M.~{Arimoto}, M.~{Axelsson} et~al., {A Decade of Gamma-Ray Bursts
  Observed by Fermi-LAT: The Second GRB Catalog}, \apj 878~(1) (2019) 52.
\newblock \href {http://arxiv.org/abs/1906.11403} {\path{arXiv:1906.11403}},
  \href {https://doi.org/10.3847/1538-4357/ab1d4e}
  {\path{doi:10.3847/1538-4357/ab1d4e}}.

\bibitem{2025arXiv250804557F}
L.~{Foffano}, M.~{Tavani}, {TeV Afterglows of Gamma-Ray Bursts: Theoretical
  Analysis and Prospects for Future Observations}, arXiv e-prints (2025)
  arXiv:2508.04557\href {http://arxiv.org/abs/2508.04557}
  {\path{arXiv:2508.04557}}, \href {https://doi.org/10.48550/arXiv.2508.04557}
  {\path{doi:10.48550/arXiv.2508.04557}}.

\bibitem{2019Natur.575..455M}
{MAGIC Collaboration}, V.~A. {Acciari}, S.~{Ansoldi} et~al., {Teraelectronvolt
  emission from the {\ensuremath{\gamma}}-ray burst GRB 190114C}, \nat
  575~(7783) (2019) 455--458.
\newblock \href {http://arxiv.org/abs/2006.07249} {\path{arXiv:2006.07249}},
  \href {https://doi.org/10.1038/s41586-019-1750-x}
  {\path{doi:10.1038/s41586-019-1750-x}}.

\bibitem{2020ApJ...890....9A}
M.~{Ajello}, M.~{Arimoto}, M.~{Axelsson} et~al., {Fermi and Swift Observations
  of GRB 190114C: Tracing the Evolution of High-energy Emission from Prompt to
  Afterglow}, \apj 890~(1) (2020) 9.
\newblock \href {http://arxiv.org/abs/1909.10605} {\path{arXiv:1909.10605}},
  \href {https://doi.org/10.3847/1538-4357/ab5b05}
  {\path{doi:10.3847/1538-4357/ab5b05}}.

\bibitem{2019Natur.575..459M}
{MAGIC Collaboration}, V.~A. {Acciari}, S.~{Ansoldi} et~al., {Observation of
  inverse Compton emission from a long {\ensuremath{\gamma}}-ray burst}, \nat
  575~(7783) (2019) 459--463.
\newblock \href {http://arxiv.org/abs/2006.07251} {\path{arXiv:2006.07251}},
  \href {https://doi.org/10.1038/s41586-019-1754-6}
  {\path{doi:10.1038/s41586-019-1754-6}}.

\bibitem{2023MNRAS.520..839K}
M.~{Klinger}, D.~{Tak}, A.~M. {Taylor}, S.~J. {Zhu}, {Probing the
  multiwavelength emission scenario of GRB 190114C}, \mnras 520~(1) (2023)
  839--849.
\newblock \href {http://arxiv.org/abs/2206.11148} {\path{arXiv:2206.11148}},
  \href {https://doi.org/10.1093/mnras/stad142}
  {\path{doi:10.1093/mnras/stad142}}.

\bibitem{2021Sci...372.1081H}
{H.~E.~S.~S. Collaboration}, H.~{Abdalla}, F.~{Aharonian} et~al., {Revealing
  x-ray and gamma ray temporal and spectral similarities in the GRB 190829A
  afterglow}, Science 372~(6546) (2021) 1081--1085.
\newblock \href {http://arxiv.org/abs/2106.02510} {\path{arXiv:2106.02510}},
  \href {https://doi.org/10.1126/science.abe8560}
  {\path{doi:10.1126/science.abe8560}}.

\bibitem{2020ApJ...903....9C}
V.~{Chand}, P.~S. {Pal}, A.~{Banerjee}, V.~{Sharma}, P.~H.~T. {Tam}, X.~{He},
  {MAGICal GRB 190114C: Implications of Cutoff in the Spectrum at sub-GeV
  Energies}, \apj 903~(1) (2020) 9.
\newblock \href {http://arxiv.org/abs/1905.11844} {\path{arXiv:1905.11844}},
  \href {https://doi.org/10.3847/1538-4357/abb5fc}
  {\path{doi:10.3847/1538-4357/abb5fc}}.

\bibitem{2019natur.575..464a}
H.~{Abdalla}, R.~{Adam}, F.~{Aharonian} et~al., {A very-high-energy component
  deep in the {\ensuremath{\gamma}}-ray burst afterglow}, \nat 575~(7783)
  (2019) 464--467.
\newblock \href {http://arxiv.org/abs/1911.08961} {\path{arXiv:1911.08961}},
  \href {https://doi.org/10.1038/s41586-019-1743-9}
  {\path{doi:10.1038/s41586-019-1743-9}}.

\bibitem{2021apj...923..135d}
E.~{Derishev}, T.~{Piran}, {GRB Afterglow Parameters in the Era of TeV
  Observations: The Case of GRB 190114C}, \apj 923~(2) (2021) 135.
\newblock \href {http://arxiv.org/abs/2106.12035} {\path{arXiv:2106.12035}},
  \href {https://doi.org/10.3847/1538-4357/ac2dec}
  {\path{doi:10.3847/1538-4357/ac2dec}}.

\bibitem{2023ApJ...952L..42L}
S.~{Lesage}, P.~{Veres}, M.~S. {Briggs} et~al., {Fermi-GBM Discovery of GRB
  221009A: An Extraordinarily Bright GRB from Onset to Afterglow}, \apjl
  952~(2) (2023) L42.
\newblock \href {http://arxiv.org/abs/2303.14172} {\path{arXiv:2303.14172}},
  \href {https://doi.org/10.3847/2041-8213/ace5b4}
  {\path{doi:10.3847/2041-8213/ace5b4}}.

\bibitem{2023Sci...380.1390L}
{LHAASO Collaboration}, Z.~{Cao}, F.~{Aharonian} et~al., {A tera-electron volt
  afterglow from a narrow jet in an extremely bright gamma-ray burst.}, Science
  380~(6652) (2023) 1390--1396.
\newblock \href {http://arxiv.org/abs/2306.06372} {\path{arXiv:2306.06372}},
  \href {https://doi.org/10.1126/science.adg9328}
  {\path{doi:10.1126/science.adg9328}}.

\bibitem{2024MNRAS.530..347D}
E.~{Derishev}, T.~{Piran}, {The contemporaneous phase of GRB afterglows -
  application to GRB 221009A}, \mnras 530~(1) (2024) 347--359.
\newblock \href {http://arxiv.org/abs/2312.01447} {\path{arXiv:2312.01447}},
  \href {https://doi.org/10.1093/mnras/stae609}
  {\path{doi:10.1093/mnras/stae609}}.

\bibitem{2023ApJ...946L..27A}
F.~{Aharonian}, F.~{Ait Benkhali}, J.~{Aschersleben} et~al., {H.E.S.S.
  Follow-up Observations of GRB 221009A}, \apjl 946~(1) (2023) L27.
\newblock \href {http://arxiv.org/abs/2303.10558} {\path{arXiv:2303.10558}},
  \href {https://doi.org/10.3847/2041-8213/acc405}
  {\path{doi:10.3847/2041-8213/acc405}}.

\bibitem{2025ApJ...988L..42A}
K.~{Abe}, S.~{Abe}, A.~{Abhishek} et~al., {GRB 221009A: Observations with LST-1
  of CTAO and Implications for Structured Jets in Long Gamma-Ray Bursts}, \apjl
  988~(2) (2025) L42.
\newblock \href {http://arxiv.org/abs/2507.03077} {\path{arXiv:2507.03077}},
  \href {https://doi.org/10.3847/2041-8213/ade4cf}
  {\path{doi:10.3847/2041-8213/ade4cf}}.

\bibitem{2025ApJS..277...24A}
M.~{Axelsson}, M.~{Ajello}, M.~{Arimoto} et~al., {GRB 221009A: The B.O.A.T.
  Burst that Shines in Gamma Rays}, \apjs 277~(1) (2025) 24.
\newblock \href {http://arxiv.org/abs/2409.04580} {\path{arXiv:2409.04580}},
  \href {https://doi.org/10.3847/1538-4365/ada272}
  {\path{doi:10.3847/1538-4365/ada272}}.

\bibitem{2007ApJ...663L.101T}
V.~{Tatischeff}, M.~{Hernanz}, {Evidence for Nonlinear Diffusive Shock
  Acceleration of Cosmic Rays in the 2006 Outburst of the Recurrent Nova RS
  Ophiuchi}, \apjl 663~(2) (2007) L101--L104.
\newblock \href {http://arxiv.org/abs/0705.4422} {\path{arXiv:0705.4422}},
  \href {https://doi.org/10.1086/520049} {\path{doi:10.1086/520049}}.

\bibitem{2010Sci...329..817A}
A.~A. {Abdo}, M.~{Ackermann}, M.~{Ajello} et~al., {Gamma-Ray Emission
  Concurrent with the Nova in the Symbiotic Binary V407 Cygni}, Science
  329~(5993) (2010) 817--821.
\newblock \href {http://arxiv.org/abs/1008.3912} {\path{arXiv:1008.3912}},
  \href {https://doi.org/10.1126/science.1192537}
  {\path{doi:10.1126/science.1192537}}.

\bibitem{2012ApJ...754...77A}
E.~{Aliu}, S.~{Archambault}, T.~{Arlen} et~al., {VERITAS Observations of the
  Nova in V407 Cygni}, \apj 754~(1) (2012) 77.
\newblock \href {http://arxiv.org/abs/1205.5287} {\path{arXiv:1205.5287}},
  \href {https://doi.org/10.1088/0004-637X/754/1/77}
  {\path{doi:10.1088/0004-637X/754/1/77}}.

\bibitem{2018A&A...609A.120F}
A.~{Franckowiak}, P.~{Jean}, M.~{Wood}, C.~C. {Cheung}, S.~{Buson}, {Search for
  gamma-ray emission from Galactic novae with the Fermi -LAT}, \aap 609 (2018)
  A120.
\newblock \href {http://arxiv.org/abs/1710.04736} {\path{arXiv:1710.04736}},
  \href {https://doi.org/10.1051/0004-6361/201731516}
  {\path{doi:10.1051/0004-6361/201731516}}.

\bibitem{2022MNRAS.517.6150S}
B.~E. {Schaefer}, {Comprehensive catalogue of the overall best distances and
  properties of 402 galactic novae}, \mnras 517~(4) (2022) 6150--6169.
\newblock \href {http://arxiv.org/abs/2210.03181} {\path{arXiv:2210.03181}},
  \href {https://doi.org/10.1093/mnras/stac2900}
  {\path{doi:10.1093/mnras/stac2900}}.

\bibitem{2015A&A...582A..67A}
M.~L. {Ahnen}, S.~{Ansoldi}, L.~A. {Antonelli} et~al., {Very high-energy
  {\ensuremath{\gamma}}-ray observations of novae and dwarf novae with the
  MAGIC telescopes}, \aap 582 (2015) A67.
\newblock \href {http://arxiv.org/abs/1508.04902} {\path{arXiv:1508.04902}},
  \href {https://doi.org/10.1051/0004-6361/201526478}
  {\path{doi:10.1051/0004-6361/201526478}}.

\bibitem{2022ApJ...935...44C}
C.~C. {Cheung}, T.~J. {Johnson}, P.~{Jean}, M.~{Kerr}, K.~L. {Page}, J.~P.
  {Osborne}, A.~P. {Beardmore}, K.~V. {Sokolovsky}, F.~{Teyssier},
  S.~{Ciprini}, G.~{Mart{\'\i}-Devesa}, I.~{Mereu}, S.~{Razzaque}, K.~S.
  {Wood}, S.~N. {Shore}, S.~{Korotkiy}, A.~{Levina}, A.~{Blumenzweig}, {Fermi
  LAT Gamma-ray Detection of the Recurrent Nova RS Ophiuchi during its 2021
  Outburst}, \apj 935~(1) (2022) 44.
\newblock \href {http://arxiv.org/abs/2207.02921} {\path{arXiv:2207.02921}},
  \href {https://doi.org/10.3847/1538-4357/ac7eb7}
  {\path{doi:10.3847/1538-4357/ac7eb7}}.

\bibitem{2022Sci...376...77H}
{H.~E.~S.~S. Collaboration}, F.~{Aharonian}, F.~{Ait Benkhali} et~al.,
  {Time-resolved hadronic particle acceleration in the recurrent nova RS
  Ophiuchi}, Science 376~(6588) (2022) 77--80.
\newblock \href {http://arxiv.org/abs/2202.08201} {\path{arXiv:2202.08201}},
  \href {https://doi.org/10.1126/science.abn0567}
  {\path{doi:10.1126/science.abn0567}}.

\bibitem{2022NatAs...6..689A}
V.~A. {Acciari}, S.~{Ansoldi}, L.~A. {Antonelli} et~al., {Proton acceleration
  in thermonuclear nova explosions revealed by gamma rays}, Nature Astronomy 6
  (2022) 689--697.
\newblock \href {http://arxiv.org/abs/2202.07681} {\path{arXiv:2202.07681}},
  \href {https://doi.org/10.1038/s41550-022-01640-z}
  {\path{doi:10.1038/s41550-022-01640-z}}.

\bibitem{2025A&A...695A.152A}
K.~{Abe}, S.~{Abe}, A.~{Abhishek} et~al., {Detection of RS Oph with LST-1 and
  modelling of its HE/VHE gamma-ray emission}, \aap 695 (2025) A152.
\newblock \href {http://arxiv.org/abs/2503.13283} {\path{arXiv:2503.13283}},
  \href {https://doi.org/10.1051/0004-6361/202452447}
  {\path{doi:10.1051/0004-6361/202452447}}.

\bibitem{2012BaltA..21...62H}
M.~{Hernanz}, V.~{Tatischeff}, {High Energy Emission of Symbiotic Recurrent
  Novae: RS Oph and V407 Cyg}, Baltic Astronomy 21 (2012) 62--67.
\newblock \href {http://arxiv.org/abs/1111.4129} {\path{arXiv:1111.4129}},
  \href {https://doi.org/10.1515/astro-2017-0359}
  {\path{doi:10.1515/astro-2017-0359}}.

\bibitem{2026NewAR.10201747R}
X.~{Rodrigues}, {Neutrinos from extreme astrophysical sources}, \nar 102 (2026)
  101747.
\newblock \href {http://arxiv.org/abs/2603.10167} {\path{arXiv:2603.10167}},
  \href {https://doi.org/10.1016/j.newar.2026.101747}
  {\path{doi:10.1016/j.newar.2026.101747}}.

\bibitem{2023JHEAp..38...22B}
W.~{Bednarek}, J.~{Sitarek}, {Gamma rays from nebulae around recurrent novae},
  Journal of High Energy Astrophysics 38 (2023) 22--31.
\newblock \href {http://arxiv.org/abs/2303.15741} {\path{arXiv:2303.15741}},
  \href {https://doi.org/10.1016/j.jheap.2023.03.004}
  {\path{doi:10.1016/j.jheap.2023.03.004}}.

\bibitem{2023JHA....54..436S}
B.~E. {Schaefer}, {The recurrent nova T CrB had prior eruptions observed near
  December 1787 and October 1217 AD}, Journal for the History of Astronomy
  54~(4) (2023) 436--455.
\newblock \href {http://arxiv.org/abs/2308.13668} {\path{arXiv:2308.13668}},
  \href {https://doi.org/10.1177/00218286231200492}
  {\path{doi:10.1177/00218286231200492}}.

\bibitem{2024RNAAS...8..272S}
J.~{Schneider}, {When will the Next T CrB Eruption Occur?}, Research Notes of
  the American Astronomical Society 8~(10) (2024) 272.
\newblock \href {https://doi.org/10.3847/2515-5172/ad8bba}
  {\path{doi:10.3847/2515-5172/ad8bba}}.

\bibitem{2023MNRAS.524.3146S}
B.~E. {Schaefer}, {The B \& V light curves for recurrent nova T CrB from
  1842-2022, the unique pre- and post-eruption high-states, the complex period
  changes, and the upcoming eruption in 2025.5 {\ensuremath{\pm}} 1.3}, \mnras
  524~(2) (2023) 3146--3165.
\newblock \href {http://arxiv.org/abs/2303.04933} {\path{arXiv:2303.04933}},
  \href {https://doi.org/10.1093/mnras/stad735}
  {\path{doi:10.1093/mnras/stad735}}.

\bibitem{2025MNRAS.540..205A}
K.~{Abe}, S.~{Abe}, J.~{Abhir} et~al., {Galactic transient sources with the
  Cherenkov Telescope Array Observatory}, \mnras 540~(1) (2025) 205--238.
\newblock \href {http://arxiv.org/abs/2405.04469} {\path{arXiv:2405.04469}},
  \href {https://doi.org/10.1093/mnras/staf655}
  {\path{doi:10.1093/mnras/staf655}}.

\bibitem{1994Natur.371...46M}
I.~F. {Mirabel}, L.~F. {Rodr{\'\i}guez}, {A superluminal source in the Galaxy},
  \nat 371~(6492) (1994) 46--48.
\newblock \href {https://doi.org/10.1038/371046a0}
  {\path{doi:10.1038/371046a0}}.

\bibitem{2006csxs.book..157M}
J.~E. {McClintock}, R.~A. {Remillard}, {Black hole binaries}, in: W.~H.~G.
  {Lewin}, M.~{van der Klis} (Eds.), Compact stellar X-ray sources, Vol.~39,
  Cambridge University Press, 2006, Ch.~4, pp. 157--213.
\newblock \href {https://doi.org/10.48550/arXiv.astro-ph/0306213}
  {\path{doi:10.48550/arXiv.astro-ph/0306213}}.

\bibitem{2004MNRAS.355.1105F}
R.~P. {Fender}, T.~M. {Belloni}, E.~{Gallo}, {Towards a unified model for black
  hole X-ray binary jets}, \mnras 355~(4) (2004) 1105--1118.
\newblock \href {http://arxiv.org/abs/astro-ph/0409360}
  {\path{arXiv:astro-ph/0409360}}, \href
  {https://doi.org/10.1111/j.1365-2966.2004.08384.x}
  {\path{doi:10.1111/j.1365-2966.2004.08384.x}}.

\bibitem{2003MNRAS.344...60G}
E.~{Gallo}, R.~P. {Fender}, G.~G. {Pooley}, {A universal radio-X-ray
  correlation in low/hard state black hole binaries}, \mnras 344~(1) (2003)
  60--72.
\newblock \href {http://arxiv.org/abs/astro-ph/0305231}
  {\path{arXiv:astro-ph/0305231}}, \href
  {https://doi.org/10.1046/j.1365-8711.2003.06791.x}
  {\path{doi:10.1046/j.1365-8711.2003.06791.x}}.

\bibitem{2003A&A...400.1007C}
S.~{Corbel}, M.~A. {Nowak}, R.~P. {Fender}, A.~K. {Tzioumis}, S.~{Markoff},
  {Radio/X-ray correlation in the low/hard state of GX 339-4}, \aap 400 (2003)
  1007--1012.
\newblock \href {http://arxiv.org/abs/astro-ph/0301436}
  {\path{arXiv:astro-ph/0301436}}, \href
  {https://doi.org/10.1051/0004-6361:20030090}
  {\path{doi:10.1051/0004-6361:20030090}}.

\bibitem{2005ASPC..340..269R}
M.~{Rib{\'o}}, {Microquasars}, in: J.~{Romney}, M.~{Reid} (Eds.), Future
  Directions in High Resolution Astronomy, Vol. 340 of Astronomical Society of
  the Pacific Conference Series, 2005, p. 269.
\newblock \href {http://arxiv.org/abs/astro-ph/0402134}
  {\path{arXiv:astro-ph/0402134}}, \href
  {https://doi.org/10.48550/arXiv.astro-ph/0402134}
  {\path{doi:10.48550/arXiv.astro-ph/0402134}}.

\bibitem{2002Sci...298..196C}
S.~{Corbel}, R.~P. {Fender}, A.~K. {Tzioumis}, J.~A. {Tomsick}, J.~A. {Orosz},
  J.~M. {Miller}, R.~{Wijnands}, P.~{Kaaret}, {Large-Scale, Decelerating,
  Relativistic X-ray Jets from the Microquasar XTE J1550-564}, Science
  298~(5591) (2002) 196--199.
\newblock \href {http://arxiv.org/abs/astro-ph/0210224}
  {\path{arXiv:astro-ph/0210224}}, \href
  {https://doi.org/10.1126/science.1075857}
  {\path{doi:10.1126/science.1075857}}.

\bibitem{2009IJMPD..18..347B}
V.~{Bosch-Ramon}, D.~{Khangulyan}, {Understanding the Very-High Emission from
  Microquasars}, International Journal of Modern Physics D 18~(3) (2009)
  347--387.
\newblock \href {http://arxiv.org/abs/0805.4123} {\path{arXiv:0805.4123}},
  \href {https://doi.org/10.1142/S0218271809014601}
  {\path{doi:10.1142/S0218271809014601}}.

\bibitem{2009A&A...497..325B}
P.~{Bordas}, V.~{Bosch-Ramon}, J.~M. {Paredes}, M.~{Perucho}, {Non-thermal
  emission from microquasar/ISM interaction}, \aap 497~(2) (2009) 325--334.
\newblock \href {http://arxiv.org/abs/0903.3293} {\path{arXiv:0903.3293}},
  \href {https://doi.org/10.1051/0004-6361/200810781}
  {\path{doi:10.1051/0004-6361/200810781}}.

\bibitem{2011A&A...528A..89B}
V.~{Bosch-Ramon}, M.~{Perucho}, P.~{Bordas}, {The termination region of
  high-mass microquasar jets}, \aap 528 (2011) A89.
\newblock \href {http://arxiv.org/abs/1101.5049} {\path{arXiv:1101.5049}},
  \href {https://doi.org/10.1051/0004-6361/201016364}
  {\path{doi:10.1051/0004-6361/201016364}}.

\bibitem{2017A&A...604A..39D}
V.~M. {de la Cita}, S.~{del Palacio}, V.~{Bosch-Ramon}, X.~{Paredes-Fortuny},
  G.~E. {Romero}, D.~{Khangulyan}, {Gamma rays from clumpy wind-jet
  interactions in high-mass microquasars}, \aap 604 (2017) A39.
\newblock \href {http://arxiv.org/abs/1701.05028} {\path{arXiv:1701.05028}},
  \href {https://doi.org/10.1051/0004-6361/201630060}
  {\path{doi:10.1051/0004-6361/201630060}}.

\bibitem{2024NatAs...8.1031V}
A.~{Veledina}, F.~{Muleri}, J.~{Poutanen} et~al., {Cygnus X-3 revealed as a
  Galactic ultraluminous X-ray source by IXPE}, Nature Astronomy 8 (2024)
  1031--1046.
\newblock \href {http://arxiv.org/abs/2303.01174} {\path{arXiv:2303.01174}},
  \href {https://doi.org/10.1038/s41550-024-02294-9}
  {\path{doi:10.1038/s41550-024-02294-9}}.

\bibitem{2004vhec.book.....A}
F.~A. {Aharonian}, {Very high energy cosmic gamma radiation : a crucial window
  on the extreme Universe}, World Scientific, 2004.
\newblock \href {https://doi.org/10.1142/4657} {\path{doi:10.1142/4657}}.

\bibitem{1985ICRC....9..407H}
A.~M. {Hillas}, {Why is Cygnus X-3 (with related sources) a highlight of
  cosmic-ray astrophysics?}, in: F.~C. {Jones} (Ed.), 19th International Cosmic
  Ray Conference (ICRC19), Volume 9, Vol.~9 of International Cosmic Ray
  Conference, 1985, p. 407.

\bibitem{2018A&A...612A..10H}
{H.~E.~S.~S. Collaboration}, H.~{Abdalla}, A.~{Abramowski} et~al., {A search
  for very high-energy flares from the microquasars GRS 1915+105, Circinus X-1,
  and V4641 Sgr using contemporaneous H.E.S.S. and RXTE observations}, \aap 612
  (2018) A10.
\newblock \href {http://arxiv.org/abs/1607.04613} {\path{arXiv:1607.04613}},
  \href {https://doi.org/10.1051/0004-6361/201527773}
  {\path{doi:10.1051/0004-6361/201527773}}.

\bibitem{2018A&A...612A..14M}
{MAGIC Collaboration}, M.~L. {Ahnen}, S.~{Ansoldi} et~al., {Constraints on
  particle acceleration in SS433/W50 from MAGIC and H.E.S.S. observations},
  \aap 612 (2018) A14.
\newblock \href {http://arxiv.org/abs/1707.03658} {\path{arXiv:1707.03658}},
  \href {https://doi.org/10.1051/0004-6361/201731169}
  {\path{doi:10.1051/0004-6361/201731169}}.

\bibitem{2018Natur.562...82A}
A.~U. {Abeysekara}, A.~{Albert}, R.~{Alfaro} et~al., {Very-high-energy particle
  acceleration powered by the jets of the microquasar SS 433}, \nat 562~(7725)
  (2018) 82--85.
\newblock \href {http://arxiv.org/abs/1810.01892} {\path{arXiv:1810.01892}},
  \href {https://doi.org/10.1038/s41586-018-0565-5}
  {\path{doi:10.1038/s41586-018-0565-5}}.

\bibitem{2024arXiv241008988L}
{LHAASO Collaboration}, {Ultrahigh-Energy Gamma-ray Emission Associated with
  Black Hole-Jet Systems}, arXiv e-prints (2024) arXiv:2410.08988\href
  {http://arxiv.org/abs/2410.08988} {\path{arXiv:2410.08988}}, \href
  {https://doi.org/10.48550/arXiv.2410.08988}
  {\path{doi:10.48550/arXiv.2410.08988}}.

\bibitem{2024ApJ...976...30A}
R.~{Alfaro}, C.~{Alvarez}, J.~C. {Arteaga-Vel{\'a}zquez} et~al., {Spectral
  Study of Very-high-energy Gamma Rays from SS 433 with HAWC}, \apj 976~(1)
  (2024) 30.
\newblock \href {http://arxiv.org/abs/2410.21796} {\path{arXiv:2410.21796}},
  \href {https://doi.org/10.3847/1538-4357/ad7e1b}
  {\path{doi:10.3847/1538-4357/ad7e1b}}.

\bibitem{2024Sci...383..402H}
{H.~E.~S.~S. Collaboration}, F.~{Aharonian}, F.~{Ait Benkhali} et~al.,
  {Acceleration and transport of relativistic electrons in the jets of the
  microquasar SS 433}, Science 383~(6681) (2024) 402--406.
\newblock \href {http://arxiv.org/abs/2401.16019} {\path{arXiv:2401.16019}},
  \href {https://doi.org/10.1126/science.adi2048}
  {\path{doi:10.1126/science.adi2048}}.

\bibitem{2025arXiv251110537A}
A.~{Acharyya}, F.~{Aharonian}, H.~{Ashkar} et~al., {Constraining the nature of
  the most extreme Galactic particle accelerator. H.E.S.S. observations of the
  microquasar V4641 Sgr}, arXiv e-prints (2025) arXiv:2511.10537\href
  {http://arxiv.org/abs/2511.10537} {\path{arXiv:2511.10537}}, \href
  {https://doi.org/10.48550/arXiv.2511.10537}
  {\path{doi:10.48550/arXiv.2511.10537}}.

\bibitem{2025MNRAS.tmp.2104O}
P.~{O'Neill}, A.~{Ingram}, E.~{Nathan}, G.~{Mastroserio}, M.~{van der Klis},
  M.~{Lucchini}, J.~{Mitchell}, {X-ray reverberation black hole mass and
  distance estimates of Cygnus X-1}, \mnras (Dec. 2025).
\newblock \href {http://arxiv.org/abs/2501.12788} {\path{arXiv:2501.12788}},
  \href {https://doi.org/10.1093/mnras/staf2232}
  {\path{doi:10.1093/mnras/staf2232}}.

\bibitem{2021Sci...371.1046M}
J.~C.~A. {Miller-Jones}, A.~{Bahramian}, J.~A. {Orosz}, I.~{Mandel}, L.~{Gou},
  T.~J. {Maccarone}, C.~J. {Neijssel}, X.~{Zhao}, J.~{Zi{\'o}{\l}kowski}, M.~J.
  {Reid}, P.~{Uttley}, X.~{Zheng}, D.-Y. {Byun}, R.~{Dodson}, V.~{Grinberg},
  T.~{Jung}, J.-S. {Kim}, B.~{Marcote}, S.~{Markoff}, M.~J. {Rioja}, A.~P.
  {Rushton}, D.~M. {Russell}, G.~R. {Sivakoff}, A.~J. {Tetarenko}, V.~{Tudose},
  J.~{Wilms}, {Cygnus X-1 contains a 21-solar mass black
  hole{\textemdash}Implications for massive star winds}, Science 371~(6533)
  (2021) 1046--1049.
\newblock \href {http://arxiv.org/abs/2102.09091} {\path{arXiv:2102.09091}},
  \href {https://doi.org/10.1126/science.abb3363}
  {\path{doi:10.1126/science.abb3363}}.

\bibitem{2017MNRAS.472.3474A}
M.~L. {Ahnen}, S.~{Ansoldi}, L.~A. {Antonelli} et~al., {Search for very
  high-energy gamma-ray emission from the microquasar Cygnus X-1 with the MAGIC
  telescopes}, \mnras 472~(3) (2017) 3474--3485.
\newblock \href {http://arxiv.org/abs/1708.03689} {\path{arXiv:1708.03689}},
  \href {https://doi.org/10.1093/mnras/stx2087}
  {\path{doi:10.1093/mnras/stx2087}}.

\bibitem{2013MNRAS.434.2380M}
D.~{Malyshev}, A.~A. {Zdziarski}, M.~{Chernyakova}, {High-energy gamma-ray
  emission from Cyg X-1 measured by Fermi and its theoretical implications},
  \mnras 434~(3) (2013) 2380--2389.
\newblock \href {http://arxiv.org/abs/1305.5920} {\path{arXiv:1305.5920}},
  \href {https://doi.org/10.1093/mnras/stt1184}
  {\path{doi:10.1093/mnras/stt1184}}.

\bibitem{2021ApJ...912L...4A}
A.~{Albert}, R.~{Alfaro}, C.~{Alvarez} et~al., {HAWC Search for High-mass
  Microquasars}, \apjl 912~(1) (2021) L4.
\newblock \href {http://arxiv.org/abs/2101.08945} {\path{arXiv:2101.08945}},
  \href {https://doi.org/10.3847/2041-8213/abf35a}
  {\path{doi:10.3847/2041-8213/abf35a}}.

\bibitem{2023ApJ...959...85R}
M.~J. {Reid}, J.~C.~A. {Miller-Jones}, {On the Distances to the X-Ray Binaries
  Cygnus X-3 and GRS 1915+105}, \apj 959~(2) (2023) 85.
\newblock \href {http://arxiv.org/abs/2309.15027} {\path{arXiv:2309.15027}},
  \href {https://doi.org/10.3847/1538-4357/acfe0c}
  {\path{doi:10.3847/1538-4357/acfe0c}}.

\bibitem{2010ApJ...718..488S}
C.~R. {Shrader}, L.~{Titarchuk}, N.~{Shaposhnikov}, {New Evidence for a Black
  Hole in the Compact Binary Cygnus X-3}, \apj 718~(1) (2010) 488--493.
\newblock \href {http://arxiv.org/abs/1005.5362} {\path{arXiv:1005.5362}},
  \href {https://doi.org/10.1088/0004-637X/718/1/488}
  {\path{doi:10.1088/0004-637X/718/1/488}}.

\bibitem{2013MNRAS.429L.104Z}
A.~A. {Zdziarski}, J.~{Mikolajewska}, K.~{Belczynski}, {Cyg X-3: a low-mass
  black hole or a neutron star.}, \mnras 429 (2013) L104--L108.
\newblock \href {http://arxiv.org/abs/1208.5455} {\path{arXiv:1208.5455}},
  \href {https://doi.org/10.1093/mnrasl/sls035}
  {\path{doi:10.1093/mnrasl/sls035}}.

\bibitem{2017MNRAS.472.2181K}
K.~I.~I. {Koljonen}, T.~J. {Maccarone}, {Gemini/GNIRS infrared spectroscopy of
  the Wolf-Rayet stellar wind in Cygnus X-3}, \mnras 472~(2) (2017) 2181--2195.
\newblock \href {http://arxiv.org/abs/1708.04050} {\path{arXiv:1708.04050}},
  \href {https://doi.org/10.1093/mnras/stx2106}
  {\path{doi:10.1093/mnras/stx2106}}.

\bibitem{2025arXiv251216638L}
{LHAASO Collaboration}, {Cygnus X-3: A variable petaelectronvolt gamma-ray
  source}, arXiv e-prints (2025) arXiv:2512.16638\href
  {http://arxiv.org/abs/2512.16638} {\path{arXiv:2512.16638}}, \href
  {https://doi.org/10.48550/arXiv.2512.16638}
  {\path{doi:10.48550/arXiv.2512.16638}}.

\bibitem{2012A&A...545A.110P}
G.~{Piano}, M.~{Tavani}, V.~{Vittorini} et~al., {The AGILE monitoring of Cygnus
  X-3: transient gamma-ray emission and spectral constraints}, \aap 545 (2012)
  A110.
\newblock \href {http://arxiv.org/abs/1207.6288} {\path{arXiv:1207.6288}},
  \href {https://doi.org/10.1051/0004-6361/201219145}
  {\path{doi:10.1051/0004-6361/201219145}}.

\bibitem{2012MNRAS.421.2947C}
S.~{Corbel}, G.~{Dubus}, J.~A. {Tomsick}, A.~{Szostek}, R.~H.~D. {Corbet},
  J.~C.~A. {Miller-Jones}, J.~L. {Richards}, G.~{Pooley}, S.~{Trushkin},
  R.~{Dubois}, A.~B. {Hill}, M.~{Kerr}, W.~{Max-Moerbeck}, A.~C.~S. {Readhead},
  A.~{Bodaghee}, V.~{Tudose}, D.~{Parent}, J.~{Wilms}, K.~{Pottschmidt}, {A
  giant radio flare from Cygnus X-3 with associated {\ensuremath{\gamma}}-ray
  emission}, \mnras 421~(4) (2012) 2947--2955.
\newblock \href {http://arxiv.org/abs/1201.3356} {\path{arXiv:1201.3356}},
  \href {https://doi.org/10.1111/j.1365-2966.2012.20517.x}
  {\path{doi:10.1111/j.1365-2966.2012.20517.x}}.

\bibitem{2010ApJ...721..843A}
J.~{Aleksi{\'c}}, L.~A. {Antonelli}, P.~{Antoranz} et~al., {Magic Constraints
  on {\ensuremath{\gamma}}-ray Emission from Cygnus X-3}, \apj 721~(1) (2010)
  843--855.
\newblock \href {http://arxiv.org/abs/1005.0740} {\path{arXiv:1005.0740}},
  \href {https://doi.org/10.1088/0004-637X/721/1/843}
  {\path{doi:10.1088/0004-637X/721/1/843}}.

\bibitem{2013ApJ...779..150A}
S.~{Archambault}, M.~{Beilicke}, W.~{Benbow} et~al., {VERITAS Observations of
  the Microquasar Cygnus X-3}, \apj 779~(2) (2013) 150.
\newblock \href {http://arxiv.org/abs/1311.0919} {\path{arXiv:1311.0919}},
  \href {https://doi.org/10.1088/0004-637X/779/2/150}
  {\path{doi:10.1088/0004-637X/779/2/150}}.

\bibitem{BarriosJiménez:2025yf}
L.~Barrios~Jiménez, E.~Molina, M.~Carretero-Castrillo, J.~Becerra~González,
  M.~Ribó, J.~M. Paredes, {Results of the historical observations of the
  microquasar Cygnus X-3 with the MAGIC telescopes.}, in: Proceedings of 39th
  International Cosmic Ray Conference {\textemdash} PoS(ICRC2025), Vol. 501,
  2025, p. 564.
\newblock \href {https://doi.org/10.22323/1.501.0564}
  {\path{doi:10.22323/1.501.0564}}.

\bibitem{2017ApJ...839...84P}
G.~{Piano}, P.~{Munar-Adrover}, F.~{Verrecchia}, M.~{Tavani}, S.~A. {Trushkin},
  {High-energy Gamma-Ray Activity from V404 Cygni Detected by AGILE during the
  2015 June Outburst}, \apj 839~(2) (2017) 84.
\newblock \href {http://arxiv.org/abs/1703.10085} {\path{arXiv:1703.10085}},
  \href {https://doi.org/10.3847/1538-4357/aa6796}
  {\path{doi:10.3847/1538-4357/aa6796}}.

\bibitem{2021MNRAS.506.6029H}
M.~{Harvey}, C.~B. {Rulten}, P.~M. {Chadwick}, {V404 Cygni with Fermi-LAT},
  \mnras 506~(4) (2021) 6029--6038.
\newblock \href {http://arxiv.org/abs/2107.09395} {\path{arXiv:2107.09395}},
  \href {https://doi.org/10.1093/mnras/stab2097}
  {\path{doi:10.1093/mnras/stab2097}}.

\bibitem{2017MNRAS.471.1688A}
M.~L. {Ahnen}, S.~{Ansoldi}, L.~A. {Antonelli} et~al., {MAGIC observations of
  the microquasar V404 Cygni during the 2015 outburst}, \mnras 471~(2) (2017)
  1688--1693.
\newblock \href {http://arxiv.org/abs/1707.00887} {\path{arXiv:1707.00887}},
  \href {https://doi.org/10.1093/mnras/stx1690}
  {\path{doi:10.1093/mnras/stx1690}}.

\bibitem{2016ApJ...831..113A}
A.~{Archer}, W.~{Benbow}, R.~{Bird} et~al., {Very High Energy Observations of
  the Binaries V 404 Cyg and 4U 0115+634 during Giant X-Ray Outbursts}, \apj
  831~(1) (2016) 113.
\newblock \href {http://arxiv.org/abs/1608.06464} {\path{arXiv:1608.06464}},
  \href {https://doi.org/10.3847/0004-637X/831/1/113}
  {\path{doi:10.3847/0004-637X/831/1/113}}.

\bibitem{2011ApJ...735L...5A}
J.~{Aleksi{\'c}}, E.~A. {Alvarez}, L.~A. {Antonelli} et~al., {A Search for Very
  High Energy Gamma-Ray Emission from Scorpius X-1 with the Magic Telescopes},
  \apjl 735~(1) (2011) L5.
\newblock \href {http://arxiv.org/abs/1103.5677} {\path{arXiv:1103.5677}},
  \href {https://doi.org/10.1088/2041-8205/735/1/L5}
  {\path{doi:10.1088/2041-8205/735/1/L5}}.

\bibitem{1994ApJS...92..567M}
I.~V. {Moskalenko}, S.~{Karakula}, {Light curves of close binaries in TeV
  energy region}, \apjs 92~(2) (1994) 567--573.
\newblock \href {https://doi.org/10.1086/192017} {\path{doi:10.1086/192017}}.

\bibitem{2010A&A...520A..83H}
{H.~E.~S.~S. Collaboration}, A.~{Abramowski}, F.~{Acero} et~al., {VHE
  {\ensuremath{\gamma}}-ray emission of PKS 2155-304: spectral and temporal
  variability}, \aap 520 (2010) A83.
\newblock \href {http://arxiv.org/abs/1005.3702} {\path{arXiv:1005.3702}},
  \href {https://doi.org/10.1051/0004-6361/201014484}
  {\path{doi:10.1051/0004-6361/201014484}}.

\bibitem{2012A&A...539A.149H}
{H.~E.~S.~S. Collaboration}, A.~{Abramowski}, F.~{Acero} et~al., {A
  multiwavelength view of the flaring state of PKS 2155-304 in 2006}, \aap 539
  (2012) A149.
\newblock \href {http://arxiv.org/abs/1201.4135} {\path{arXiv:1201.4135}},
  \href {https://doi.org/10.1051/0004-6361/201117509}
  {\path{doi:10.1051/0004-6361/201117509}}.

\bibitem{2017A&A...598A..39H}
{H.~E.~S.~S. Collaboration}, H.~{Abdalla}, A.~{Abramowski} et~al.,
  {Characterizing the {\ensuremath{\gamma}}-ray long-term variability of PKS
  2155-304 with H.E.S.S. and Fermi-LAT}, \aap 598 (2017) A39.
\newblock \href {http://arxiv.org/abs/1610.03311} {\path{arXiv:1610.03311}},
  \href {https://doi.org/10.1051/0004-6361/201629419}
  {\path{doi:10.1051/0004-6361/201629419}}.

\bibitem{2009ApJ...696L.150A}
F.~{Aharonian}, A.~G. {Akhperjanian}, G.~{Anton} et~al., {Simultaneous
  Observations of PKS 2155-304 with HESS, Fermi, RXTE, and Atom: Spectral
  Energy Distributions and Variability in a Low State}, \apjl 696~(2) (2009)
  L150--L155.
\newblock \href {http://arxiv.org/abs/0903.2924} {\path{arXiv:0903.2924}},
  \href {https://doi.org/10.1088/0004-637X/696/2/L150}
  {\path{doi:10.1088/0004-637X/696/2/L150}}.

\bibitem{2025arXiv251004803N}
L.~{Nikoli{\'c}}, G.~{Verna}, M.~{Manganaro}, G.~{Bonnoli}, I.~{Agudo},
  G.~{Silvestri}, D.~{Cerasole}, F.~{Schiavone}, F.~{Podobnik},
  J.~{Otero-Santos}, {Recent observations of PKS 2155-304 with MAGIC and LST-1
  in a multi-wavelength context}, arXiv e-prints (2025) arXiv:2510.04803\href
  {http://arxiv.org/abs/2510.04803} {\path{arXiv:2510.04803}}, \href
  {https://doi.org/10.48550/arXiv.2510.04803}
  {\path{doi:10.48550/arXiv.2510.04803}}.

\bibitem{2026JHEAp..5000472B}
A.~M. {Bharathan}, C.~S. {Stalin}, S.~{Sahayanathan}, B.~{Mathew}, {Clues on
  the X-ray emission mechanism of blazars PKS 2155‑304 and 3C 454.3 through
  polarization studies}, Journal of High Energy Astrophysics 50 (2026) 100472.
\newblock \href {http://arxiv.org/abs/2509.16976} {\path{arXiv:2509.16976}},
  \href {https://doi.org/10.1016/j.jheap.2025.100472}
  {\path{doi:10.1016/j.jheap.2025.100472}}.

\bibitem{2010ApJ...718..455S}
E.~{Striani}, S.~{Vercellone}, M.~{Tavani} et~al., {The Extraordinary Gamma-ray
  Flare of the Blazar 3C 454.3}, \apj 718~(1) (2010) 455--459.
\newblock \href {http://arxiv.org/abs/1005.4891} {\path{arXiv:1005.4891}},
  \href {https://doi.org/10.1088/0004-637X/718/1/455}
  {\path{doi:10.1088/0004-637X/718/1/455}}.

\bibitem{2011ApJ...736L..38V}
S.~{Vercellone}, E.~{Striani}, V.~{Vittorini} et~al., {The Brightest Gamma-Ray
  Flaring Blazar in the Sky: AGILE and Multi-wavelength Observations of 3C
  454.3 During 2010 November}, \apjl 736~(2) (2011) L38.
\newblock \href {http://arxiv.org/abs/1106.5162} {\path{arXiv:1106.5162}},
  \href {https://doi.org/10.1088/2041-8205/736/2/L38}
  {\path{doi:10.1088/2041-8205/736/2/L38}}.

\bibitem{2011ApJ...733L..26A}
A.~A. {Abdo}, M.~{Ackermann}, M.~{Ajello} et~al., {Fermi Gamma-ray Space
  Telescope Observations of the Gamma-ray Outburst from 3C454.3 in November
  2010}, \apjl 733~(2) (2011) L26.
\newblock \href {http://arxiv.org/abs/1102.0277} {\path{arXiv:1102.0277}},
  \href {https://doi.org/10.1088/2041-8205/733/2/L26}
  {\path{doi:10.1088/2041-8205/733/2/L26}}.

\bibitem{2010ApJ...712..405V}
S.~{Vercellone}, F.~{D'Ammando}, V.~{Vittorini} et~al., {Multiwavelength
  Observations of 3C 454.3. III. Eighteen Months of Agile Monitoring of the
  ``Crazy Diamond''}, \apj 712~(1) (2010) 405--420.
\newblock \href {http://arxiv.org/abs/1002.1020} {\path{arXiv:1002.1020}},
  \href {https://doi.org/10.1088/0004-637X/712/1/405}
  {\path{doi:10.1088/0004-637X/712/1/405}}.

\bibitem{2016MNRAS.458..354C}
R.~T. {Coogan}, A.~M. {Brown}, P.~M. {Chadwick}, {Localizing the
  {\ensuremath{\gamma}}-ray emission region during the 2014 June outburst of 3C
  454.3}, \mnras 458~(1) (2016) 354--365.
\newblock \href {http://arxiv.org/abs/1601.07180} {\path{arXiv:1601.07180}},
  \href {https://doi.org/10.1093/mnras/stw199}
  {\path{doi:10.1093/mnras/stw199}}.

\bibitem{2020ApJS..248....8D}
A.~K. {Das}, R.~{Prince}, N.~{Gupta}, {Gamma-Ray Flares in the Long-term Light
  Curve of 3C 454.3}, \apjs 248~(1) (2020) 8.
\newblock \href {http://arxiv.org/abs/2003.08266} {\path{arXiv:2003.08266}},
  \href {https://doi.org/10.3847/1538-4365/ab80c3}
  {\path{doi:10.3847/1538-4365/ab80c3}}.

\bibitem{2009A&A...498...83A}
H.~{Anderhub}, L.~A. {Antonelli}, P.~{Antoranz} et~al., {MAGIC upper limits to
  the VHE gamma-ray flux of 3C 454.3 in high emission state}, \aap 498~(1)
  (2009) 83--87.
\newblock \href {http://arxiv.org/abs/0811.1680} {\path{arXiv:0811.1680}},
  \href {https://doi.org/10.1051/0004-6361/200811326}
  {\path{doi:10.1051/0004-6361/200811326}}.

\bibitem{2010ApJ...723L.207A}
J.~{Aleksi{\'c}}, L.~A. {Antonelli}, P.~{Antoranz} et~al., {Detection of Very
  High Energy {\ensuremath{\gamma}}-ray Emission from the Perseus Cluster
  Head-Tail Galaxy IC 310 by the MAGIC Telescopes}, \apjl 723~(2) (2010)
  L207--L212.
\newblock \href {http://arxiv.org/abs/1009.2155} {\path{arXiv:1009.2155}},
  \href {https://doi.org/10.1088/2041-8205/723/2/L207}
  {\path{doi:10.1088/2041-8205/723/2/L207}}.

\bibitem{2017A&A...603A..25A}
M.~L. {Ahnen}, S.~{Ansoldi}, L.~A. {Antonelli} et~al., {First multi-wavelength
  campaign on the gamma-ray-loud active galaxy IC 310}, \aap 603 (2017) A25.
\newblock \href {http://arxiv.org/abs/1703.07651} {\path{arXiv:1703.07651}},
  \href {https://doi.org/10.1051/0004-6361/201630347}
  {\path{doi:10.1051/0004-6361/201630347}}.

\bibitem{2014Sci...346.1080A}
J.~{Aleksi{\'c}}, S.~{Ansoldi}, L.~A. {Antonelli} et~al., {Black hole lightning
  due to particle acceleration at subhorizon scales}, Science 346~(6213) (2014)
  1080--1084.
\newblock \href {http://arxiv.org/abs/1412.4936} {\path{arXiv:1412.4936}},
  \href {https://doi.org/10.1126/science.1256183}
  {\path{doi:10.1126/science.1256183}}.

\bibitem{2019MNRAS.485.3277G}
J.~A. {Graham}, A.~M. {Brown}, P.~M. {Chadwick}, {Fermi-LAT observations of
  extreme spectral variability in IC 310}, \mnras 485~(3) (2019) 3277--3287.
\newblock \href {http://arxiv.org/abs/1903.07897} {\path{arXiv:1903.07897}},
  \href {https://doi.org/10.1093/mnras/stz588}
  {\path{doi:10.1093/mnras/stz588}}.

\bibitem{2009MNRAS.395L..29G}
D.~{Giannios}, D.~A. {Uzdensky}, M.~C. {Begelman}, {Fast TeV variability in
  blazars: jets in a jet}, \mnras 395~(1) (2009) L29--L33.
\newblock \href {http://arxiv.org/abs/0901.1877} {\path{arXiv:0901.1877}},
  \href {https://doi.org/10.1111/j.1745-3933.2009.00635.x}
  {\path{doi:10.1111/j.1745-3933.2009.00635.x}}.

\bibitem{2016ApJ...818...50H}
K.~{Hirotani}, H.-Y. {Pu}, {Energetic Gamma Radiation from Rapidly Rotating
  Black Holes}, \apj 818~(1) (2016) 50.
\newblock \href {http://arxiv.org/abs/1512.05026} {\path{arXiv:1512.05026}},
  \href {https://doi.org/10.3847/0004-637X/818/1/50}
  {\path{doi:10.3847/0004-637X/818/1/50}}.

\bibitem{2011ApJ...730..123L}
A.~{Levinson}, F.~{Rieger}, {Variable TeV Emission as a Manifestation of Jet
  Formation in M87?}, \apj 730~(2) (2011) 123.
\newblock \href {http://arxiv.org/abs/1011.5319} {\path{arXiv:1011.5319}},
  \href {https://doi.org/10.1088/0004-637X/730/2/123}
  {\path{doi:10.1088/0004-637X/730/2/123}}.

\bibitem{2010ApJ...724.1517B}
M.~V. {Barkov}, F.~A. {Aharonian}, V.~{Bosch-Ramon}, {Gamma-ray Flares from Red
  Giant/Jet Interactions in Active Galactic Nuclei}, \apj 724~(2) (2010)
  1517--1523.
\newblock \href {http://arxiv.org/abs/1005.5252} {\path{arXiv:1005.5252}},
  \href {https://doi.org/10.1088/0004-637X/724/2/1517}
  {\path{doi:10.1088/0004-637X/724/2/1517}}.

\bibitem{2012ApJ...749..119B}
M.~V. {Barkov}, F.~A. {Aharonian}, S.~V. {Bogovalov}, S.~R. {Kelner},
  D.~{Khangulyan}, {Rapid TeV Variability in Blazars as a Result of Jet-Star
  Interaction}, \apj 749~(2) (2012) 119.
\newblock \href {http://arxiv.org/abs/1012.1787} {\path{arXiv:1012.1787}},
  \href {https://doi.org/10.1088/0004-637X/749/2/119}
  {\path{doi:10.1088/0004-637X/749/2/119}}.

\bibitem{2010A&A...522A..97A}
A.~T. {Araudo}, V.~{Bosch-Ramon}, G.~E. {Romero}, {Gamma rays from cloud
  penetration at the base of AGN jets}, \aap 522 (2010) A97.
\newblock \href {http://arxiv.org/abs/1007.2199} {\path{arXiv:1007.2199}},
  \href {https://doi.org/10.1051/0004-6361/201014660}
  {\path{doi:10.1051/0004-6361/201014660}}.

\bibitem{2013MNRAS.436.3626A}
A.~T. {Araudo}, V.~{Bosch-Ramon}, G.~E. {Romero}, {Gamma-ray emission from
  massive stars interacting with active galactic nuclei jets}, \mnras 436~(4)
  (2013) 3626--3639.
\newblock \href {http://arxiv.org/abs/1309.7114} {\path{arXiv:1309.7114}},
  \href {https://doi.org/10.1093/mnras/stt1840}
  {\path{doi:10.1093/mnras/stt1840}}.

\bibitem{2018MNRAS.481.1455K}
D.~{Khangulyan}, V.~{Bosch-Ramon}, Y.~{Uchiyama}, {Inverse Compton emission
  from relativistic jets in binary systems}, \mnras 481~(2) (2018) 1455--1468.
\newblock \href {http://arxiv.org/abs/1808.09628} {\path{arXiv:1808.09628}},
  \href {https://doi.org/10.1093/mnras/sty2356}
  {\path{doi:10.1093/mnras/sty2356}}.

\bibitem{2019A&A...627A.100H}
{H.~E.~S.~S. Collaboration}, H.~{Abdalla}, F.~{Aharonian} et~al., {H.E.S.S. and
  Suzaku observations of the Vela X pulsar wind nebula}, \aap 627 (2019) A100.
\newblock \href {http://arxiv.org/abs/1905.07975} {\path{arXiv:1905.07975}},
  \href {https://doi.org/10.1051/0004-6361/201935458}
  {\path{doi:10.1051/0004-6361/201935458}}.

\bibitem{2017MNRAS.470.2539P}
C.~C. {Popescu}, R.~{Yang}, R.~J. {Tuffs}, G.~{Natale}, M.~{Rushton},
  F.~{Aharonian}, {A radiation transfer model for the Milky Way: I. Radiation
  fields and application to high-energy astrophysics}, \mnras 470~(3) (2017)
  2539--2558.
\newblock \href {http://arxiv.org/abs/1705.06652} {\path{arXiv:1705.06652}},
  \href {https://doi.org/10.1093/mnras/stx1282}
  {\path{doi:10.1093/mnras/stx1282}}.

\bibitem{1996MNRAS.278..525A}
A.~M. {Atoyan}, F.~A. {Aharonian}, {On the mechanisms of gamma radiation in the
  Crab Nebula}, \mnras 278~(2) (1996) 525--541.
\newblock \href {https://doi.org/10.1093/mnras/278.2.525}
  {\path{doi:10.1093/mnras/278.2.525}}.

\bibitem{2011Sci...331..736T}
M.~{Tavani}, A.~{Bulgarelli}, V.~{Vittorini} et~al., {Discovery of Powerful
  Gamma-Ray Flares from the Crab Nebula}, Science 331~(6018) (2011) 736.
\newblock \href {http://arxiv.org/abs/1101.2311} {\path{arXiv:1101.2311}},
  \href {https://doi.org/10.1126/science.1200083}
  {\path{doi:10.1126/science.1200083}}.

\bibitem{2011Sci...331..739A}
A.~A. {Abdo}, M.~{Ackermann}, M.~{Ajello} et~al., {Gamma-Ray Flares from the
  Crab Nebula}, Science 331~(6018) (2011) 739.
\newblock \href {http://arxiv.org/abs/1011.3855} {\path{arXiv:1011.3855}},
  \href {https://doi.org/10.1126/science.1199705}
  {\path{doi:10.1126/science.1199705}}.

\bibitem{2012ApJ...749...26B}
R.~{Buehler}, J.~D. {Scargle}, R.~D. {Blandford}, L.~{Baldini}, M.~G. {Baring},
  A.~{Belfiore}, E.~{Charles}, J.~{Chiang}, F.~{D'Ammando}, C.~D. {Dermer},
  S.~{Funk}, J.~E. {Grove}, A.~K. {Harding}, E.~{Hays}, M.~{Kerr},
  F.~{Massaro}, M.~N. {Mazziotta}, R.~W. {Romani}, P.~M. {Saz Parkinson}, A.~F.
  {Tennant}, M.~C. {Weisskopf}, {Gamma-Ray Activity in the Crab Nebula: The
  Exceptional Flare of 2011 April}, \apj 749~(1) (2012) 26.
\newblock \href {http://arxiv.org/abs/1112.1979} {\path{arXiv:1112.1979}},
  \href {https://doi.org/10.1088/0004-637X/749/1/26}
  {\path{doi:10.1088/0004-637X/749/1/26}}.

\bibitem{2020ApJ...897...33A}
M.~{Arakawa}, M.~{Hayashida}, D.~{Khangulyan}, Y.~{Uchiyama}, {Detection of
  Small Flares from the Crab Nebula with Fermi-LAT}, \apj 897~(1) (2020) 33.
\newblock \href {http://arxiv.org/abs/2005.07958} {\path{arXiv:2005.07958}},
  \href {https://doi.org/10.3847/1538-4357/ab9368}
  {\path{doi:10.3847/1538-4357/ab9368}}.

\bibitem{2014RPPh...77f6901B}
R.~{B{\"u}hler}, R.~{Blandford}, {The surprising Crab pulsar and its nebula: a
  review}, Reports on Progress in Physics 77~(6) (2014) 066901.
\newblock \href {http://arxiv.org/abs/1309.7046} {\path{arXiv:1309.7046}},
  \href {https://doi.org/10.1088/0034-4885/77/6/066901}
  {\path{doi:10.1088/0034-4885/77/6/066901}}.

\bibitem{2011ApJ...737L..40U}
D.~A. {Uzdensky}, B.~{Cerutti}, M.~C. {Begelman}, {Reconnection-powered Linear
  Accelerator and Gamma-Ray Flares in the Crab Nebula}, \apjl 737~(2) (2011)
  L40.
\newblock \href {http://arxiv.org/abs/1105.0942} {\path{arXiv:1105.0942}},
  \href {https://doi.org/10.1088/2041-8205/737/2/L40}
  {\path{doi:10.1088/2041-8205/737/2/L40}}.

\bibitem{2012ApJ...746..148C}
B.~{Cerutti}, D.~A. {Uzdensky}, M.~C. {Begelman}, {Extreme Particle
  Acceleration in Magnetic Reconnection Layers: Application to the Gamma-Ray
  Flares in the Crab Nebula}, \apj 746~(2) (2012) 148.
\newblock \href {http://arxiv.org/abs/1110.0557} {\path{arXiv:1110.0557}},
  \href {https://doi.org/10.1088/0004-637X/746/2/148}
  {\path{doi:10.1088/0004-637X/746/2/148}}.

\bibitem{2014A&A...562L...4H}
{H.~E.~S.~S. Collaboration}, A.~{Abramowski}, F.~{Aharonian} et~al., {H.E.S.S.
  observations of the Crab during its March 2013 GeV gamma-ray flare}, \aap 562
  (2014) L4.
\newblock \href {http://arxiv.org/abs/1311.3187} {\path{arXiv:1311.3187}},
  \href {https://doi.org/10.1051/0004-6361/201323013}
  {\path{doi:10.1051/0004-6361/201323013}}.

\bibitem{2014ApJ...781L..11A}
E.~{Aliu}, S.~{Archambault}, T.~{Aune} et~al., {A Search for Enhanced Very High
  Energy Gamma-Ray Emission from the 2013 March Crab Nebula Flare}, \apjl
  781~(1) (2014) L11.
\newblock \href {http://arxiv.org/abs/1309.5949} {\path{arXiv:1309.5949}},
  \href {https://doi.org/10.1088/2041-8205/781/1/L11}
  {\path{doi:10.1088/2041-8205/781/1/L11}}.

\bibitem{2020A&A...634A..25M}
{MAGIC Collaboration}, M.~L. {Ahnen}, S.~{Ansoldi} et~al., {Statistics of VHE
  {\ensuremath{\gamma}}-rays in temporal association with radio giant pulses
  from the Crab pulsar}, \aap 634 (2020) A25.
\newblock \href {http://arxiv.org/abs/1911.00634} {\path{arXiv:1911.00634}},
  \href {https://doi.org/10.1051/0004-6361/201833555}
  {\path{doi:10.1051/0004-6361/201833555}}.

\bibitem{2021Sci...372..187E}
T.~{Enoto}, T.~{Terasawa}, S.~{Kisaka} et~al., {Enhanced x-ray emission
  coinciding with giant radio pulses from the Crab Pulsar}, Science 372~(6538)
  (2021) 187--190.
\newblock \href {http://arxiv.org/abs/2104.03492} {\path{arXiv:2104.03492}},
  \href {https://doi.org/10.1126/science.abd4659}
  {\path{doi:10.1126/science.abd4659}}.

\bibitem{2021MNRAS.501..337M}
E.~{Mestre}, E.~{de O{\~n}a Wilhelmi}, D.~{Khangulyan}, R.~{Zanin}, F.~{Acero},
  D.~F. {Torres}, {The Crab nebula variability at short time-scales with the
  Cherenkov telescope array}, \mnras 501~(1) (2021) 337--346.
\newblock \href {http://arxiv.org/abs/2011.08586} {\path{arXiv:2011.08586}},
  \href {https://doi.org/10.1093/mnras/staa3599}
  {\path{doi:10.1093/mnras/staa3599}}.

\bibitem{2013apj...774...61k}
S.~R. {Kelner}, F.~A. {Aharonian}, D.~{Khangulyan}, {On the Jitter Radiation},
  \apj 774~(1) (2013) 61.
\newblock \href {http://arxiv.org/abs/1304.0493} {\path{arXiv:1304.0493}},
  \href {https://doi.org/10.1088/0004-637X/774/1/61}
  {\path{doi:10.1088/0004-637X/774/1/61}}.

\end{thebibliography}






\end{document}